\documentclass[reprint,nofootinbib,aps,nobalancelastpage,twocolumn,longbibiliography]{revtex4-2}
\usepackage{amsmath}
\usepackage{amsfonts}
\usepackage{amssymb}
\usepackage{mathtools}
\usepackage{graphicx}
\usepackage[dvipsnames]{xcolor}
\usepackage[colorlinks=True,citecolor=brown,linkcolor=myRed,urlcolor=brown]{hyperref}
\usepackage{hypcap}
\usepackage{braket}
\usepackage{yfonts}
\usepackage{bm}
\usepackage{soul}
\usepackage{dsfont}

\newcommand\eq[1]{\begin{align}#1\end{align}}

\newcommand{\nh}{N_\mathcal{H}}

\newcommand{\pd}{{\phantom\dagger}}

\newcommand{\comment}[1]{}

\newcommand{\E}{\mathcal{E}}
\newcommand{\ssh}[2][I]{s^{\scriptscriptstyle{#1}}_{#2}}

\definecolor{myBlue}{RGB}{31,119,180}
\definecolor{myOrange}{RGB}{255,127,14}
\definecolor{myGreen}{RGB}{44,160,44}
\definecolor{myRed}{RGB}{214,39,40}
\definecolor{myPurple}{RGB}{148,103,189}

\makeatletter
\def\p@figure{\color{myBlue}}
\def\p@equation{\color{myRed}}
\def\p@section{\color{myGreen}}
\makeatother

\begin{document}

\title{
The Fock-space landscape of many-body localisation
}

\author{Sthitadhi Roy}
\email{sthitadhi.roy@icts.res.in}
\affiliation{International Centre for Theoretical Sciences, Tata Institute of Fundamental Research, Bengaluru 560089, India}

\author{David E. Logan}
\email[Corresponding author~]{david.logan@chem.ox.ac.uk}
\affiliation{Physical and Theoretical Chemistry, University of Oxford, Oxford OX13QZ, United Kingdom\\and Department of Physics, Indian Institute of Science, Bengaluru 560012, India}

\date{\today}

\begin{abstract}

This article reviews recent progress in understanding the physics of many-body localisation (MBL) in disordered and interacting quantum many-body systems, from the perspective of ergodicity breaking on the associated Fock space. This approach to MBL is underpinned by mapping the dynamics of the many-body system onto that of a fictitious single particle on the high-dimensional, correlated and disordered Fock-space graph; yet, as we  elaborate, the problem is fundamentally different from that of conventional Anderson localisation on high-dimensional or hierarchical graphs. We discuss in detail the nature of eigenstate correlations on the Fock space, both static and dynamic, and in the ergodic and many-body localised phases as well as in the vicinity of the MBL transition. The latter in turn sheds light on the nature of the  transition, and motivates a scaling theory for it in terms of Fock-space based quantities. We also illustrate how these quantities can be concretely connected to real-space observables.  An overview is given of several analytical and numerical techniques which have proven important in developing a comprehensive picture. Finally, we comment on some open questions in the field of  MBL where the Fock-space approach is likely to prove insightful.

\end{abstract}

\maketitle
\onecolumngrid
\tableofcontents
\twocolumngrid


\section{Introduction \label{sec:intro}}

The presence of randomness, or disorder, is as ubiquitous in nature as it is fundamental to a host of physical phenomena in a wide variety of complex systems. These range from the physics of glasses and exotic magnetic phases of matter, to the statistical mechanics of biological systems. In the context of quantum systems, one of the most remarkable such phenomenon is that of disorder-induced localisation. In a seminal  1958 paper, Anderson~\cite{anderson1958absence} showed that non-interacting quantum particles in a lattice with disordered on-site potentials fail to show diffusion for strong disorder.  Although the importance of dimensionality was appreciated early on~\cite{mott1961theory}, it was a further two decades before the advent of the celebrated scaling theory of localisation~\cite{abrahams1979scaling}; with the realisation that in one and two dimensions, single-particle eigenstates are exponentially localised in space even for  arbitrarily weak disorder, whereas in three and higher dimensions a critical disorder strength is required for localisation to set in. In the ensuing half century or so, 
disorder-induced localisation has  very much formed a cornerstone of modern condensed matter and statistical physics~\cite{Abrahams50years}.

Yet Anderson localisation \emph{per se} is a one-body phenomenon, so a natural question arises: what is the fate
of localisation in the presence of interactions between the quantum particles? Although long recognised as an important and many-sided problem~\cite{MottAdvPhys1967,FleishmanPWA1980,lee1985disordered,Altshuler1985}, 
some twenty years ago the field acquired renewed  impetus under the umbrella of `many-body localisation' (MBL),
after two important works~\cite{gornyi2005interacting,basko2006metal} showed that interacting fermions on a disordered chain exhibit a metal-insulator transition at weak interactions, but at strictly finite temperatures; suggesting the possibility of a stable localised phase in the presence of interactions, and a novel localisation 
transition. Soon after, it was realised that this localisation in interacting, disordered systems was not restricted to weak interactions or low temperatures. In fact, 
for an isolated system in one spatial dimension, it was found numerically that excited eigenstates at arbitrarily high energy densities -- in other words eigenstates from the centre of the many-body spectrum -- could exhibit localised behaviour at sufficiently strong disorder~\citep{oganesyan2007localisation,znidaric2008many,pal2010many}. The most important outcome of this was 
the realisation that eigenstates of many-body localised systems violate the eigenstate thermalisation hypothesis (ETH)~\cite{srednicki1994chaos,deutsch1991quantum,deutsch2018eigenstate,dalessio2016from}, and therefore fall outside the paradigm of equilibrium statistical mechanics and thermodynamics. This naturally raised fundamental questions about the universal, defining features of many-body localised systems, and started the field of MBL as we know it today~\cite{nandkishore2015many,abanin2017recent,alet2018many,abanin2019colloquium}.

The central implication of the ETH is that the remainder of an isolated system acts as an effective bath for any subsystem within it, and hence local observables supported on the subsystem thermalise to a Gibbs ensemble characterised by the energy of the initial state. In this way, each eigenstate acts in effect as
a proxy for an ensemble. On the other hand, in MBL systems different parts of the system do not communicate effectively with each other, reflecting the non-ergodic nature of the eigenstates.
As a result, the rest of the system fails to act as a bath for any subsystem and hence the latter fails to thermalise. This leads to fundamental differences in the phenomenology of MBL systems compared to generic ergodic systems, as reflected in a plethora of spectral and dynamical properties~\cite{nandkishore2015many,abanin2017recent,abanin2019colloquium,alet2018many},
and now briefly reprised.

\subsection{Phenomenology of the many-body localised phase \label{sec:introphenom}}

While it is widely accepted that the MBL phase is stable in one spatial dimension~\cite{Imbrie2016PRL,imbrie2016many,imbrie2017local}, the situation in higher dimensions is thus far unresolved and remains a topic of debate~\cite{deroeck2017stability,wahl2017signatures,yao2023observation}. As such, in this review we restrict ourselves exclusively to one dimension.

Since the MBL phase and the transition to it from the ergodic phase 
occurs at the level of single eigenstates -- or eigenstates lying within a small microcanonical energy window -- the most useful diagnostics for it centre on properties of the many-body spectrum and eigenstates at arbitrary 
non-zero energy densities above the ground state. For instance, in ergodic systems the many-body energy levels repel each other, just like those in random matrices~\cite{mehta2014random}, leading to Wigner-Dyson surmises for the nearest-neighbour level-spacing distributions. In the MBL phase by contrast, level repulsion is absent,  and the level-spacings in consequence follow a Poisson distribution~\cite{oganesyan2007localisation,pal2010many,luitz2015many,atas2013distribution}.  Matrix elements of local observables in the eigenbasis of the Hamiltonian follow the form stipulated by ETH in ergodic systems, whereby the diagonal elements are smooth functions of energy and the off-diagonals are exponentially suppressed in system size~\cite{dalessio2016from}. In the MBL phase by contrast, the matrix elements are rather erratic functions of energies, as eigenstates nearby in energy can be drastically different. In a similar vein, eigenstates at arbitrary energy densities in ergodic systems typically have long-ranged, volume-law entanglement, whereas those in MBL systems have short-ranged, area-law entanglement~\cite{bauer2013area}.

As the above distinctions between ergodic and MBL systems are reflected in arbitrarily excited eigenstates, their signatures are naturally manifest dynamically as well. Initially out-of-equilibrium states evolve under ergodic dynamics such that local observables take on expectation values given by thermal Gibbs ensembles, with temperature set by the energy of the initial state. 
Dynamics in the MBL phase  by contrast ensures the system retains memory of the initial state for an 
arbitrarily long time, a fact closely related to the absence of transport of conserved
quantities~\cite{znidaric2008many,lev2015absence}.
Despite the absence of transport, entanglement in MBL systems does however grow unboundedly under 
dynamics~\cite{bardarson2012unbounded,serbyn2013universal,huang2021extensive},  albeit anomalously slowly -- logarithmically in time~\cite{bardarson2012unbounded,serbyn2013universal}, in contrast to ergodic systems where entanglement growth is usually ballistic~\cite{kim2013ballistic,lezama2019powerlaw}.

Many of the above features of MBL systems can be explained phenomenologically by hypothesising that MBL systems host an extensive number of mutually commuting local integrals of motion (LIOM)~\cite{serbyn2013local,huse2014phenomenology,ros2015integrals}. These LIOMs can be thought of as (weakly) dressed versions of the Anderson localised orbitals for weak interactions, or the trivially localised orbitals at infinitely strong disorder or vanishing interactions~\cite{imbrie2016many,imbrie2017local}; such that their support decays exponentially away from the real-space site on which they are localised. While theories based on the existence of LIOMs have led to several important insights about the MBL phase~\cite{bardarson2012unbounded,serbyn2013local,serbyn2013universal,huse2014phenomenology,bauer2013area,serbyn2014quantum}, more recent work suggests they may not paint the full picture. Several numerical results, seemingly at odds with the LIOM picture, can in fact be understood if one allows for eigenstates not just to be LIOM configurations, but also to contain resonances between them~\cite{morningstar2021avalanches,garratt2021resonances,garratt2022resonant,crowley2022constructive,long2023phenomenology}.

Within the purview of phenomenological theories, progress was also made towards understanding the nature of the MBL transition, occurring via the so-called avalanche mechanism~\cite{deroeck2017stability,thiery2018many-body}.
The essential idea is to consider an MBL system in the vicinity of the transition as a patchwork of LIOMs and local ergodic inclusions, with the latter inevitably present on entropic grounds. This insight was then bootstrapped into a coarse-grained RG flow, where either the ergodic inclusions subsume the localised regions -- leading to an ergodic phase -- or they do not and the MBL phase remains stable.  A remarkable outcome of these works was that the MBL transition was understood to be Kosterlitz-Thouless like~\cite{goremykina2019analytically,dumitrescu2018kosterlitz,morningstar2019renormalization,morningstar2020manybody}. This was also found to be consistent with numerical results and analytical arguments based on eigenstate correlations on the Fock space, which we will discuss in detail later in the article.

What is clear from the above is that the MBL phase in disordered, interacting quantum systems is a novel dynamical phase. As it breaks ergodicity in a robust fashion, fundamental questions arise pertaining to the `universal' properties of the MBL phase, as well the dynamical phase transition 
to it from an ergodic phase. Phenomenological theories, based on real-space observables, have led to key insights into the physics. At the same time, it is naturally desirable to have theories which are 
wholly rooted in the microscopic description of the systems and use the most basic quantities, such as eigenstate correlations, as building blocks of the theory. 

This is precisely where viewing the problem of many-body localisation in the landscape of Fock space emerges, not just as a useful complementary viewpoint but almost as a basic necessity. This approach 
to understanding the MBL phase and the transition has seen substantial progress in the last few years,
and an overview of where the field stands in this regard is the focus of the review.


\subsection{The Fock-space view}
The many-body Fock space provides a natural setting to treat interacting, disordered systems 
of the kind commonly encountered in questions of thermalisation or localisation. 
At heart, this arises simply because any quantum state -- be it an eigenstate of the Hamiltonian or a time-evolving state -- can be expressed as
\eq{
\ket{\psi} = \sum_{I=1}^{\nh} \psi_I\ket{I}\,,
\label{eq:state-general}
}
where $\{\ket{I}\}$ is an appropriate set of many-body basis states for the $\nh$-dimensional Fock space. The most common models for studying many-body localisation comprise interacting, disordered spins or fermions (Sec.\ \ref{sec:models}); convenient basis choices for the former include classical spin configurations polarised along a fixed direction, and, for the latter, the occupation number states.
More importantly, once the basis is chosen the Hamiltonian of such a system can be expressed quite generally as 
\eq{
H = \sum_{I}\mathcal{E}_I^{\pd}\ket{I}\bra{I}+\sum_{I\neq J}\mathcal{T}_{IJ}^{\pd}\ket{I}\bra{J}\,.
\label{eq:fs-ham-general}
}
An approach of this sort was in fact employed long ago, in the context of quantum ergodicity and
 energy localisation in high-dimensional Fermi-resonant systems~\cite{logan1990quantum,LeitnerReview2015}, as well as in regard to quasiparticle decay in disordered quantum dots~\cite{altshuler1997quasiparticle}. In Sec.~\ref{sec:models} we will discuss explicitly the (exact) mapping to the form 
Eq.~\ref{eq:fs-ham-general} for an interacting, disordered spin or fermionic Hamiltonian. For now, however, it is sufficient to realise that Eq.~\ref{eq:fs-ham-general} is nothing but a tight-binding Hamiltonian for a fictitious single particle on the associated Fock-space graph (or `lattice'). The 
many-body basis states form the nodes (or `sites') of the graph, with associated  on-site energies 
$\mathcal{E}_I$. The matrix elements of the Hamiltonian which couple different basis states, $\mathcal{T}_{IJ}$, generate edges on the graph, which amount to hopping amplitudes for the effective single 
particle. Finally, a wavefunction for this fictitious particle is of course precisely of form Eq.~\ref{eq:state-general}, so is indeed an actual many-body state of interest. This overall way of thinking 
implies that spectral and dynamical properties of the interacting, disordered system can be understood in terms of the effective single-particle problem, albeit at the cost of studying the latter on the rather complex `lattice' that is the Fock-space graph.

At this stage, it may be tempting to equate the problem on Fock space to that of Anderson (one-body) localisation on high-dimensional or hierarchical graphs, with much about the latter having been understood since early pioneering work on Anderson localisation on tree-like graphs~\cite{abou-chacra1973self}. However, a not insignificant part of this review, Sec.~\ref{sec:fscorrs}, will focus on why this temptation should be resisted, since many-body localisation on Fock-space is fundamentally quite distinct from standard Anderson localisation on high-dimensional graphs. 

At the root of this difference lies the fact that the Fock-space graph encodes the locality of the underlying real-space, locally interacting Hamiltonian. This has important consequences, the most significant of which is that
the matrix elements of the Fock-space Hamiltonian, $\{\mathcal{E}_I, \mathcal{T}_{IJ}\}$, are very strongly correlated; which in fact lies at the heart of the stability of an MBL phase~\cite{logan2019many,roy2020fock,roy2020localisation}. In addition, the distribution of the Fock-space site energies $\mathcal{E}_I$, as well as the effective connectivities on the Fock-space graph generated by the hoppings $\mathcal{T}_{IJ}$, also scale non-trivially with the system size~\cite{welsh2018simple} and hence in turn with the Fock-space dimension $\nh$. This has important consequences for the nature of an MBL phase on the Fock space, and how it is distinct from an Anderson localised phase on high-dimensional graphs~\cite{tikhonov2016anderson,tikhonov2019critical,tikhonov2019statistics}. In the latter, the eigenstates are exponentially localised on rare $O(1)$ number of branches of the graph~\cite{garciamata2017scaling,garciamata2020two}.  In the MBL phase by contrast, the eigenstates show \emph{multifractal} statistics~\cite{deluca2013ergodicity,luitz2015many,mace2019multifractal,detomasi2020rare,roy2021fockspace,tikhonov2021eigenstate},  and with strong inhomogeneity across the Fock-space graph~\cite{roy2021fockspace}. The latter can be interpreted as the analogue of the Anderson localised states living on rare branches, but with the difference that the number of such branches on the Fock-space graph can scale non-trivially with the system size. A detailed account of the eigenstate anatomy on the Fock-space graph is given in Sec.~\ref{sec:anatomy}. We will also discuss how this anatomy of eigenstate correlations is directly related to real-space local observables~\cite{roy2021fockspace} and entanglement measures~\cite{roy2022hilbert}.

From a theoretical perspective, quantities that powerfully encode ergodicity breaking on the Fock space
are the associated propagators, both local and non-local. The local propagators quantify the rate of loss of probability from a given basis state~\cite{Economoubook}, and hence provide a natural diagnostic for localisation -- indeed their one-body counterparts have long been used successfully
in the study of Anderson localisation~\cite{anderson1958absence,economous1972existence,abou-chacra1973self,ThoulessReview1974,licciardello1975study,Heinrichs1977,Stein1979,logan1987dephasing,*logan1987dipolar,logan1985anderson,*DELPGWPRB1984}
-- while non-local propagators provide a  complementary quantitative measure of how the states spread out on the Fock-space graph.  A detailed discussion of these will  be given in Sec.~\ref{sec:props}.  In particular, we will discuss
how a self-consistent mean-field theory for the local propagator~\cite{logan2019many,roy2020fock,roy2019self} was 
important in demonstrating the centrality of Fock-space correlations for the stability of the MBL phase. 
We will also discuss several approaches for computing non-local propagators, such as the Forward Scattering Approximation~\cite{pietracaprina2016forward} as well as an exact Fock-space decimation scheme~\cite{MonthusGarel2010PRB}, and how results from them revealed important features of MBL systems.

This will naturally segue into consideration (Sec.~\ref{sec:scaling}) of the MBL transition, from the perspective of  
the eigenstate anatomy on the Fock space. Specifically, we will review a broad selection of results which indicate a discontinuity in multifractal exponents across the transition~\cite{mace2019multifractal,roy2021fockspace,detomasi2020rare,sutradhar2022scaling}, and the emergence of a correlation volume on the Fock-space which diverges with an essential singularity on approaching the transition from the ergodic side~\cite{mace2019multifractal,roy2021fockspace,sutradhar2022scaling,MonthusGarel2010PRB} -- the latter suggesting consonance of these results with phenomenological theories predicting a Kosterlitz-Thouless like scenario.

Finally, Sec.~\ref{sec:conclusion} contains concluding remarks together with a brief outlook for the future.


\section{Models and their Fock-space graphs \label{sec:models}}

We begin by considering a selection of models commonly employed in the field, and relevant features of their associated Fock-space graphs.


\subsection{Standard models of many-body localisation}

The `standard model'
model for many-body localisation is often regarded as the disordered, interacting spinless fermion chain, described by the Hamiltonian~\cite{basko2006metal,gornyi2005interacting,oganesyan2007localisation,znidaric2008many,pal2010many}
\eq{
	H = t\sum_{i=1}^L[c_i^\dagger c_{i+1}^{\phantom{\dagger}} + \mathrm{h.c.}]+ \sum_{i=1}^L\epsilon_{i}^{\pd}\hat{n}_{i}^{\pd} + \sum_{i=1}^LV_{i}^{\pd}  \hat{n}_i^{\phantom{\dagger}}\hat{n}_{i+1}^{\phantom{\dagger}}\,,
	\label{eq:ham-fermion}
}
with $c_i^\dagger (c_{i}^{\pd})$ the fermionic creation (annihilation) operator at site $i$, 
and $\hat{n}_{i}=c_{i}^{\dagger}c_{i}^{\pd}$ the corresponding number operator. The disordered on-site potentials $\{\epsilon_i\}$, and the density-density interaction strengths $\{V_i\}$, are drawn from uniform distributions $[-W_\epsilon,W_\epsilon]$ and $[V-W_V,V+W_V]$ respectively, with $W_{\epsilon}$ and $W_V$ the corresponding disorder strengths (while it is usual to take $V_i$ to be constant, for generality we allow it here to be random). The model in Eq.~\ref{eq:ham-fermion} is of course nothing but the conventional model of Anderson localisation in 1D, but with an additional nearest-neighbour interaction of strength $V_i$.

Using a Jordan-Wigner transformation, the spinless fermion model can be mapped onto the disordered XXZ spin chain (modulo an additive constant),
\eq{
	H_\mathrm{XXZ} = J\sum_{i=1}^L[\sigma^x_i\sigma^x_{i+1}+\sigma^y_i\sigma^y_{i+1}]\nonumber\\
	+\sum_{i=1}^L [J_i^{\pd}\sigma^z_i\sigma^z_{i+1} + h_i\sigma^z_i]\,
	\label{eq:ham-xxz}
}
(the corresponding Heisenberg model is the particular case $J_{i}=J$). Here,
$\sigma^\mu_i$ denotes the Pauli matrix~\footnote{The Pauli matrices we use in practice satisfy the commutation relations $[\sigma_{i}^{\alpha},\sigma_{i}^{\beta}]=2 i\epsilon_{\alpha\beta\gamma}\sigma_{i}^{\gamma}$; such that $\sigma^x_i\sigma^x_{i+1}+\sigma^y_i\sigma^y_{i+1}=2(\sigma_{i}^{+}\sigma_{i+1}^{-}+\sigma_{i}^{-}\sigma_{i+1}^{+})$, and $[\sigma_{i}^{z}]^{2}=\mathbb{I}$ (rather than $\tfrac{1}{4}\mathbb{I}$).} for the spin-1/2 on site $i$, and  $J_i\in [J_z-W_{J_{z}},J_z+W_{J_{z}}]$ and $h_i\in [-W,W]$.  The parameters of the two equivalent models are related by $J=t/2$, $J_z=V/4$, $W_{J_{z}}=W_V/4$, and $W = W_\epsilon/2$. The spinless fermion Hamiltonian Eq.~\ref{eq:ham-fermion} conserves the total number of fermions, $N = \sum_{i=1}^L \hat{n}_i$; which translates for the XXZ model to conservation of total $z$-magnetisation, $S^z_\mathrm{tot}=\sum_{i=1}^L\sigma^z_i$ (the two being related via $S^z_\mathrm{tot}=2N-L$). While it is customary to work at half-filling $N = L/2$ -- or equivalently in the $S^z_\mathrm{tot}=0$ sector -- much of what we discuss generalises readily to arbitrary filling fractions $\nu = N/L$.

Another spin model which has gained prominence in the recent past  is the random tilted-field Ising (TFI) chain (also referred to as the mixed-field Ising chain in the literature), described by~\cite{imbrie2016many,imbrie2017local,abanin2021distinguishing}
\eq{
	H_\mathrm{TFI} = \Gamma\sum_{i=1}^L \sigma^x_{i} + \sum_{i=1}^L [J_i\sigma^z_{i}\sigma^z_{i+1} + h_i\sigma^z_i]\,,
	\label{eq:ham-tfi}
}
with the random $J_{i}$ and $h_{i}$ again as specified above (and $\sigma_{i}^{x}=\sigma_{i}^{+}+\sigma_{i}^{-}$). Unlike the XXZ model, the TFI chain does not have a representation in terms of locally interacting fermions, and neither does it possess any explicit conservation law except total energy. 

State-of-the-art numerical studies using exact diagonalisation have estimated the critical disorder strengths for the MBL transitions in the two models to be $W_c^\mathrm{XXZ}/J \approx 3.5-5.5$ (for $S_z^\mathrm{tot}=0$ and $J_i=J$ for all sites $i$)~\cite{luitz2015many,mace2019multifractal,sierant2020polfed} and $W_c^\mathrm{TFI}/\Gamma \approx 4$ (for $J_{z}=1$ and $W_{J_{z}}=0.2$)~\cite{abanin2021distinguishing} -- obviously with substantial error bars, reflecting the modest system sizes accessible to numerics. Arguments based on resonances in the MBL phase, and numerical simulations of real-time dynamics, have however suggested that the critical disorder strengths could be substantially larger~\cite{morningstar2021avalanches,sels2022bath,sierant2022challenges,long2023phenomenology,SierantPRL2020A}. In fact some works have proposed the absence of a strict MBL phase altogether~\cite{suntajs2020quantum,SuntajsPRB2020,SirkerPRL2020,SelsPolkPRE2021,SirkerPRB2021}, for the particular parameter regimes  studied, although they have been rebutted to varying degrees~\cite{abanin2021distinguishing,LuitzBarLevPRB2020,SierantPRL2020A,sierant2020polfed,scocco2024thermalizations}. We doubt that these disagreements are likely to be settled with currently available computational techniques and resources. However, we take the view that while these issues raise questions about the actual values of $W_c$ for the specific models, the proven existence of an MBL phase in 1D~\cite{imbrie2016many,Imbrie2016PRL,imbrie2017local,DeRoeck2024arXiv} is not in question. Even if the critical disorder strength in the thermodynamic limit is larger than that predicted from numerical studies on finite-sized systems, our view is that the features seen in this finite-size MBL regime will persist in a thermodynamically large system in a genuine MBL phase.


\subsection{The Fock-space graphs\label{sec:FSGs}}

\begin{figure}
\includegraphics[width=\linewidth]{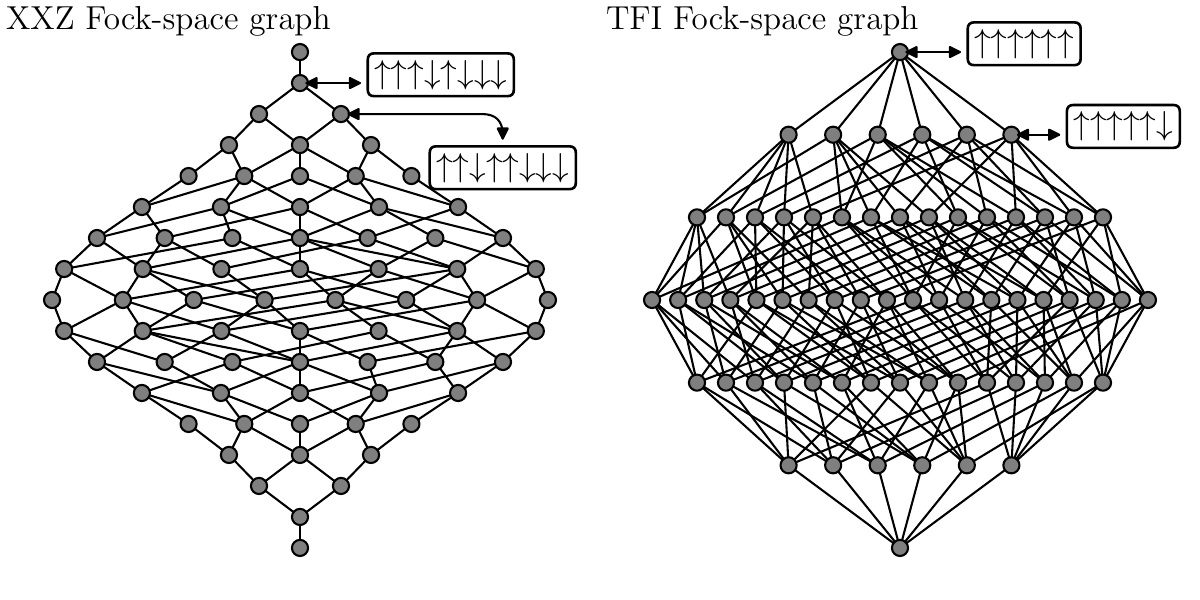}
\caption{Fock-space graphs for the XXZ model (left, illustrated for $L=8$ spins and $S_{z}^{\mathrm{tot}}=0$), and the TFI model (right, shown for $L=6$). Filled circles denote the Fock-space sites/nodes, which are $\sigma^z$-configurations as illustrated. Grey lines denote the edges generated by the spin-flip terms in $H$. For the XXZ model, the apical site has been chosen as 
$|$$\uparrow\uparrow\uparrow\uparrow\downarrow\downarrow\downarrow\downarrow\rangle$. For 
the TFI model the choice of apical site is immaterial, as the graph is a representation of a $2^{L}$-dimensional hypercube, in which all corners (the nodes) are equivalent.
}
\label{fig:fs-graphs}
\end{figure}

We turn now to consider the Fock-space graphs of the XXZ and TFI models, which are illustrated in Fig.~\ref{fig:fs-graphs}. The origin of the tight-binding form for the Fock-space Hamiltonian, Eq.~\ref{eq:fs-ham-general}, is itself quite trivial. For any Hamiltonian separated as $H=H_{0}+H_{1}$, the former is by definition diagonal in its own basis, $H_{0}=\sum_{I}\mathcal{E}_{I}|I\rangle\langle I|$, while the latter
is generally non-diagonal, $H_{1}=\sum_{I\neq J}\mathcal{T}_{IJ}|I\rangle\langle J|$. The freedom here 
resides of course in the choice of $H_{0}$ and its associated basis $\{|I\rangle\}$.
 
For both the XXZ and TFI models, Eq.~\ref{eq:ham-xxz} and Eq.~\ref{eq:ham-tfi}, the disorder couples to the $z$-components of the spins. Hence, in the trivially localised limit of $W\to\infty$, the eigenstates are $\sigma^z$-product states. This motivates $\sigma^z$-product states as natural basis states, $\ket{I} = \ket{\{s^I_i\}}$ where $s_i^I=\pm 1$ denotes the $z$-component of the spin at real-space site $i$ in basis state $\ket{I}$ (reflecting the $H_{0}\equiv \sum_{i}[J_{i}\sigma_{i}^{z}\sigma_{i+1}^{z} +h_{i}\sigma_{i}^{z}]$ common to both models). These basis states form the nodes of the Fock-space graph, referred to henceforth as Fock-space sites; and their number defines the Fock-space dimension, $\nh$. For the TFI model, the absence of explicit conservation laws implies all the $\{s^I_i\}$ configurations form a part of the Fock-space graph, whence $\nh^\mathrm{TFI}=2^L$. On the other hand, for the XXZ chain the dimension depends on the total $z$-component of spin, and for $S_{z}^{\mathrm{tot}}=0$,
$\nh^\mathrm{XXZ} =\binom{L}{L/2}$
 is given asymptotically by $\nh^\mathrm{XXZ}\simeq 2^L\sqrt{2/\pi L}$. The evident upshot of this is that the Fock-space dimension is exponentially large in the system size $L$.

Each of these Fock-space sites has an on-site energy given by
\eq{
	\mathcal{E}_I^{\pd} = \sum_{i=1}^L [J_i^{\pd} s_i^I s_{i+1}^I + h_i^{\pd} s_i^I]\,.
	\label{eq:FS-site-energy}
}
Their distribution -- over Fock-space sites as well as over disorder realisations --
generates the effective Fock-space disorder in the tight-binding Hamiltonian.
 Moreover, the  Fock-space site energies must be mutually correlated, rather than independent; as is evident because their number, $\nh$, is exponentially large in $L$, yet there are only $\mathcal{O}(L)$ independent random variables (the $J_i$'s and $h_i$'s). We will dwell more on these points in Sec.~\ref{sec:fscorrs}. At this stage, however, it is sufficient to realise that the one-body marginal distribution of the Fock-space site energies is a Gaussian, with a variance $\mu_{\mathcal{E}}^{2}$ that scales linearly with the number of real-space sites $L$~\cite{welsh2018simple},
\eq{
	\mu_{\mathcal{E}} ~\sim ~ \sqrt{L}~W_\mathrm{FS}\,,
}
where $W_\mathrm{FS}$ is a function of $W$, $W_{J_{z}}$, and $J_z$. This is readily understood from the form in Eq.~\ref{eq:FS-site-energy}: since each $\mathcal{E}_I$ is a sum of $L$ uncorrelated random numbers,  the central limit theorem mandates that $\mathcal{E}_I$ has a normal distribution with a   standard deviation $\mu_{\mathcal{E}}$ scaling as $\sqrt{L}$.

The next point of consideration are the edges, or links, of the graph (Fig.~\ref{fig:fs-graphs}). 
These are generated by the spin-flip terms in the Hamiltonians in Eq.~\ref{eq:ham-xxz} and Eq.~\ref{eq:ham-tfi}, which correspond to $H_{1}=H-H_{0} \equiv \sum_{I\neq J}\mathcal{T}_{IJ}|I\rangle\langle J|$
and represent hopping between the Fock-space sites in the effective tight-binding problem. 
Unlike the Fock-space site energies, there is no disorder in the hoppings $\{\mathcal{T}_{IJ}\}$;
indeed, for either model, all non-zero $\mathcal{T}_{IJ}$'s are coincident.
From its structure, it is clear that in the XXZ model any Fock-space site $I$ is connected to Fock-space sites $K$ which differ from $I$ by the flip of a single nearest-neighbour antiparallel spin pair. Similarly, for the TFI model, a Fock-space site $I$ is connected to all sites $K$ which differ from $I$ by a single spin flip. To make this quantitative we need to consider the distribution of the connectivities of the Fock-space sites.
Formally, the connectivity of Fock-space site $I$, denoted by $Z_I$, is the number of Fock-space sites $J$ for which $\mathcal{T}_{IJ}\neq 0$. In general, it depends on $I$ and therefore has a distribution over the Fock-space sites (a property intrinsic to the Fock-space graph, and not dependent on the
disorder realisation).

For the TFI model, computing the distribution is trivial, as each edge on the Fock-space graph corresponds to a single spin flip. Any Fock-space site is thus connected to exactly $L$ other sites, each corresponding to a flip of one of the $L$ spins in the chain. The distribution of connectivities
is then simply  $P_Z^\mathrm{TFI}(Z) =\delta(Z-L)$. For the XXZ chain the corresponding distribution is somewhat non-trivial, as edges correspond to flipping of domain walls, so the connectivity of Fock-space site $I$ is $Z_I = \sum_i (1-s_i^I s_{i+1}^I)/2$.  The distribution of $Z$ is thus determined by that of $\sum_i s_i^I s_{i+1}^I$; and the latter can be computed for finite $L$ and arbitrary $S_{z}^{\mathrm{tot}}$
using basic combinatorics~\cite{welsh2018simple}. But the central point here is that for $L\gg 1$ 
-- the domain of obvious interest -- the distribution $P_Z^\mathrm{XXZ}(Z)$ is a Gaussian, $Z\sim \mathcal{N}(\overline{Z}, \mu_Z)$, with a mean $\overline{Z}=L/2$ and a standard deviation $\mu_Z=\sqrt{L}/2$ in the half-filled sector.\footnote{The mean and standard deviation for arbitrary $S_{z}^{\mathrm{tot}}$ are given by $\overline{Z} = \tfrac{1}{2}(1-[S_{z}^{\mathrm{tot}}/L]^{2})L$ and $\mu_Z = \overline{Z}/\sqrt{L}$~\cite{welsh2018simple}.} 
Further,  in the thermodynamic limit the standard deviation is negligible compared to the mean, whence 
$P_Z^\mathrm{XXZ}(Z)\sim \delta(Z-L/2)$ is essentially $\delta$-distributed on $Z=L/2$.

\begin{figure}
\includegraphics[width=\linewidth]{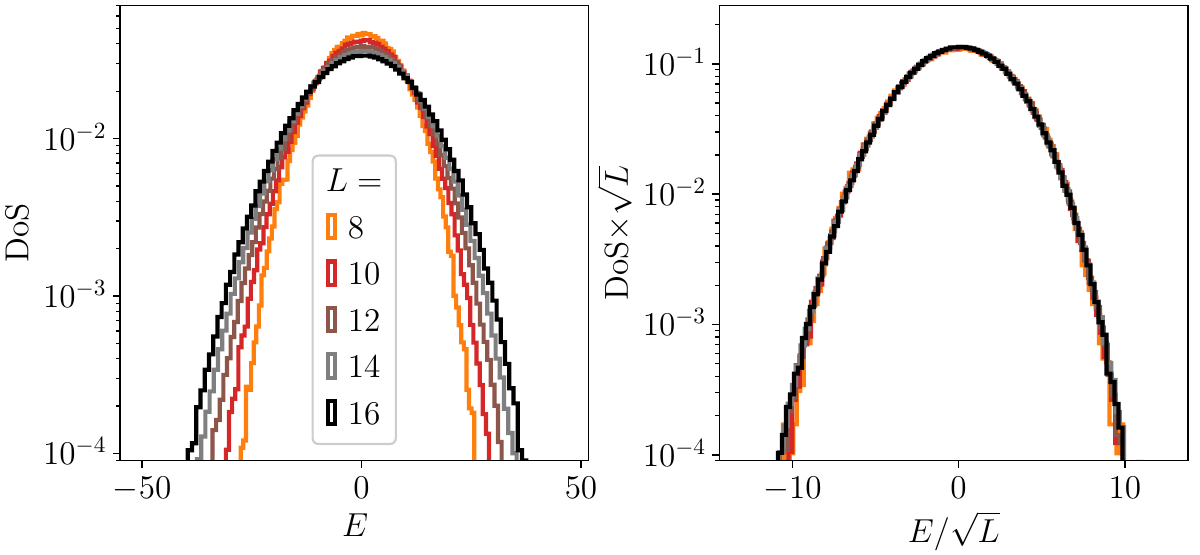}
\caption{Many-body density of states (DoS) for the XXZ model ($S_{z}^{\mathrm{tot}}=0$), shown for a range of systems sizes $L=8-16$  (with $J_{z}=1=J$, $W=4$ and $W_{J_{z}}=0$).
Left panel: shown \emph{vs} energy $E$ (relative to its mean for each disorder realisation, $\mathrm{tr}[H]$). Right panel: with energy rescaled by $\sqrt{L}$, the DoS rapidly converges to an $L$-independent
Gaussian.
}
\label{fig:DoSXXZ}
\end{figure}

The linear in $L$ scaling of the variance of $\mathcal{E}_I$, as well as that of the mean connectivity, has an important consequence -- the many-body density of states (DoS) is also a Gaussian with a variance $\mu_{E}^{2}$ proportional to $L$, as illustrated in Fig.~\ref{fig:DoSXXZ}. To see this, note that 
\begin{equation}
	\mu_E^2 = \overline{\mathrm{tr}\big[\big(H -\mathrm{tr}[H]	\big)^{2}\big]}
	=\overline{\mathrm{tr}[H^2]} - \overline{(\mathrm{tr}[H])^2}\,,
	\label{eq:muEsq}
\end{equation}
where $\mathrm{tr}[\cdot\cdot] = \nh^{-1}\sum_I \braket{I|\cdot\cdot|I}$  denotes an average over the Fock-space sites and the overline denotes an average over disorder realisations. 
A potential pitfall to beware of here is that it is not in general correct to define the variance of the many-body DoS via $\overline{\mathrm{tr}[H^2]} - \big(\overline{\mathrm{tr}[H]}\big)^2$. This is because the centre of gravity of the spectrum  ($\mathrm{tr}[H]$) fluctuates across disorder realisations on an energy scale $\propto \sqrt{L}$, which can lead to spurious contributions to the variance as well as incorrect identification of the centre of the spectrum and mobility edges~\cite{welsh2018simple}. To remove these, it is important to shift the spectrum for each realisation  relative to its mean $\mathrm{tr}[H]$, and then compute the variance -- as in Eq.~\ref{eq:muEsq}~\cite{welsh2018simple}.

Using the tight-binding form of the Hamiltonian,  Eq.~\ref{eq:fs-ham-general}, the variance is easily
obtained, 
\eq{
	\mu_E^2 = 
	\mu_{\mathcal{E}}^{2}+ \nh^{-1}\sum_{I\neq J}\overline{|\mathcal{T}_{IJ}|^2}\,.
}
Since the Fock-space hopping matrix elements are uniform in the two models ($J$ for XXZ and $\Gamma$ for the TFI model), we readily have
\eq{
	\mu_E^2 = \begin{cases}
	L(W_\mathrm{FS}^2 + J^2/2) & :\mathrm{XXZ}\\
	L(W_\mathrm{FS}^2 + \Gamma^2) & :\mathrm{TFI}
	\end{cases}\,.
}
This scaling of $\mu_E$ with $\sqrt{L}$ has the consequence that for any calculation to admit a well-defined thermodynamic limit, all energy scales must be rescaled by $\sqrt{L}$. One example is given in Fig.~\ref{fig:DoSXXZ}, showing that on recaling by $\sqrt{L}$ the DoS rapidly converges to an $L$-independent Gaussian. This also has fundamental implications for the structure of the disorder correlations (discussed in Sec.~\ref{sec:fscorrs}), as well for the scaling of Fock-space propagators (see Sec.~\ref{sec:props})
which in turn carry information about the characteristic features of the MBL and ergodic phases.

So far, we have established quantitatively the distributions of the on-site energies and connectivities on the Fock-space graph. The final ingredient needed to complete the description is a notion of distance on the graph; and since each node is a spin-1/2 configuration, the Hamming distance offers a  natural and convenient choice. The Hamming distance between two Fock-space sites $I$ and $J$ is defined as the number of real-space sites on which the spin configuration in the two Fock-space sites is different; mathematically,
\eq{
	r_{IJ}^{\pd} = \frac{1}{2}\Big(L - \sum_{i=1}^Ls_i^Is_i^J\Big)\,.
	\label{eq:hamming-dist}
} 
For the TFI Fock-space graph, the Hamming distance coincides with the graph distance, the latter defined as the shortest path on the graph between two nodes, given the set of available edges. For any given Fock-space site $I$, the number of sites $J$ at (Hamming or graph) distance $r$ from $I$ is 
$N_r = \binom{L}{r}$; it is simply the number of ways of choosing $r$ out of $L$ spins to flip. 

In general, however, the graph distance may not coincide with the Hamming distance. The XXZ chain itself is a case in point. The Hamming distance remains a useful measure of distance, particularly in making connections between Fock-space observables and real-space observables via Eq.~\ref{eq:hamming-dist} (Sec.~\ref{sec:anatomy}).  But the Hamming and graph distances between two Fock-space sites can be entirely different; for example, the states $\ket{\uparrow\uparrow\cdot\cdot\uparrow\uparrow\downarrow\downarrow\cdot\cdot\downarrow\downarrow}$ and 
$\ket{\downarrow\downarrow\cdot\cdot\downarrow\downarrow\uparrow\uparrow\cdot\cdot\uparrow\uparrow}$
-- the apical and bottom nodes on the XXZ Fock-space graph (Fig.~\ref{fig:fs-graphs}) -- are separated by a Hamming distance of $L$, but the graph distance is $\tfrac{1}{4}L^{2}$. The graph distance nevertheless remains useful in the context of perturbative, locator expansions, since it is the leading order at which two Fock-space sites are connected under perturbation theory.

This concludes our overview of the salient features of the Fock-space graphs and associated tight-binding Hamiltonian (Eq.~\ref{eq:fs-ham-general}), for two standard models used for studying MBL with short-ranged interactions. Before proceeding, however, we touch briefly upon long-ranged and all-to-all interacting models.


\subsection{Long-ranged and all-to-all models}

 Beyond short-ranged models, the physics of localisation is also popularly studied in models with long-ranged interactions, especially those with power-law interactions~\cite{burin2006energy,yao2014manybody,burin2015localisation,burin2015manybody,tikhonov2018manybody,roy2019self,nag2019manybody,detomasi2019algebraic,maksymov2020many,2022fermionic,cheng2023many}. They find relevance in the context of Coulomb or dipolar interactions~\cite{choi2017observation}, as well as in trapped ion systems. One such model is a power-law interacting version of the XXZ chain in Eq.~\ref{eq:ham-xxz}, described by the Hamiltonian
 \eq{
 	H_\mathrm{LR-XXZ}^{\pd} = \sum_{i>j}&\frac{J}{(i-j)^\alpha}[\sigma^x_i\sigma^x_{j}+\sigma^y_i\sigma^y_{j}]\nonumber\\
	&+\sum_{i>j}\frac{J_z}{(i-j)^\beta}\sigma^z_i\sigma^z_{j} + \sum_ih_i\sigma^z_i\,,
	\label{eq:hamLR-XXZ}
 }
 where $\alpha$ and $\beta$ are power-law exponents controlling respectively the decay of the pairwise spin-flip terms and the interaction terms. From the Fock-space point-of-view, one of the crucial differences 
between the short-ranged models and that of Eq.~\ref{eq:hamLR-XXZ} lies in the scaling of various distributions. The Fock-space site-energy distribution continues to be Gaussian, but its 
standard deviation scales with $L$ as~\cite{roy2019self}
 \eq{
 \mu_\mathcal{E}^{\pd}\sim\begin{cases}
 						\sqrt{L}\,\quad& :\beta>1/2\\
 						\sqrt{L \ln L}\,\quad& :\beta=1/2\\
 						L^{1-\beta}\,\quad& :\beta<1/2
 					\end{cases}\,.
 }
As far as the connectivities are concerned, in an absolute sense each Fock-space site is connected 
on an average to $O(L^2)$ sites. However, as the strength of the matrix element decays algebraically with the distance, a more relevant quantity is the effective connectivity -- a weighted sum, $\overline{Z} \sim \sum_r \overline{z(r)}/r^{2\alpha}$, where $\overline{z(r)} = L(L-r)/2(L-1)$ is the average connectivity corresponding to a pair of spin-flips at separation $r$ on the chain. This average effective connectivity 
scales with $L$ as~\cite{roy2019self}
\eq{
	\overline{Z} \sim\begin{cases}
 						L\,;\quad& \alpha>1/2\\
 						L \ln L\,;\quad& \alpha=1/2\\
 						L^{2-2\alpha}\,;\quad& \alpha<1/2
 					\end{cases}\,.
}
These different scalings of $\mu_\mathcal{E}$ and $\overline{Z}$ with $L$,
compared to the standard models discussed above, place fundamental constraints on the parameter space in which the model Eq.~\ref{eq:hamLR-XXZ} can host an MBL phase; we shall review this briefly in Sec.~\ref{sec:props}.

Another class of models, which are devoid of any spatial structure but may show localisation transitions, are all-to-all interacting models known as $p$-spin models. These have had a long history in the study of spin glasses~\cite{derrida1980random,sherrington1975solvable,kirkpatrick1987pspin,wolynes1987pspin,goldschmidt1990solvable,nieuwenhuizen1998quantum}, and more recently have attracted attention in the context of ergodicity breaking~\cite{laumann2014many,baldwin2016manybody,baldwin2017clustering,burin2017localization,mukherjee2018many}. This family of models is described by 
\eq{
H_{p\text{-spin}}^{\pd}
	= \frac{1}{L^{\frac{p-1}{2}}}\sum_{i_1,\cdots,i_p}J_{i_1\cdots i_p}^{\pd}\sigma^z_{i_1}\cdots\sigma^z_{i_p}+\Gamma\sum_i \sigma^x_i\,,
	\label{eq:ham-pspin}
}
where the couplings $J_{i_1\cdots i_p}$ are independent Gaussian random variables (including across different permutations of $\{i_1,\cdots,i_p\}$),  with zero  mean and standard deviation $W_{J}$. 
A second category of all-to-all models,
likewise lacking any real-space locality and offering
an interesting setting to study Fock-space localisation (or its absence) in its own right~\cite{altshuler1997quasiparticle,micklitz2019nonergodic,dieplinger2021SYK,monteiro2021minimal,monteiro2021quantum,herre2023ergodicity}, fall under the purview of perturbed Sachdev-Ye-Kitaev models, and are closely related to fermionic quantum dot models.
The general form of their Hamiltonian is
\eq{
H = \sum_i\epsilon_i^{\pd} c_i^\dagger c_i^{\pd} +\sum_{i,j,k,l}V_{ijkl}^{\pd}c_i^\dagger c_j^\dagger c_k^{\pd} c_l^{\pd}\,,
}
where the relative variances of the distributions of the $\{\epsilon_i\}$ and of the $\{V_{ijkl}\}$ 
in effect tune a localisation transition. 
Substantive discussion of these all-to-all models is however beyond the scope of this review, as our primary focus will be on many-body localisation in locally interacting models.


\section{Eigenstate anatomy \label{sec:anatomy}}

So far, we have established that the problem of many-body localisation in disordered, interacting systems can be mapped onto that of a fictitious single particle hopping on the complex, correlated Fock-space graph. This automatically leads to the intepretation of the many-body eigenstates' wavefunctions as those of the effective single particle on the graph, see Eq.~\ref{eq:state-general}. The statistics of these eigenstate amplitudes on the Fock-space graph are 
therefore of natural interest, in particular from the viewpoint of ergodicity breaking on Fock space. This section reviews several aspects of such, namely the scaling of inverse participation ratios and participation entropies, dynamical eigenstate correlations, and connections between the eigenstate anatomy and local observables.


\subsection{Ergodicity breaking on Fock space}

\subsubsection{Scaling of IPRs and PEs}
\begin{figure}
\includegraphics[width=\linewidth]{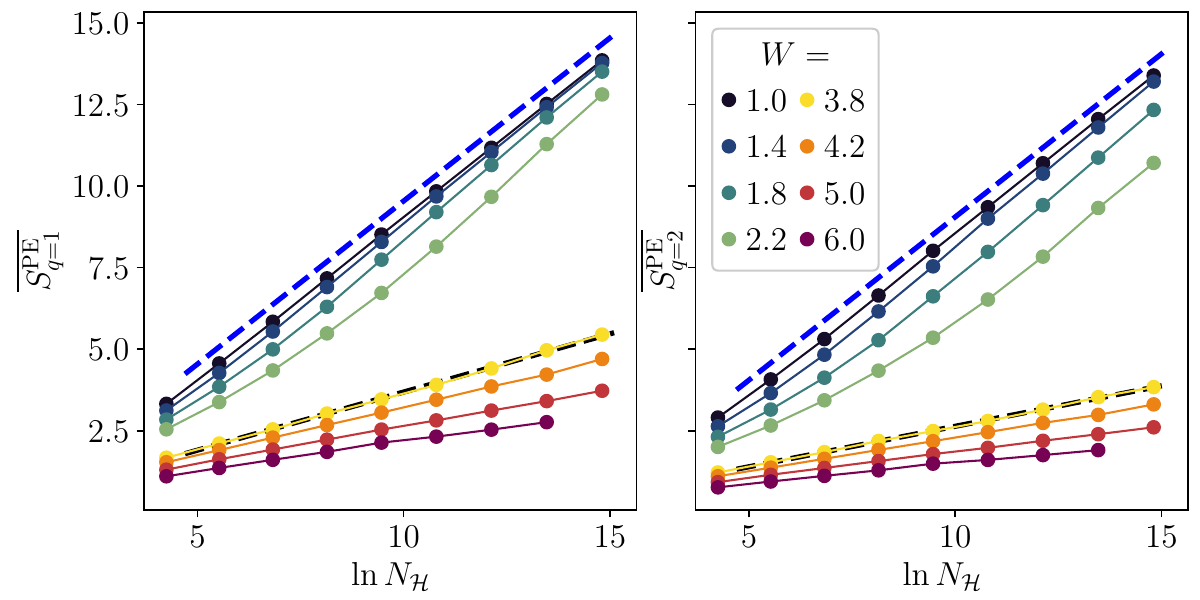}
\caption{ 
Participation entropies, $S^\mathrm{PE}_q$, for $q=1$ (left) and $q=2$ (right) for the disordered XXZ chain  \eqref{eq:ham-xxz}, show linear behaviour as a function of $\ln\nh$. In the ergodic phase, the slope is unity (indicated by the blue dashed lines), implying $D_q=1$; whereas in the MBL phase the slope is strictly less than 1, indicating fractal scaling. The data is adapted from Ref.~\cite{mace2019multifractal} with parameters $J=1=J_{z}$
and $W_{J_{z}}=0$.}
\label{fig:ipr-xxz}
\end{figure}

One of the most common measures for quantifying a wavefunction's ergodicity, or lack of it, is the inverse participation ratio (IPR), or equivalently the participation entropy (PE). The generalised $q^\mathrm{th}$-IPR of an eigenstate $\ket{\psi} = \sum_{I}\psi_I\ket{I}$ in the basis $\{\ket{I}\}$ is defined as 
\eq{
    \mathcal{L}_q^{\pd} = \sum_{I}|\psi_I^{\pd}|^{2q} \sim \nh^{-\tau_q}\,,
    \label{eq:IPR}
} 
with $\tau_{q}$ referred to as the (multi)fractal exponent. It is also common to define $\tau_q = D_q(q-1)$, 
with $D_q$ called the fractal dimension. Similarly, the $q^\mathrm{th}$-PE is defined 
via $\ln\mathcal{L}_q$ as 
\eq{
	 S^\mathrm{PE}_q = -\frac{1}{q-1}\ln\mathcal{L}_{q}^{\pd}\sim D_q^{\pd}\ln\nh+\cdots\,
    \label{eq:PE}
}
(such that $ S^\mathrm{PE}_1=-\sum_{I}|\psi_{I}|^{2}\ln|\psi_{I}|^{2}$ is a Shannon entropy).
The natural choice of basis is provided by the `trivially' localised, product-state eigenstates of $H_0$. For perfectly ergodic states, uniformly extended over the Fock space on average, $|\psi_I|^2\sim \nh^{-1}$ such that $\mathcal{L}_q\sim\nh^{-(q-1)}$, or equivalently $D_q=1$. Localised states {\it \`a la} Anderson would by contrast reside within a finite volume of the Fock space, whence their IPR would not scale with the Fock-space dimension, corresponding to $D_{q}=0$. The key point to realise here is that, for any finite disorder strength, MBL eigenstates are {\it not} Anderson localised on the Fock-space graph; instead, their IPRs scale anomalously with the Fock-space dimension, $0<D_q<1$ with the inequalities being strict. The upshot is that MBL eigenstates are in fact fractal on the Fock space~\cite{deluca2013ergodicity,luitz2015many,mace2019multifractal,roy2021fockspace,detomasi2020rare} and the MBL transition is  a transition from ergodicity to multifractality on the Fock space. Clear numerical evidence for this~\cite{mace2019multifractal}  is provided in Fig.~\ref{fig:ipr-xxz} for the disordered XXZ chain~\eqref{eq:ham-xxz}, in both the ergodic and MBL phases. In the former, $\overline{S^\mathrm{PE}_q}\sim \ln\nh$ for both $q=1$ and $q=2$, implying $D_q=1$, whereas in the MBL phase $0<D_q<1$, with $D_q$ depending on $q$ and $W$. To avoid repetition we have shown results only for the XXZ chain, but these remain qualitatively the same for the TFI chain (and indeed for any locally interacting model).

It is also worth mentioning that $D_q$ in the MBL phase is not universal, in the sense that its precise value depends explicitly on the basis choice as well as the models, disorder strengths etc.  What is however robust is that MBL eigenstates remain multifractal in any local basis that is not fine-tuned.

\subsubsection{Origins of multifractality in the MBL phase}

The fact that MBL eigenstates are fractal and not Anderson localised on the Fock-space graph, is a crucial point of distinction between MBL and conventional Anderson localisation in high-dimensional graphs~\cite{abou-chacra1973self,tikhonov2016anderson,garciamata2017scaling,garciamata2020two}. This has arguably been a point of contention in regard to treating the MBL problem on Fock space in a manner akin to Anderson localisation on high-dimensional graphs~\cite{tikhonov2021anderson}. In this section we clarify the issue, and point out that the MBL eigenstates cannot be Anderson localised on the Fock space. To show this, we prove below that eigenstates of the non-interacting but \emph{many-body} Anderson insulator ($V_i=0$ limit of the model in Eq.~\ref{eq:ham-fermion}) cannot be Anderson localised in the occupation number basis~\cite{Creeddphil}. On physical grounds it follows naturally that adding interactions cannot further localise the many-body eigenstate on Fock-space, whence eigenstates cannot be Anderson localised in the presence of interactions either. As such, they can only be either fractal ($0<D_{q}<1$), or ergodic ($D_{q}=1$).

To consider the non-interacting many-body problem, recall that single-particle eigenstates can be expressed as 
\eq{
	\ket{\phi_\alpha} = \sum_{i=1}^L \phi_{\alpha,i}\ket{i}\,,
}
where $\ket{i}$ denotes a real-space, site-localised orbital. The many-body wavefunction, for both an eigenstate $\ket{\psi_0}$ as well as a basis state $\ket{I}$, can be written as a Slater determinant (where the subscript in $\ket{\psi_0}$ denotes the non-interacting nature of the problem). Using this, the wavefunction density can be written as 
\eq{
	|\braket{\psi_0|I}|^2 = |\mathrm{det}[\mathcal{M}]|^2\,,
}
where $\cal{M}$ is a $\nu L\times \nu L$  matrix with elements $\mathcal{M}_{i\alpha}=\phi_{i,\alpha}$. The $\alpha$ index runs over the $\nu L$ ($=N$) occupied Anderson-localised orbitals in $\ket{\psi_0}$, and the $i$ index runs over the $\nu L$ occupied real-space sites in $\ket{I}$. By Hadamard's inequality one has
\eq{
	|\mathrm{det}[\mathcal{M}]|^2 \le \prod_{\alpha=1}^{\nu L}||v_\alpha||^2 = \prod_{\alpha=1}^{\nu L}\sum_{i \in I}|\phi_{i\alpha}|^2\,,
	\label{eq:SDM}
}
where $||v_{\alpha}||$ denotes the norm of the $\alpha^\mathrm{th}$ column of $\mathcal{M}$. 
Normalisation of the single-particle orbitals implies $\sum_i|\phi_{i,\alpha}|^2=1$; but since each sum in Eq.~\ref{eq:SDM} runs over only a subset of real-space sites, $|\braket{\psi_0|I}|^2$ is a product of $\nu L$ numbers each of which is strictly less than unity. This trivially implies that $|\braket{\psi_0|I}|^2$ is exponentially small in $L$. The key point here is that the above argument is true for all $I$, which precludes Anderson localisation of the many-body wavefunction on the Fock space. The non-interacting, many-body eigenstates can thus only be fractal, ergodic eigenstates being precluded for $W\neq 0$ in the 1-dimensional case of interest. In the presence of interactions, some further (if limited) progress can be made using H\"older inequalities~\cite{Creeddphil}, but the 
main point here is the physical one mentioned above: that interactions can only further delocalise the state on Fock space, and
as such cannot remove the fractality to make way for Anderson localisation on the Fock space.

 A physical implication of the multifractality of MBL eigenstates is that they can be viewed as 
in effect living on a fragmented Fock space~\cite{pietracaprina2021hilbert,prelovsek2018reduced}, which allows for numerically efficient computations. It also provides motivation  for the study of classical percolation models in Fock space, as proxies for the MBL transition~\cite{roy2018exact,roy2018percolation}.


\subsection{Fock-spatial correlations}

Having discussed the statistics of eigenstate amplitudes for single Fock-space sites, a natural next step is to understand two-point correlations on the Fock space. For a given eigenstate $\ket{\psi}$, the simplest and most natural
such correlation is~\cite{roy2021fockspace} 
\eq{
	F_\psi(r)^{\pd} = \sum_{\substack{I,J: \\ r_{IJ}=r}}|\psi_I^{\pd}|^2|\psi_J^{\pd}|^2\,,
	\label{eq:Fr}
}
with the sum over all sites $I,J$ separated by a fixed Hamming distance $r_{IJ}=r$ (itself defined in Eq.~\ref{eq:hamming-dist}). Note that normalisation of the eigenstates implies the sum rule $\sum_{r=0}^L F_\psi(r)=1$. Together with the fact that $F_{\psi}(r)\ge 0$, this leads directly to the  interpretation of $F_\psi(r)$ as a probability distribution over the $r$-space, moments of which will be discussed in the next subsection.

In the ergodic phase, at least sufficiently deep inside it, the eigenstates can be considered as random Gaussian vectors such that each $\psi_I$ is a Gaussian random number with zero mean and a variance $\nh^{-1}$, {\it i.e.}, $\overline{|\psi_I|^2}\sim \nh^{-1}$ and $\overline{|\psi_I|^4}\sim 3\nh^{-2}$. This also entails that $|\psi_I|^2|\psi_J|^2$ is, on an average, independent of the Hamming distance $r_{IJ}$. Using the above in the definition Eq.~\ref{eq:Fr} 
gives for the ergodic phase
\eq{
	\overline{F_{\rm erg}}(r) =\frac{2}{\nh}\delta_{r,0}^{\pd} +  \frac{N_r}{\nh}\,,
 \label{eq:Ferg}
}
where the overline denotes the average over disorder realisations (and a selection of eigenstates
at the energy of interest, usually the middle of the spectrum). As previously, $N_{r}$ denotes the number of Fock-space sites at Hamming distance $r$ from a given one;  with $N_{r}=\binom{L}{r}$ for the TFI model Eq.~\ref{eq:ham-tfi}, which we consider explicitly in the following.

In the MBL phase by contrast, the physical intuition is that the two-point correlation between  arbitrarily chosen Fock-space sites decays  exponentially with their mutual Hamming distance, $\overline{|\psi_I|^2|\psi_J|^2}\sim e^{-r_{IJ}/\xi_F}$. It is important to stress here that this exponential decay does not imply Anderson localisation of eigenstates on the Fock-space graph; in fact, as discussed shortly, the decay scale $\xi_F$ can be explicitly related to the multifractal exponent in the MBL phase. Accounting for the normalisation of $F_{\psi}(r)$, we have 
\eq{
\overline{F_{\rm MBL}}(r) = \frac{1}{(1+e^{-1/\xi_F})^L}\binom{L}{r} e^{-r/\xi_F}\,.
\label{eq:FMBL}
}
It is useful to realise that $\overline{F}(r)$ in both phases (Eq.~\ref{eq:Ferg} and Eq.~\ref{eq:FMBL}) can be written as a binomial expansion $\overline{F}(r) = N_r p^r (1-p)^{L-r}$, with 
\eq{
p=(1+e^{1/\xi_F})^{-1}
\label{eq:pdef}
}
in the MBL phase and $p=1/2$ in the ergodic phase. Note that the latter, $p=1/2$,  is consistent with a divergent $\xi_F$, as one would expect physically in an ergodic phase.

\begin{figure}
\includegraphics[width=\linewidth]{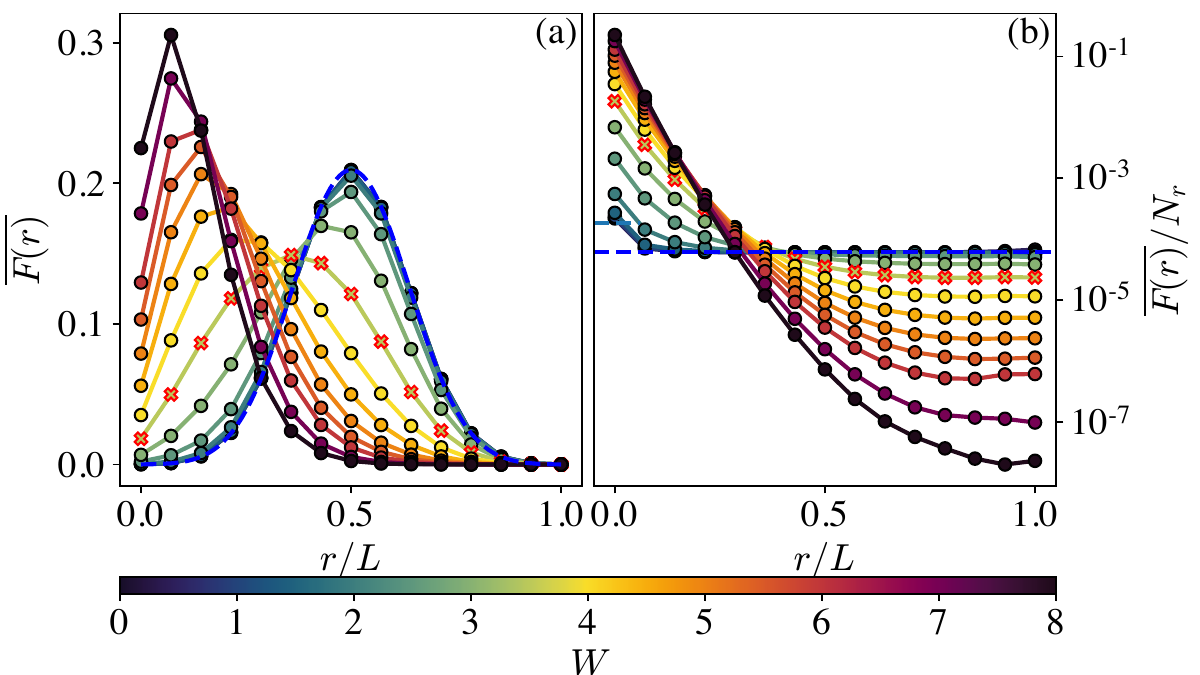}
\caption{(a) The Fock-spatial correlation, $\overline{F}(r)$$\equiv \overline{F_{\psi}}(r)$ (Eq.~\ref{eq:Fr}), for band-centre eigenstates of the TFI model, shown  \emph{vs} $r/L$ for various disorder strengths coded by colour.
 The blue dashed line shows the binomial profile centred at $r=L/2$, which is the expected result in the ergodic phase. (b) Same data as in (a), but rescaled by $N_r$. The two horizontal dashed lines correspond respectively to $3/\nh$ (light blue) and $1/\nh$ (dark blue), indicating the expected result in the ergodic phase (Eq.~\ref{eq:Ferg}). 
In the MBL phase on the other hand, the expected exponential decay is also evident for $r/L \lesssim 1/2$ (finite-size effects naturally arise in the vicinity of $r\approx L$). Results are shown for $L = 14$. The figure is taken from Ref.~\cite{roy2021fockspace} where the parameters used are $J=1$ and $W_{J_{z}}=0.2$.
}
\label{fig:Fr}
\end{figure}

For the TFI model (Eq.~\ref{eq:ham-tfi}), numerical results for $\overline{F}(r)\equiv \overline{F_{\psi}}(r)$ are shown 
in Fig.~\ref{fig:Fr}(a) over a range of disorder strengths $W$ across the MBL transition~\cite{roy2021fockspace}. The binomial form for the function in either phase is clearly visible.  Fig.~\ref{fig:Fr}(b) shows the  data rescaled as $\overline{F}(r)/N_{r}$, from which is seen both the essentially flat $r$-dependence in the ergodic phase and the exponential decay in the MBL phase.

Let us now discuss the relation between the Fock-space correlation length $\xi_F$, and the multifractal statistics of eigenstates in the MBL phase. This hinges on the simple observation from Eq.~\ref{eq:Fr} that $F_\psi(r=0) = \sum_{I}|\psi_I|^4$, which is nothing but the $\mathcal{L}_2$ IPR for the eigenstate $\ket{\psi}$. Eq.~\ref{eq:FMBL} thus  gives
\eq{
\overline{\mathcal{L}_2^{\rm MBL}} = (1+e^{-1/\xi_F})^{-L}\sim \nh^{-\tau_2}\,,
}
from which (recalling that $\nh =2^{L}$) the multifractal exponent can be read off directly  as
\eq{
\tau_2^{\pd} = \ln(1+e^{-1/\xi_F})/\ln 2\,.
\label{eq:tau2-xi}
}
As $\xi_F$ is finite and strictly positive in the MBL phase, this implies $0<\tau_2<1$ (with the inequalities being strict), indicating the fractal nature of the eigenstates. It is also evident that the fractal character of MBL eigenstates on the Fock-space emerges out of the competition between an exponential decay of the Fock-spatial correlation with Hamming distance $r$, and the exponential growth of the number of Fock-space sites with $r$. 
Finally here, note that Eq.~\ref{eq:tau2-xi} is also consistent with the ergodic phase, where a divergent $\xi_F$ implies $\tau_2=1$.

While the above results in the MBL phase were obtained numerically and argued for on physical grounds, we add that
they can also be derived explicitly  in the $J_i=0$ limit of the TFI model (Eq.~\ref{eq:ham-tfi}). 
This limit is manifestly MBL, as it corresponds to a set of non-interacting spins (while remaining fully
connected on the Fock-space graph), and was referred to as MBL\textsubscript{0} in Ref.~\cite{roy2021fockspace}. In this case, all results discussed above for the MBL phase go through with  $p=(1/2W)\tan^{-1}(W)$ explicitly.

\begin{figure}
\includegraphics[width=\linewidth]{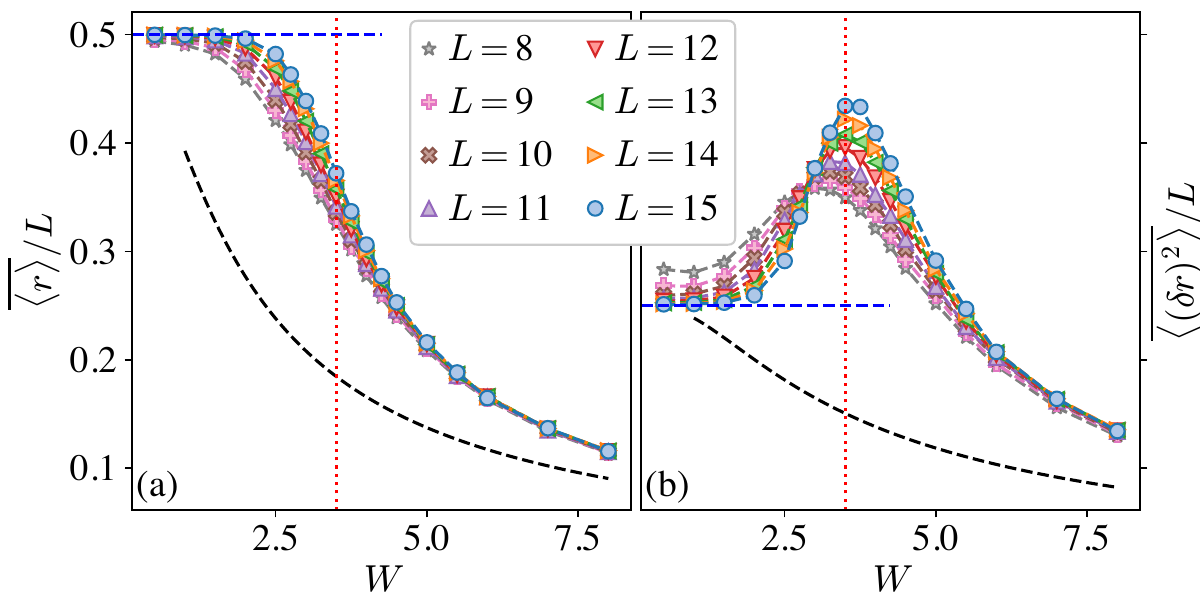}
\caption{The moments $\overline{\braket{r}}$ and $\overline{\braket{(\delta r)^2}}$ (Eq.~\ref{eq:rrsq-def})
of $F(r)\equiv F_{\psi}(r)$ for the TFI model, shown as function of disorder strength $W$ for various system sizes
$L$. Blue horizontal dashed lines show  the expected results in the ergodic phase, $\overline{\braket{r}}=L/2$ and $\overline{\braket{(\delta r)^2}}=L/4$ (Eq.~\ref{eq:rrsq}). Black dashed lines show results for  the MBL\textsubscript{0} limit. Vertical dotted line indicates the MBL transition at $W_{\mathrm{c}}\simeq 3.5$~\cite{abanin2021distinguishing}.
The figure is taken from Ref.~\cite{roy2021fockspace}, with parameters $J=1$ and $W_{J_{z}}=0.2$.}
\label{fig:r-rsq}
\end{figure}

As pointed out previously, $F_{\psi}(r)$ (Eq.~\ref{eq:Fr}) can be interpreted as a probability distribution.
Its moments are thus of natural interest, specifically
\eq{
\braket{r} = \sum_{r=0}^L r {F_\psi^{\pd}}(r)\,,~\braket{(\delta r)^2} = \sum_{r=0}^L r^2 F_\psi^{\pd}(r)-\braket{r}^2\,
\label{eq:rrsq-def}
}
from which, using the binomial form of $\overline{F}_{\psi}(r)$, one has
\eq{
\overline{\braket{r}} = pL\,,~~~~\overline{\langle(\delta r)^2\rangle} = p(1-p)L\,.
\label{eq:rrsq}
}
Fig.~\ref{fig:r-rsq} shows results for the two moments over a range of disorder strengths, $W$. In the ergodic phase, where $p=1/2$, these are indeed consistent with the prediction of Eq.~\ref{eq:rrsq}, with $\overline{\braket{r}}/L$ saturating to 1/2 with increasing $L$, and $\overline{\langle(\delta r)^2\rangle}/L$ to 1/4. In the MBL phase too, the data clearly shows linear scaling of both moments with $L$; and with $p$ (Eq.~\ref{eq:pdef}) decaying with increasing $W$, consistent with the physical expectation that the correlation length $\xi_{F}$ should diminish steadily with increasing $W$. In the vicinity of the MBL transition, however, $\overline{\langle(\delta r)^2\rangle}/L$ 
is seen to grow with $L$ (and in fact does so linearly with $L$); suggesting that fluctuations in the wavefunction amplitudes on the Fock-space graph diverge at the MBL transition.

In the next subsection we will discuss the implications of these behaviours of the moments for real-space local observables.  But before that, we mention a quantity related to $F_{\psi}(r)$, introduced in Ref.~\cite{detomasi2020rare} and referred to as the radial probability distribution. For an eigenstate $\ket{\psi}$, it is defined as 
\eq{
	\Pi_\psi^{\pd}(r) = \sum_{J: r_{I_0J}^{\pd}=r}|\psi_J^{\pd}|^2\,,
	\label{eq:Pir}
}
where $I_0$  is the Fock-space site on which the wavefunction has maximum amplitude, {\it i.e.} $|\psi_{I_0}|^2=\sup_I|\psi_I|^2$. Physically, $\Pi_\psi(r)$ gives the total weight of the eigenstate on Fock-space sites at Hamming distance $r$ from the site ($I_{0}$) on which it is site-localised at $\Gamma=0$. Similarly to $F_{\psi}(r)$, $\Pi_{\psi}(r)$ can also be interpreted as a probability distribution, so one can again define a mean distance and associated variance, as
\eq{
	\braket{d}=\sum_{r=0}^L r\Pi_{\psi}^{\pd}(r)\,,
	~\braket{(\delta d)^2}=\sum_{r=0}^L r^2\Pi_{\psi}^{\pd}(r)-\braket{d}^2\,.
	\label{eq:d-dsq}
}
The form of $\Pi_{\psi}(r)$ bears many similarities to that of $F_{\psi}(r)$~\cite{detomasi2020rare}
(and we return to it again below). In particular, $\overline{\Pi_{\psi}}(r)$ is also  described by a binomial distribution $\overline{\Pi_{\psi}}(r)=\binom{L}{r}p_\Pi^r(1-p_\Pi)^{L-r}$ with $p_\Pi = (1+e^{1/\xi_\Pi})^{-1}$;
and with $\xi_{\Pi}$ a Fock-space length scale which is likewise finite throughout the MBL phase, with
$\overline{\Pi_{\psi}}(r)\propto N_r e^{-r/\xi_\Pi}$ (analogous to $\overline{F_{\psi}}(r)$, Eq.~\ref{eq:FMBL}).


\subsection{Eigenstate anatomy and real-space observables\label{sec:anatomyRSO}}

We turn now to connections between the anatomy of eigenstates on the Fock space, and real-space observables. A key result in this context is an explicit relation~\cite{roy2021fockspace}  between the local $\sigma^z$-polarisation in eigenstates and the moments of $F_{\psi}(r)$ discussed above. The $\sigma^z$-polarisation in an eigenstate $\ket{\psi}$ for a given sample $S$ is defined as 
\eq{
    {\cal M}_S^{\pd} = L^{-1}\sum_{i=1}^L\braket{\psi|\sigma^z_i|\psi}^2\,,
\label{eq:MS-tau}		
}
where the summand is simply the infinite-time limit of the local spin autocorrelation $\langle \psi |\sigma_{i}^{z}(t)\sigma_{i}^{z}|\psi\rangle$. Physically, $\mathcal{M}_{S}$ provides a measure of how polarised  the spins are along the $z$-direction, and hence how `close' the eigenstate is to  a $\sigma^z$-product state. Using the definitions of the Hamming distance in Eq.~\ref{eq:hamming-dist} and $\braket{r}$ in Eq.~\ref{eq:rrsq-def}, it can be shown that~\cite{roy2021fockspace}
\eq{
{\cal M}_S^{\pd} = 1 - 2\frac{\braket{r}}{L}\,,
\label{eq:msrconn}
}
such that upon disorder averaging we have (Eq.~\ref{eq:rrsq}) $\overline{{\cal M}_{S}} = 1 - 2p$. In the ergodic phase, $p=1/2$ thus automatically implies the vanishing polarisation characteristic of that phase; while in the MBL phase  (via Eq.~\ref{eq:pdef})
\eq{\overline{{\cal M}_{S}} = (e^{1/\xi_F}-1)/(e^{1/\xi_F}+1)\,,
\label{eq:MSMBL}
} 
which is strictly non-zero for any finite $\xi_{F}$,  indicative of the persistent memory of initial conditions that is symptomatic of the MBL phase. One upshot of  this concrete relation  is that understanding the critical scaling of the IPRs and PEs (see Sec.~\ref{sec:scaling}) provides insights into the critical behaviour of $\xi_F$, which in turns provides a direct understanding of the scaling of real-space local observables in the vicinity of the transition.

We add further that explicit relations between quantities on real space and those on Fock space, are not  confined to the mean polarisations and mean $\braket{r}$. One can also consider higher-point correlation functions of eigenstate amplitudes, which are related to the fluctuations of the polarisation or even the entanglement structure of eigenstates~\cite{roy2022hilbert}. 
As a concrete example, inter-sample fluctuations of ${\cal M}_S$, defined as $\chi_{\rm inter}=\overline{{\cal M}_S^2}-\overline{{\cal M}_S}^2$, can be shown to be 
$\chi_{\rm inter} = \sum_{r,s}rs~ C_F(r,s)$, with
\eq{
C_F(r,s)^{\pd}=\overline{F_{\psi}(r)F_{\psi}(s)}-\overline{F_{\psi}(r)}\,~\overline{F_{\psi}(s)}\,;
}
and where the structure of the Fock-space correlation $C_F(r,s)$ is qualitatively different in the two phases~\cite{roy2021fockspace}.

As a further illustration of the connections between quantities on Fock- and real-space, we discuss briefly and heuristically the relation between the radial probability distribution $\Pi(r)$~\cite{detomasi2020rare}  and the $\ell$-bit (or LIOM) picture~\cite{serbyn2013local,huse2014phenomenology,ros2015integrals}. The basic premise of the  latter is that, sufficiently deep in the MBL phase, eigenstates are adiabatically connected to the $W\to\infty$ eigenstates via quasilocal unitary transformations: $\ket{\psi}=U\ket{I_0}$ with $U$ the quasilocal unitary in question (and $I_{0}$ again as above). The trivially localised operators in the $W\to\infty$ limit, $\sigma^z_i$, are also dressed by the quasilocal $U$. They are referred to as $\ell$-bits and denoted by $\tau^z_i=U\sigma_{i}^{z}U^{\dagger}$; with MBL eigenstates simultaneous eigenstates of both the $\sigma^z_i$ and $\tau^z_i$ operators. It can be shown~\cite{detomasi2020rare}  that the average distance $\braket{d}$, defined in Eq.~\ref{eq:d-dsq} can be expressed as
\eq{
	\frac{\overline{\braket{d}}}{L}= (1+e^{1/\xi_{\Pi}^{\pd}})^{-1}=\frac{1}{2}-\frac{1}{2L}\sum_{i}\overline{\braket{\psi|\sigma^z_i\tau^z_i|\psi}}\,.
	\label{eq:d-sigmatau}
}
The $\ell$-bit operator can be decomposed as 
\eq{
	\tau^z_i = \frac{1}{\mathcal{N}}\left[\frac{ h_i^{\pd}\sigma^z_i+\Gamma\sigma^x_i}{\sqrt{h_i^2+\Gamma^2}}\mathrm{sgn}(h_i)+\sum_{j\neq i,\mu}J_{j,\mu}^{\pd}e^{-\vert i-j\vert/\zeta}\sigma^\mu_j+\cdots\right]\,
	\label{eq:tauz}
}
where the first term is just the on-site rotation of the operator, and the second  (with $\braket{J_{j,\mu}^2}=J^2$) denotes the exponentially decaying support of the $\ell$-bit operators with a lengthscale $\zeta\equiv\zeta(W,J)$
(such that only the first term in Eq.\ \ref{eq:tauz} survives in the extreme MBL limit $\zeta \to 0$).

Using Eqs.~\ref{eq:tauz},\ref{eq:d-sigmatau}, and estimating the normalisation factor  $\mathcal{N}$ by ensuring that the operator norm of $\tau_i^z$ is conserved, an explicit relation between the Fock-space lengthscale $\xi_\Pi$ and the 
real-space $\ell$-bit localisation length $\zeta$ can then be obtained, as
\eq{
	1+\frac{J^2}{1-e^{-2/\zeta}} = g_W^2\left(\frac{e^{1/\xi_\Pi^{\pd}}+1}{e^{1/\xi_\Pi^{\pd}}-1}\right)^2\,
	\label{eq:xi-zeta}
}
where $g_W = (\sqrt{W^2+1}-1)/W$. This direct connection between a Fock-space localisation length $\xi_{\Pi}$ and the $\ell$-bit localisation length $\zeta$, again implies that the critical behaviour of  Fock-space lengthscales can provide a direct handle on that of the $\ell$-bit operators.


\subsection{Dynamical eigenstate correlations}

So far, we have reviewed properties of single eigenstates from the viewpoint of their anatomy on the Fock space. However, since many-body localisation is a fundamentally dynamical phenomenon, it is of natural interest to consider dynamical eigenstate correlations, {\it i.e.} correlations between different eigenstates resolved by their energy differences. To this end, following Refs.~\cite{tikhonov2021eigenstate,tikhonov2021anderson}, 
we define a dynamical eigenstate correlation by
\eq{
G(\omega) = \big\langle\sum_{I}|\psi_I^{\pd}|^2|\phi_I^{\pd}|^2\delta(E_\psi^{\pd}-E_\phi^{\pd}-\omega)\big\rangle\,,
\label{eq:Gomega}
}
where $\braket{\cdots}$ denotes an average over disorder realisations and a selection of eigenstate pairs $(\psi,\phi)$ from the middle of the spectrum. 

\begin{figure}
\includegraphics[width=\linewidth]{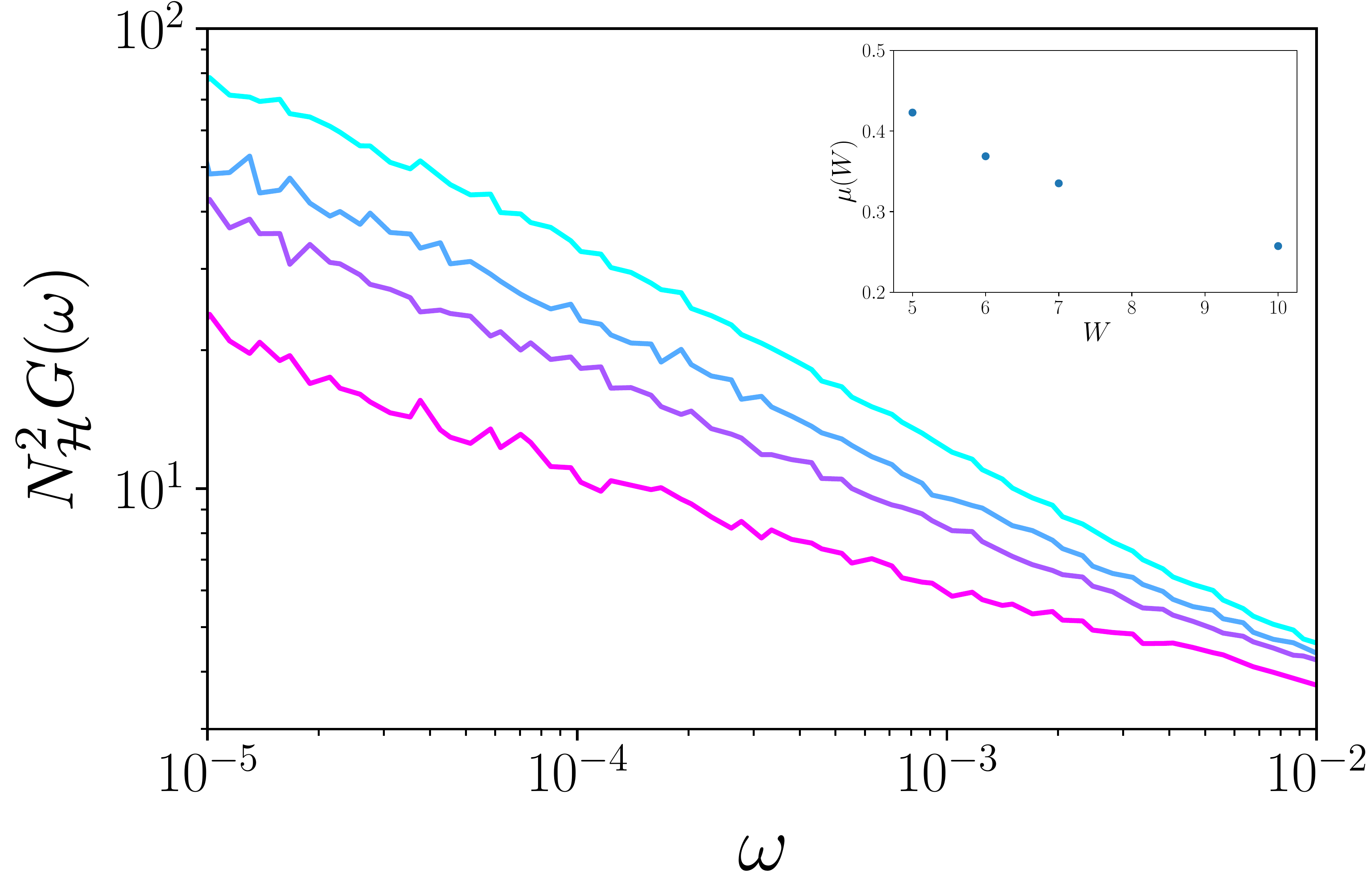}
\caption{
Dynamical eigenstate correlation function $G(\omega)$ (Eq.~\ref{eq:Gomega}) for the disordered XXZ chain with $ L = 16$~\cite{tikhonov2021eigenstate}, shown in the MBL phase for disorder strengths $W = 5, 6, 7, 10$ (from top to bottom).
The power-law decay  $G(\omega)\sim \omega^{-\mu}$ is evident. Inset shows the power-law exponent $\mu$ as a function of disorder strength $W$. The figure is taken from Ref.~\cite{tikhonov2021eigenstate}, with parameters $J=1=J_{z}$ and
$W_{J_{z}}=0$.
}
\label{fig:dyncorr}
\end{figure}

In the ergodic phase, numerical results~\cite{tikhonov2021eigenstate} show that $G(\omega)$ is $\omega$-independent at small  $\omega$; while it decays with $\omega$ above an energy scale $\omega_\xi$, which itself vanishes as the MBL transition is approached from the ergodic phase. In the MBL phase, however, a power-law decay arises on the lowest energy scales,
\eq{
G(\omega)\sim \omega^{-\mu(W)}\,,
\label{eq:Gomegapower}
}
where the exponent $\mu(W)$ decreases with increasing $W$, as shown in Fig.~\ref{fig:dyncorr} taken from Ref.~\cite{tikhonov2021eigenstate}. This power-law behaviour  can be understood via counting of resonances between many-body states.
The essential idea is that resonance counting estimates  $G(\omega)$ by~\cite{tikhonov2021eigenstate}
\eq{
\omega G(\omega) \sim (\nh \Delta) p_{\mathrm{res}}^{\pd}(\omega) O_{\rm res}^{\pd}\,,
\label{eq:Gomega-formula}
}
where $\Delta \sim \sqrt{L}/\nh$ is the mean level spacing, $p_{\mathrm{res}}(\omega)$ is  the number of resonances for a given state within the frequency window $[\omega,2\omega]$, and $O_{\rm res}$ is the resonant overlap. 

For any many-body eigenstate $\ket{\psi}$, the number of possible resonances that can occur via $r^{\rm th}$-order processes is estimated as $N_{L,r}^{\pd} \sim L m^r$, with $m\sim O(1)$; where the factor of $m^r$ arises from the number of possible ways of reorganising $r$ contiguous local degrees of freedom, and the factor of $L$ reflects the fact that this region of $r$ spins can be anywhere on the chain. The effective matrix element for a such a process should be exponentially suppressed in $r$, $M(r)\sim e^{-r/\zeta_{\rm res}}$. For the process to be resonant, the energy difference between the two states should then be  $\omega\sim M(r)$, which allows one to define a frequency-dependent resonance  lengthscale as $r(\omega)\sim \zeta_{\rm res}\ln(1/\omega)$. The probability of such resonances within a frequency window $[\omega, 2\omega]$ is then estimated by
\eq{
p_{\rm res}^{\pd}(\omega)\sim \omega N_{L,r(\omega)} = \omega L m^{r(\omega)} = L\omega^{1-\zeta_{\rm res}\ln m}\,.
}
Due to the multifractal character of MBL eigenstates, the resonant overlap can also  be estimated to be $O_{\rm res}\sim \nh^{-\tau_2}$. Putting this together in Eq.~\ref{eq:Gomega-formula} shows that
\eq{
G(\omega) 
\propto \omega ^{-\zeta_{\rm res}\ln m}\,,
}
so from Eq.~\ref{eq:Gomegapower} we can identify $\mu(W) = \zeta_{\rm res}(W)\ln m$. 

Oversimplified though it is, the above picture successfully recovers the power-law behaviour $G(\omega)\propto \omega^{-\mu}$, and the fact that the exponent $\mu$ decreases with increasing $W$ (on the natural assumption that $\zeta_{\rm res}$ does likewise).


\section{Disorder-correlations on the Fock-space graph \label{sec:fscorrs}}

We turn now to correlations between the effective on-site energies, $\mathcal{E}_I$, on the Fock-space graph,  following  Ref.~\cite{roy2020fock}
(as mentioned in Sec.\ \ref{sec:FSGs}, there is no disorder in the hoppings $\{\mathcal{T}_{IJ}\}$).
While we  pointed out above that the one-body distribution of the $\mathcal{E}_{I}$'s is normal, with a variance scaling linearly with $L$, this section is dedicated to the joint distribution of $\{\mathcal{E}_I\}$ or a subset of them.
In particular, the joint distribution will be shown to be a multivariate Gaussian, with a covariance matrix that displays what we refer to as {\it maximal correlations} between the Fock-space site energies. 
We will make this concrete shortly but, in words, the site energies of two Fock-space sites are 
defined to be maximally correlated if their covariance equals the Fock-space site energy variance.
This will be seen to hold for any two Fock-space sites whose mutual Hamming distance (Eq.~\ref{eq:hamming-dist}) is subextensive, $r_{IJ}/L \overset{L\to\infty}{=} 0$. Physically, all such Fock-space site energies are thus in effect slaved to  each other, the ramifications of which will be considered in Secs.\ \ref{sec:props} ff. That the Fock-space site-energies are strongly correlated is one important facet of the problem that renders it very different from standard one-body localisation on high-dimensional graphs.

Before making the discussion quantitative, however, let us argue on general qualitative grounds for the presence of correlations and the multivariate Gaussian nature of the joint distributions. While the Fock-space site energies $\mathcal{E}_I$ are random, there are exponentially many (in $L$)  of them; but they generically arise from only an extensive ($\mathcal{O}(L)$) number of random terms, the $\{J_i\}$ and $\{h_i\}$. This is because $H_0$ is a sum of $\mathcal{O}(L)$ terms, each having support on only $\mathcal{O}(1)$ real-space sites.\footnote{For long-ranged and all-to-all models, the number of independent random terms in the Hamiltonian, while no longer extensive in $L$, is only polynomially large in $L$.
Since this remains exponentially smaller than $\nh$, the same arguments go through.}
 The $\nh$ Fock-space site energies are made up of weighted sums of the random couplings and fields 
(Eq.\ \ref{eq:FS-site-energy}). As such, $H_0$ can be viewed as a map which takes $\mathcal{O}(L)$ 
random numbers as inputs and produces the $\nh$ random Fock-space sites energies as output. Since the number of outputs is exponentially larger than the number of inputs, it is inevitable that the outputs are correlated.

These correlations naturally mean that the most fundamental, albeit non-trivial, object to study is the 
$\nh$-dimensional joint distribution of the Fock-space site energies $\{\mathcal{E}_I\}$. This can be considered to be the distribution of an $\nh$-dimensional vector, $\bm{\mathcal{E}}$, each component of which corresponds to a particular $\mathcal{E}_I$. As each site energy is a sum of $\mathcal{O}(L)$ independent random numbers, $\bm{\mathcal{E}}$ can be represented as a sum of $\mathcal{O}(L)$ independent random vectors. The multivariate version of the central limit theorem (CLT) then implies that the $\nh$-dimensional joint distribution must be a multivariate Gaussian.


\subsection{Joint distributions of Fock-space site energies \label{sec:fscorrsjoint}}

To make the above ideas concrete but in a somewhat general setting, consider $H_0$ of the form
\begin{equation}
    H_0^{\pd} = \sum_{i=1}^L\mathcal{J}_i^{\pd}\hat{O}(\sigma^z_{i},\cdots,\sigma^z_{i+m})
    \label{eq:hdiag-zz}
\end{equation}
where each $\mathcal{J}_i$ denotes the independent random couplings and fields associated with real-space site $i$, and $\hat{O}(\sigma^z_{i},\cdots,\sigma^z_{i+m})$ is an operator made up of $\sigma^z$-operators on sites $i$ through $i+m$ with a finite $m$. For the specific cases of the XXZ and the TFI models discussed in Sec.~\ref{sec:models}, $\mathcal{J}_i$ encodes $(J_i,h_i)$ and $\mathcal{J}_i\,\hat{O}(\sigma^z_{i},\cdots,\sigma^z_{i+m}) = J_i\sigma^z_i\sigma^z_{i+1}+h_i\sigma^z_i$. 
It is useful to realise at this stage that there is a distribution over both disorder realisations and over Fock-space sites. The latter is present even in the non-disordered, translationally-invariant limit, and stems from the configurational disorder due to the random configurations of the spins over all $I$. In the following, we will carefully 
distinguish between the two.

With the notation in Eq.~\ref{eq:hdiag-zz}, the vector $\bm{\mathcal{E}}$ can then be expressed as
\begin{equation}
    \bm{\mathcal{E}} = \begin{pmatrix}\mathcal{E}_1^{\pd}\\ \mathcal{E}_2^{\pd}\\ \vdots\\ \mathcal{E}_I^{\pd}\\\vdots\\ \mathcal{E}_{\nh}^{\pd}\end{pmatrix} = \sum_{i=1}^L \mathbf{v}_i^{\pd}=
    \sum_{i=1}^{L} \begin{pmatrix}\mathcal{J}_i^{\pd} O(s_{i}^{(1)},\cdots,s_{i+m}^{(1)})\\ \mathcal{J}_i^{\pd} O(s_{i}^{(2)},\cdots,s_{i+m}^{(2)})\\ \vdots\\ \mathcal{J}_i^{\pd} O(s_{i}^{(I)},\cdots,s_{i+m}^{(I)})\\\vdots\\ \mathcal{J}_i^{\pd} O(s_{i}^{(\nh)},\cdots,s_{i+m}^{(\nh)})\end{pmatrix},
    \label{eq:mvg-example}
\end{equation}
where each of the $\nh$-dimensional vectors $\mathbf{v}_i$ is independent due to the statistical independence of the $\mathcal{J}_i$; whence the distribution of $\bm{\mathcal{E}}$ over disorder realisations is a multivariate Gaussian via the multivariate CLT. We denote this distribution as 
\begin{equation}
\mathcal{P}_{\nh}^d(\bm{\mathcal{E}}) = \frac{1}{\sqrt{(2\pi)^{\nh}\vert \mathbf{C}_d\vert}}\exp\left[-\frac{1}{2}\bm{\mathcal{E}^\prime}{}^\mathrm{T} \cdot \mathbf{C}_d^{-1}\cdot \bm{\mathcal{E}^{\prime }}\right]\,,
\label{eq:mvg-disorder}
\end{equation}
where the superscript $d$ denotes that the distribution is over disorder realisations, 
and the subscript $\nh$ that it is an $\nh$-dimensional distribution.\footnote{
Note that, since any marginal distribution of a multivariate Gaussian remains of the same form, the distribution $\mathcal{P}_{\mathcal{D}}^{d}$ over some  $\mathcal{D}$-dimensional subspace  is also a multivariate Gaussian.}
$\mathbf{C}_d$ denotes the corresponding covariance matrix (with determinant $\vert \mathbf{C}_d\vert$).
The primed column/row vectors on the right side of Eq.\ \ref{eq:mvg-disorder} have components
\eq{
\mathcal{E}_I^\prime = \mathcal{E}_I^{\pd}-\mathcal{E}_I^0\,,
\label{eq:mvg-eizero}}
where $\mathcal{E}_I^0$ is the contribution to $\mathcal{E}_I$ from the 
non-disordered part and $\mathcal{E}_I^\prime$ is the contribution purely from the disordered part. Specifically, for the XXZ and TFI models, we have $\mathcal{E}_I^0=J_z\sum_i s_i^{I}s_{i+1}^I$ and $\mathcal{E}_I^\prime = \sum_i [J_i^\prime s_i^{I}s_{i+1}^I+h_i s_i^I]$ with $J_i^\prime = J_i-J_z$.

Eq.~\ref{eq:mvg-disorder} is the joint distribution of the site-energies with contributions solely from the external disorder. However the $\{\mathcal{E}_I^0\}$ --   contributions from the non-disordered part -- are themselves distributed over the Fock-space, reflecting the aforementioned configurational disorder. This motivates a $\mathcal{D}$-dimensional distribution \emph{over the Fock space}, of the $\{\mathcal{E}_I^0\}$ for a $\mathcal{D}$-dimensional subspace consisting of a Fock-space site and all its neighbours lying within a certain Hamming distance. Naturally, $\mathcal{D}\ll\nh$ for there to be a meaningful distribution. Physically, this can be thought of in the following terms: choosing a set of $\mathcal{D}$ neighbouring Fock-space sites generates a $\mathcal{D}$-dimensional vector of the corresponding $\mathcal{E}_I^0$'s. Since $\mathcal{D}\ll\nh$, this patch of Fock-space sites can be moved all over the Fock-space graph, leading to a collection of such $\mathcal{D}$-dimensional vectors. The distribution of these vectors over the Fock-space generates the $\mathcal{D}$ dimensional distribution of the $\mathcal{E}_I^0$'s; we denote it by $\mathcal{P}_\mathcal{D}^F$, where the superscript $F$ stands for Fock space and the subscript denotes that it is $\mathcal{D}$-dimensional.

Analogous to the argument for $\mathcal{P}^d$ following Eq.~\ref{eq:mvg-example}, an argument can again be constructed for $\mathcal{P}_\mathcal{D}^F$; where instead of the statistical independence of the $\mathcal{J}_{i}$'s, one invokes the statistical independence of the $O(s_{i}^{I},\cdots,s_{i+m}^{I})$ 
over Fock-space sites in different $\mathcal{D}$-dimensional patches of the Fock-space graph. This naturally relies on $\mathcal{D}\ll \nh$, and the local nature of the models become important. Such an argument leads to the conclusion that the $\mathcal{D}$-dimensional distribution of $\{\mathcal{E}_{I}^{0}\}$ is also a multivariate Gaussian,
\begin{equation}
\mathcal{P}_{\mathcal{D}}^F(\bm{\mathcal{E}}^0) = \frac{\exp\left[-\frac{1}{2}(\bm{\mathcal{E}}^0-\overline{\mathcal{E}^0})^{\mathrm{T}} \cdot \mathbf{C}_F^{-1}\cdot (\bm{\mathcal{E}}^0-\overline{\mathcal{E}^0})\right]}{\sqrt{(2\pi)^{\mathcal{D}}\vert \mathbf{C}_F\vert}}\,,
\label{eq:mvg-fs}
\end{equation}
with  $\mathbf{C}_F$ the corresponding covariance matrix and $\overline{\mathcal{E}^0}=\nh^{-1}\sum_{I}\mathcal{E}_I^0$ denotes the mean of $\mathcal{E}_{I}^{0}$ over the Fock space.

Having established the nature of the two distributions, we next need to put them together. Note that 
$\mathcal{P}^d_\mathcal{D}$ is a conditional distribution for the $\mathcal{E}_I$'s \emph{given} a set $\{\mathcal{E}_I^0\}$, while $\mathcal{P}_\mathcal{D}^F$ in Eq.~\ref{eq:mvg-fs} is itself a distribution of the $\mathcal{E}_{I}^0$'s. Hence a $\mathcal{D}$-dimensional distribution over both disorder realisations and the Fock space is simply a convolution,
\begin{align}
\mathcal{P}_\mathcal{D}^{\pd}(\bm{\mathcal{E}}) =& (\mathcal{P}_\mathcal{D}^d\ast\mathcal{P}^F_\mathcal{D})(\bm{\mathcal{E}})\nonumber\\
=&\frac{\exp\left[-\frac{1}{2}\bm{\mathcal{E}}^{\mathrm{T}} \cdot \mathbf{C}^{-1}\cdot \bm{\mathcal{E}}\right]}{\sqrt{(2\pi)^{\mathcal{D}}\vert \mathbf{C}\vert}},
\label{eq:mvg-disorder-fs}
\end{align}
where we use the fact that the convolution of two multivariate Gaussians is again a multivariate Gaussian whose covariance matrix is the sum of the covariance matrices of the two, {\it i.e.}
\begin{equation}
\mathbf{C}=\mathbf{C}_{d,\mathcal{D}}+\mathbf{C}_F
\label{eq:cov-conv}
\end{equation}
where $\mathbf{C}_{d,\mathcal{D}}$ is the $\mathcal{D}$-dimensional submatrix of $\mathbf{C}_d$ appearing in Eq.~\ref{eq:mvg-disorder}.

 Note that the covariance matrix $\mathbf{C}$ itself scales as $L$ (although shown explicitly in the following subsection, this is readily understood from the fact that the vector $\bm{\E}$ in Eq.~\ref{eq:mvg-example} is a sum of $L$ independent vectors); so that it is $\mathbf{C}/L$ whose elements remain finite as $L\to \infty$. Hence, for the distribution $\mathcal{P}_\mathcal{D}$ in Eq.~\ref{eq:mvg-disorder-fs} to remain well defined in the thermodynamic limit, one needs to consider the joint distribution of $\bm{\E}/\sqrt{L}$ (rather than of $\bm{\E}$ itself); and in terms of which rescaled energies all elements of the covariance matrix are $O(1)$. This provides further evidence for the fact -- and re-emphasises the point made earlier -- that for any computation to remain well defined in the thermodynamic limit, all 
energies must be scaled by $\sqrt{L}$.


\subsection{Covariance matrix}
The final ingredient required for full characterisation of the Fock-space site-energy distributions is the explicit structure of the covariance matrix. This of course depends on the microscopic details of 
$H_0$, and here we consider the TFI and XXZ models, for which the Fock-space site energy is given by Eq.~\ref{eq:FS-site-energy}. Consider two Fock-space sites $I$ and $J$ separated by a Hamming distance $r_{IJ}$. 

Let us first compute the contribution to the covariance arising from the disordered component of $H_0$, which we denote by $C^{IJ}_d$ (this is simply the $(I,J)$ component of the correlation matrix $\mathbf{C}_d$). In the following, $\langle\cdots\rangle$ denotes a disorder average.
Recalling from above that   $\mathcal{E}_I^\prime = \sum_i [J_i^\prime s_i^{I}s_{i+1}^I+h_i s_i^I]$ with $J_i^\prime = J_i-J_z$, and recognising that $\braket{\mathcal{E}_I^\prime}=0$, then  $C_{d}^{IJ} \equiv \langle \mathcal{E}_{I}^{\prime}\mathcal{E}_{J}^{\prime}\rangle$ is given by
\begin{equation}
\begin{aligned}
C_d^{IJ} &=\sum_{i,j}[\braket{J_i^{\prime} J_j^{\prime}}s^{I}_{i}s^{I}_{i+1}s^{J}_{j}s^{J}_{j+1}
+
\braket{h_i^{\pd} h_j^{\pd}}s_i^{I}s_j^{J}]\\
&=\tfrac{1}{3}\sum_{i}[\underbrace{W_{J_{z}}^2(s^{I}_{i}s^{J}_{i})(s^{I}_{i+1}s^{J}_{i+1})}_{C_{d,\mathrm{int}}^{IJ}}
+
\underbrace{W^2s_i^{I}s_i^{J}}_{C_{d,\mathrm{fields}}^{IJ}}]\,,
\end{aligned}
\label{eq:cdIK-tfi}
\end{equation}
on using $\braket{J_i^\prime J_j^\prime}=W_{J_{z}}^2\delta_{ij}/3$ and $\braket{h_i h_j}=W^2\delta_{ij}/3$.  We next calculate the separate contributions due to the disordered couplings ($C_{d,\mathrm{int}}^{IJ}$) and fields ($C_{d,\mathrm{fields}}^{IJ}$).

$C_{d,\mathrm{int}}^{IJ}$ in Eq.~\eqref{eq:cdIK-tfi} can be determined as follows. There are $r_{IJ}$ real-space sites with $\ssh[I]i\ssh[J]i =-1$, corresponding to the $r_{IJ}$ sites where the spin configurations are different, whereas $\ssh[I]i\ssh[J]i =+1$ for the remaining
$L-r_{IJ}$ sites. Now consider a site $i$ where  $\ssh[I]i\ssh[J]i =-1$. Then the probability that on any other site $j\neq i$ we have $\ssh[I]{j}\ssh[J]{j} =-1$ is $\phi = (r_{IJ}-1)/(L-1)$, since the $r_{IJ}-1$ remaining sites where the configurations are different are uniformly distributed over the remaining $L-1$ sites; the probability that $\ssh[I]{j}\ssh[J]{j} = +1$ is then $1-\phi$.
Similarly, if $\ssh[I]i\ssh[J]i =+1$, then the probability that $\ssh[I]{j}\ssh[J]{j} =-1$ on a site $j\neq i$ is $r_{IJ}/(L-1)$, and the probability that $\ssh[I]{j}\ssh[J]{j} = +1$ is $(L-r_{IJ}-1)/(L-1)$.
$C_{d,\mathrm{int}}^{IJ}$ follows directly using these probabilities, and to leading order as $L\to \infty$ is given by
\begin{equation}
C_{d,\mathrm{int}}^{IJ} = L\frac{W_{J_{z}}^2}{3}\left(1-2\frac{r_{IJ}^{\pd}}{L}\right)^2.
\label{eq:tfi-cdrJ}
\end{equation}
This is $\propto L$ as expected, whence it is $\tilde{C}_{d,\mathrm{int}}^{IJ}=C_{d,\mathrm{int}}^{IJ}/L$ which has a well-defined thermodynamic limit. As derived above, Eq.\ \ref{eq:tfi-cdrJ} is in effect an average over all Fock-space site-pairs with fixed $r_{IJ}$. But corresponding fluctuations can also be calculated, shown to be
$\mathrm{std}(\tilde{C}_{d,\mathrm{int}}^{IJ})\propto 1/\sqrt{L}$ for any finite $r_{IJ}/L$,
and thus vanish as $L\to \infty$.

The second term in Eq.~\ref{eq:cdIK-tfi} containing the contributions of the disordered fields, $C_{d,\mathrm{fields}}^{IJ}$, is readily obtained. Since there are $r_{IJ}$ sites with
$\ssh[I]i\ssh[J]i =-1$ and $L-r_{IJ}$ where it is +1, it follows trivially that
\begin{equation}
C_{d,\mathrm{fields}}^{IJ}=L\frac{W^2}{3}\left(1-2\frac{r_{IJ}^{\pd}}{L}\right)
\label{eq:tfi-cdrh}.
\end{equation}
$C_{d}^{IJ}=C_{d,\mathrm{int}}^{IJ}+C_{d,\mathrm{fields}}^{IJ}$ then follows. An important consequence of this result is that $C_{d}^{IJ}$ depends only on the Hamming distance $r_{IJ}$ between the Fock-space sites $I$ and $J$, and not on the details of the two configurations. As such, any two Fock-space sites separated by a Hamming distance $r$ have the same covariance $C_d(r)$, where
\begin{equation}
C_{d}^{\pd}(r)= L\frac{W_{J_{z}}^2}{3}\left(1-2\frac{r}{L}\right)^2+L\frac{W^2}{3}\left(1-2\frac{r}{L}\right)\,.
\label{eq:tfi-Cdr}
\end{equation}
 The covariance matrix $C_F(r)$ can likewise be obtained (the requisite calculation is in fact 
similar to that of Eq.\ \ref{eq:tfi-cdrJ}~\cite{roy2020fock}), and is found to be
$C_F(r)=LJ_{z}^{2}(1-2\tfrac{r}{L})^{2}$.

The final expression for the total variance, $C(r)=C_{d}(r)+C_{F}(r)$,  thus follows as 
\begin{equation}
C(r)=L\left[\big(J_z^2+\tfrac{1}{3}W_{J_{z}}^2
\big)\left(1-2\frac{r}{L}\right)^2 + \tfrac{1}{3}W^{2}
\left(1-2\frac{r}{L}\right)\right]\,
\label{eq:tfi-cr}
\end{equation}
(from which can be read off the effective Fock-space disorder strength $W_\mathrm{FS}$, given by
$C(0)=\mu_{\mathcal{E}}^{2}=W_\mathrm{FS}^{2}L$). The most important feature of Eq.\ \ref{eq:tfi-cr} is that the covariance is a function of $r/L$.  In consequence, the covariance between the site energies 
of any pair of Fock-space sites which are separated by a subextensive distance ($r/L\overset{L\to\infty}=0$) is the same as the variance of the Fock-space site energies -- this is precisely the meaning of {\it maximal correlations} in the effective Fock-space disorder.

\begin{figure}
\includegraphics[width=\linewidth]{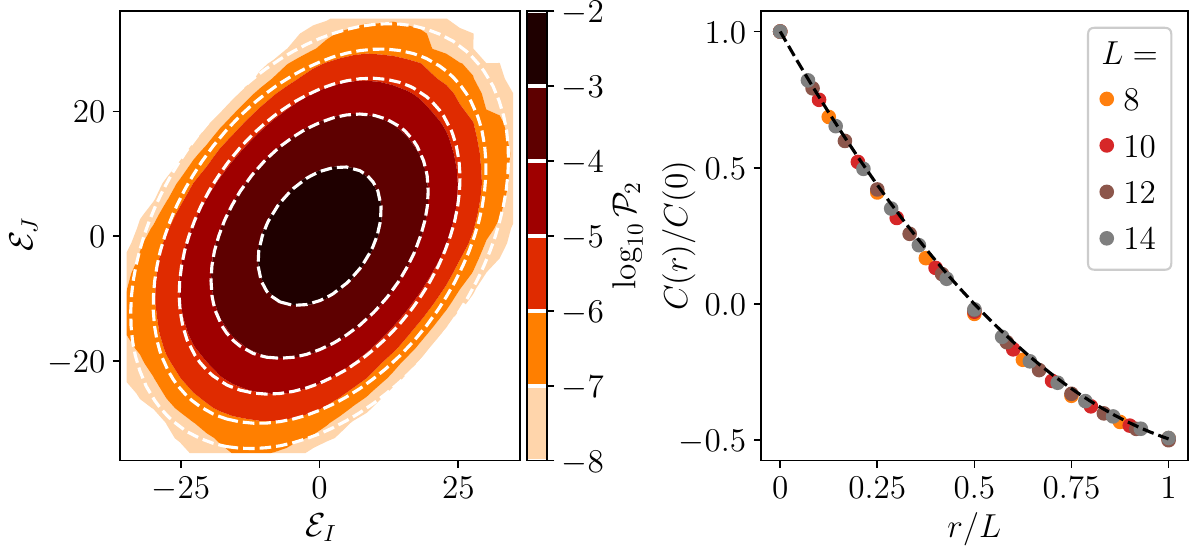}
\caption{Correlations in the effective Fock-space disorder for the Fock-space site energies given by Eq.~\ref{eq:FS-site-energy}. Left: 
2-body marginal distribution $\mathcal{P}_{2}$ over all pairs of $\mathcal{E}_I$ and $\mathcal{E}_J$ with
Hamming distance $r_{IJ}=4$ (shown for $L=14$ with $W=3,J_{z}=1, W_{J_{z}}=0.2$). The heatmap shows the numerically obtained distribution; white dashed lines show the corresponding contours for a bivariate Gaussian with a covariance given by Eq.~\ref{eq:tfi-cr}. Right:  Covariance $C(r)/C(0)$ \emph{vs} 
$r/L$ for several system sizes $L$, compared to the analytical result Eq.~\ref{eq:tfi-cr} (black dashed line). Numerical data is for the TFI model; corresponding data for the XXZ model is indistinguishable from that shown.
}
\label{fig:fs-corrs}
\end{figure}

Fig.~\ref{fig:fs-corrs} corroborates some  of these results numerically. Since a multivariate distribution is difficult to visualise, we consider (left panel) the $\mathcal{D}=2$-body marginal distribution $\mathcal{P}_{2}$ of the Fock-space site energies over all pairs of sites separated by a fixed Hamming distance $r$; comparing the numerically-determined heatmap for such with the corresponding contours for a bivariate Gaussian with a covariance given by Eq.~\ref{eq:tfi-cr}. The right panel compares 
directly the numerically computed $C(r)/C(0)$ for different system sizes $L$, with the expression in 
Eq.~\ref{eq:tfi-cr}. Excellent agreement is evident in both cases.\\

Before concluding this section, let us summarise the salient features of the Fock-space site energy distributions: (i) The joint distribution of the Fock-space site energies is 
 a multivariate Gaussian. (ii) The covariance of this distribution scales linearly with $L$. (iii) The covariance depends only on the Hamming distance $r$ between the Fock-space sites, and in fact is a function of $r/L$ such that the site-energies for Fock-space sites separated by a subextensive distance are maximally correlated.


\section{Fock-space propagators and matrix elements \label{sec:props}}

Given the perspective of MBL as corresponding to localisation of a fictitious particle on the high-dimensional, correlated and disordered Fock-space graph, the Fock-space propagators  become natural quantities of importance. Their counterparts in conventional Anderson localisation have likewise long served as insightful quantities~\cite{anderson1958absence,abou-chacra1973self,economous1972existence,Economoubook}.
This section reviews various analytical and numerical approaches for obtaining the propagators, and
some of the insights they provide.

Of particular significance is the local propagator on the Fock-space, defined as
\eq{
	G_{II}^{\pd}(\omega) = \braket{I|(\omega^+-H)^{-1}|I} = \sum_{\psi}\frac{|\braket{\psi|I}|^2}{\omega^+-E_\psi}
	\label{eq:GII}
}
with $\omega^{+}=$ $\omega $$+$$i\eta$ and $\eta$ the regulator. It is related to two important (and themselves related)
probes of localisation, the local density of states (LDoS) and the imaginary part of the self-energy. We discuss the latter in some detail shortly, but here comment briefly on the former.

The LDoS for site $I$ is given by
\eq{
	{\cal D}_I^{\pd}(\omega) = -\frac{1}{\pi}{\rm Im}[G_{II}^{\pd}(\omega)] = \sum_{\psi}|\braket{\psi|I}|^2\delta(E_\psi-\omega),
	\label{eq:fs-ldos}
}
and measures the density of eigenstates at energy $\omega$ which overlap the given site $I$.
In the context of single-particle localisation (where $I$ is replaced by a real-space site), 
$D_I(\omega)$ as a function of $\omega$ is continuous in a delocalised phase in the thermodynamic limit, 
while it is pure point like in a localised phase. This is easily  understood physically, since  delocalised  
states at all energies overlap any given site; whereas in a localised phase only a few  eigenstates at specific energies, localised in the vicinity of the given site, can overlap it (because  localised states of the same energy cannot coexist in the same region of space). The  corresponding situation in the MBL problem on the Fock-space graph is by contrast more subtle. In the ergodic phase, eigenstates are  extended over the graph, and the LDoS forms a continuum. But in the MBL phase, the multifractal character of MBL eigenstates (Sec.~\ref{sec:anatomy}) implies they overlap an exponentially large number  of Fock-space sites (albeit a vanishing fraction),
and consequently the LDoS forms a continuum in this case too.


\subsubsection{Scaling of propagators with system size}

Recall from Sec.~\ref{sec:fscorrs} that the covariance matrix for the effective Fock-space disorder scales
as $\mu_{\mathcal{E}}^{2} \propto L$ (reflecting at heart both the locality of the many-body Hamiltonian and the fact that it `lives' on energy scales $\propto \sqrt{L}$). As a result,  for the joint distribution of the Fock-space site energies to have a well-defined thermodynamic limit, the site energies must be rescaled by $\sqrt{L}$.
This is one example of the general principle that for any calculation on the Fock space to admit a well-defined thermodynamic limit, all energies must be rescaled by $\mu_{\mathcal{E}} \propto\sqrt{L}$; a further, explicit illustration of which was given in Fig.~\ref{fig:DoSXXZ} for the many-body density of states.
The same requirement naturally holds also for the  local propagator, together with its associated
self-energy, which is considered in the following subsections.


\subsection{Self-consistent mean-field theory \label{sec:self-consistent}}

We now outline what is arguably one of the simplest, but analytically tractable, theories for MBL in terms of the Fock-space propagators. Rather than focussing on the  propagator directly, a self-consistent mean-field theory 
is constructed for the local self-energy~\cite{logan2019many}. The self-energy of Fock-space site $I$, 
$S_I(\omega)$, is defined via the inverse of the local propagator \eqref{eq:GII} as
\eq{
G_{II}^{\pd}(\omega) = \frac{1}{\omega^+-{\cal E}_I^{\pd} - S_I^{\pd}(\omega)}\,,
\label{eq:self-energy-def}
}
with $S_I(\omega) = X_I(\omega)-i \Delta_I(\omega)$.  Attention naturally centres on the imaginary part, 
$\Delta_I(\omega)$, which, physically, is proportional to the rate of loss of probability from site $I$ into eigenstates of energy $\omega$ (it can be thought of roughly as a local linewidth or broadening).
As such, it constitutes a probabilistic order parameter for a localisation transition. In the context of single-particle problems, with unit probability $\Delta_I(\omega)$ is finite in the delocalised phase and vanishes $\propto \eta$ in the localised phase. As discussed above, however, in the many-body problem all many-body energy scales (including $\Delta_I$) must be rescaled by $\mu_{\mathcal{E}} \propto \sqrt{L}$ for them to admit a well-defined thermodynamic limit. 

The self-consistent mean-field theory can be set up by expressing the self-energy in terms of the  Feenberg renormalised perturbation series (RPS)~\footnote{Strictly speaking, the $G_{JJ}$ in Eq.~\ref{eq:SI-RPS} are propagators for sites $J$ with site $I$ removed from the problem, but the difference 
leads to neglectable $\mathcal{O}(1/\mu_{\mathcal{E}}^{2}) \propto 1/L$ corrections
to $\tilde{\Delta}_{I}$.}~\cite{feenberg1948note,watson1957multiple,Economoubook},
\eq{
	S_I^{\pd}(\omega) &= \sum_J{\cal T}_{IJ}^2 G_{JJ}^{\pd}(\omega)+\cdots\nonumber\\&=\sum_J\frac{{\cal T}_{IJ}^2}{\omega^+-{\cal E}_J^{\pd}-S_J^{\pd}(\omega)}+\cdots\label{eq:SI-RPS}\,,
}
with $\mathcal{T}_{IJ}$ and $\mathcal{E}_{J}$ defined in Eq.~\ref{eq:fs-ham-general}.
The form of this equation remains unchanged upon rescaling all many-body energy scales by $\sqrt{L}$, 
viz.\ $e \to \tilde{e} =e/\sqrt{L}$ where $e$ denotes any energy scale.
 With this notation, the rescaled imaginary part of the self-energy can be written as 
\eq{
	\tilde{\Delta}_I^{\pd}(\tilde{\omega}) = \sum_{J}\frac{\tilde{\cal T}_{IJ}^2 (\tilde{\eta}+\tilde{\Delta}_J^{\pd}(\tilde\omega))}{(\tilde{\omega}-\tilde{\cal E}_J^{\pd}-\tilde{X}_J^{\pd}(\omega))^2+(\tilde{\eta}+\tilde{\Delta}_J^{\pd}(\tilde\omega))^2}\,,
	\label{eq:Delta-RPS}
}
where the RPS has been truncated at leading order. With these rescalings, we expect $\tilde{\Delta}_I$ to be finite in the ergodic phase and to vanish in the MBL phase, both with unit probability, analogously to the single-particle case.

A self-consistent mean-field treatment  consists of the following three steps~\cite{logan2019many,roy2019self,roy2020fock}:
\begin{enumerate}
	 \item $\tilde X_J$ and $\tilde\Delta_J$ on the right-hand side of Eq.~\ref{eq:Delta-RPS} are replaced by their typical values, $\tilde{X_t}$ and $\tilde\Delta_t$; this constitutes the mean-field aspect of the theory.
	\item By averaging over the joint distribution of $\{\tilde{\cal E}_J\}$ (Sec.~\ref{sec:fscorrsjoint}), 
	a probability distribution for $\tilde{\Delta}$ is obtained from Eq.~\ref{eq:Delta-RPS}, which depends parametrically on $\tilde{\Delta}_t$;  this is denoted as $F_{\tilde \Delta}(\tilde{\Delta},\tilde\Delta_t)$.
	\item Self-consistency is then imposed by requiring that the typical value,  $\tilde{\Delta}_t$, obtained from $F_{\tilde \Delta}(\tilde{\Delta},\tilde\Delta_t)$, coincides with that which parametrically enters the distribution,
	{\it i.e.}
	\eq{
		\ln\tilde{\Delta}_t^{\pd} = \int_0^\infty d\tilde{\Delta}~\ln\tilde\Delta~F_{\tilde \Delta}^{\pd} (\tilde{\Delta},\tilde\Delta_t)\,.
	}
\end{enumerate}
The self-consistent $\tilde{\Delta}_t$ thus obtained is the `order parameter', which is then studied as a function of disorder strength (as detailed in~\cite{logan2019many,roy2019self,roy2020fock}).
 In particular, the self-consistent theory can be studied independently in either phase. 
In the ergodic phase, $\tilde\Delta_I\sim O(1)$, whence the limit of $\tilde{\eta}\to 0^+$ can be taken with impunity in Eq.~\ref{eq:Delta-RPS}, and  one works with the resulting expression in implementing steps 1-3 above.
In the MBL phase on the other hand, we expect $\tilde{\Delta}_I\propto\tilde{\eta}$ and hence it is $y_I\equiv \tilde{\Delta_I}/\tilde{\eta}$ that is the relevant quantity, with  Eq.~\ref{eq:Delta-RPS} taking the form
\eq{
	y_I^{\pd}(\omega) = \sum_{J}\frac{\tilde{\cal T}_{IJ}^2 (1+y_J^{\pd}(\tilde\omega))}{(\tilde{\omega}-\tilde{\cal E}_J^{\pd}-\tilde{X}_J^{\pd})^2}\,,
}
and self-consistency  imposed on $y_t=\tilde{\Delta}_{t}/\tilde{\eta}$. The breakdown of the self-consistency condition in either phase indicates the transition,  provided the breakdown occurs at the same disorder strength on approach from each phase. In all cases to which the theory has been applied~\cite{logan2019many,roy2019self,roy2020fock}, this is indeed the case, indicating an internal consistency of the approach. 

While the theory just sketched is relatively simple, a key point is that it is capable of explicitly taking into account the maximal correlations in the Fock-space site energies (Sec.~\ref{sec:fscorrsjoint}), as inherent in step 2 of the schema above. We now illustrate the consequences of this
in the context of locally interacting models.


\subsubsection{Self-consistent mean-field theory for short-ranged systems}

Consider locally interacting models, such as the disordered XXZ or TFI chains described in Sec.~\ref{sec:models}. In such  models, ${\cal T}_{IJ}$ is non-zero only for ($I,J$) site-pairs that differ by an $O(1)$ number of spin flips; and for such pairs,
\eq{
	\tilde{\cal E}_J^{\pd} = \tilde{\cal E}_I^{\pd} + \frac{W\times O(1)}{\sqrt L} \overset{L\to\infty}{\sim } \tilde{\cal E}_I^{\pd}\,.
	\label{eq:EJ=EI}
}
This reflects the maximal correlations: the reduced covariance for the Fock-space site energies,
$\rho(r)=C(r)/C(0)$ ($\propto C(r)/L$), has its maximum value of $1$ for all finite Hamming distances $r$,
whence the site energies $\tilde{\cal E}_J$ and $\tilde{\cal E}_I$ are effectively slaved to each other.
Coupled to the fact that the spin-flip terms in the disordered XXZ and TFI chains are of constant amplitude ($\mathcal{T}_{IJ}=J$ and $\Gamma$ respectively), Eq.~\ref{eq:EJ=EI} directly implies that the mean-field $\tilde{\Delta}_I(\tilde{\omega})$  arising from Eq.~\ref{eq:Delta-RPS} is
\eq{
	\tilde{\Delta}_I^{\pd}(\tilde{\omega}) = \Gamma^2\frac{ (\tilde{\eta}+\tilde{\Delta}_t^{\pd}(\tilde\omega))}{(\tilde{\omega}-\tilde{\cal E}_I^{\pd}-\tilde{X}_t^{\pd}(\tilde{\omega}))^2+(\tilde{\eta}+\tilde{\Delta}_t^{\pd}(\tilde\omega))^2}\,;
	\label{eq:Delta-RPS-corr}
}
analysis of which (via steps 2,3 above) shows an MBL phase to occur above a finite critical disorder strength~\cite{logan2019many,roy2020fock}.

A key feature of Eq.~\ref{eq:Delta-RPS-corr}, reflecting the maximal correlations, is that it now contains a single term -- as opposed to a sum of $O(L)$ independent terms, as would have arisen in the absence of maximal correlations in the Fock-space disorder. This is central to the origins of MBL~\cite{roy2020fock}. To exemplify that, consider the opposite limit, where the Fock-space site energies are completely
uncorrelated ($\rho(r)=0$ for all non-zero $r$); as embodied in the quantum random energy model (QREM)~\cite{derrida1980random,laumann2014many,baldwin2016manybody}, where each  $\tilde{\cal E}_I\sim{\cal N}(0,\sqrt{L}W)$ is an independent Gaussian random variable with standard deviation $\sqrt{L}W$. In this case,  the self-consistent theory yields~\cite{roy2020fock} a critical disorder of $W_c\sim \sqrt{L}\times\Gamma$ $\overset{L\to \infty}{\sim}\infty$;~\footnote{It is also worth noting what such a simple theory predicts for single-particle localisation on high-dimensional  graphs with connectivity $K$. Since the QREM is an effective single-particle problem on a graph with connectivity $K \sim L$, and uncorrelated disorder of strength $\sqrt{L}W$, the requisite answer follows simply on replacing $\sqrt{L}W \to W$. This gives the one-body critical $W_{c} \propto K$; which,
modulo logarithmic corrections, agrees with the large-$K$ result for a Bethe lattice~\cite{abou-chacra1973self}.} 
in other words, it predicts the \emph{absence} of an MBL phase at any finite disorder strength, 
in agreement with other studies of the QREM~\cite{laumann2014many,baldwin2016manybody}.
More generally, considering covariances $\rho(r)$ which for finite $r$ satisfy $0<\rho(r) <1$,
the self-consistent theory predicts that these too fall into the QREM class, such that an MBL phase
is not possible~\cite{roy2020fock}.

\begin{figure}
\includegraphics[width=\linewidth]{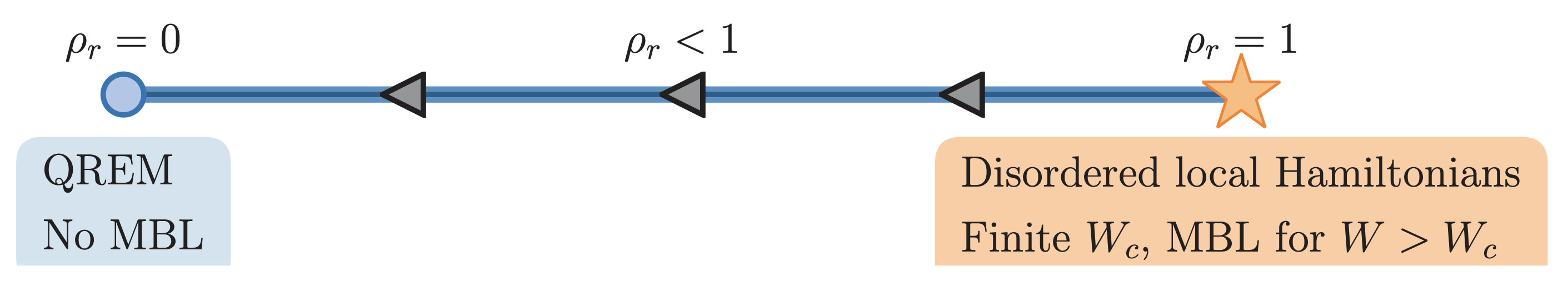}
\caption{Schematic of the central result from the self-consistent mean-field theory. Models with maximally correlated Fock-space disorder, where the rescaled covariance $\rho(r)=C(r)/C(0)$ of the Fock-space disorder
satisfies $\rho(r)=1$ for finite $r$, may host an MBL  phase above a finite disorder strength $W_{c}$. Locally interacting models fall in this category. However, any deviation from maximal correlations leads to destabilisation of MBL. Figure taken from Ref.~\cite{roy2020fock}.}
\label{fig:sc-mft-corrs}
\end{figure}

To summarise, the  most important outcome of the self-consistent mean-field theory is that maximal correlations in the Fock-space site energies is a  necessary condition for an MBL phase to be stable, as depicted schematically in Fig.~\ref{fig:sc-mft-corrs}. This prediction was tested  and supported numerically for specific models, both with and without the correlations,  using standard diagnostics computed using exact diagonalisation~\cite{roy2020fock}. It is of course important to realise that maximal correlations are not a sufficient condition for MBL to exist; several other agents can render MBL unstable, such as non-abelian symmetries~\cite{potter2016symmetry}, interplay of constraints and disorder~\cite{sierant2021constraint}, dimensional considerations~\cite{deroeck2017stability}, and sufficient long-rangedness in interactions.


\subsubsection{Self-consistent mean-field theory for long-ranged systems}
Following on from the final point above, let us briefly reprise the  outcome of the self-consistent mean-field theory applied to a long-ranged model; specifically, the power-law interacting, disordered XXZ chain given by Eq.~\ref{eq:hamLR-XXZ}~\cite{roy2019self}. 
Several works on similar models use arguments based on simple resonance counting and breakdown of the locator expansion, to suggest that MBL is not possible in 1D models where the interaction decays more slowly than $r^{-2}$~\cite{burin2006energy,yao2014manybody,burin2015manybody,burin2015localisation,tikhonov2018manybody}. 
Such arguments can however be contested, both on the grounds that simple resonance counting does not account for Fock-space correlations, and that the breakdown of the {\it bare} locator expansion does not itself guarantee delocalisation (a bare locator expansion cannot in fact ever converge in a thermodynamically large system).
Indeed, experimental setups where power-law interactions appear naturally~\cite{smith2015many,choi2017observation,rovny2018observation}, seem to suggest the presence of an MBL phase where such simplistic arguments would imply otherwise.  There is thus impetus to address this question using a theory  in which Fock-space correlations are explicitly taken into account.

\begin{figure*}
\includegraphics[width=\linewidth]{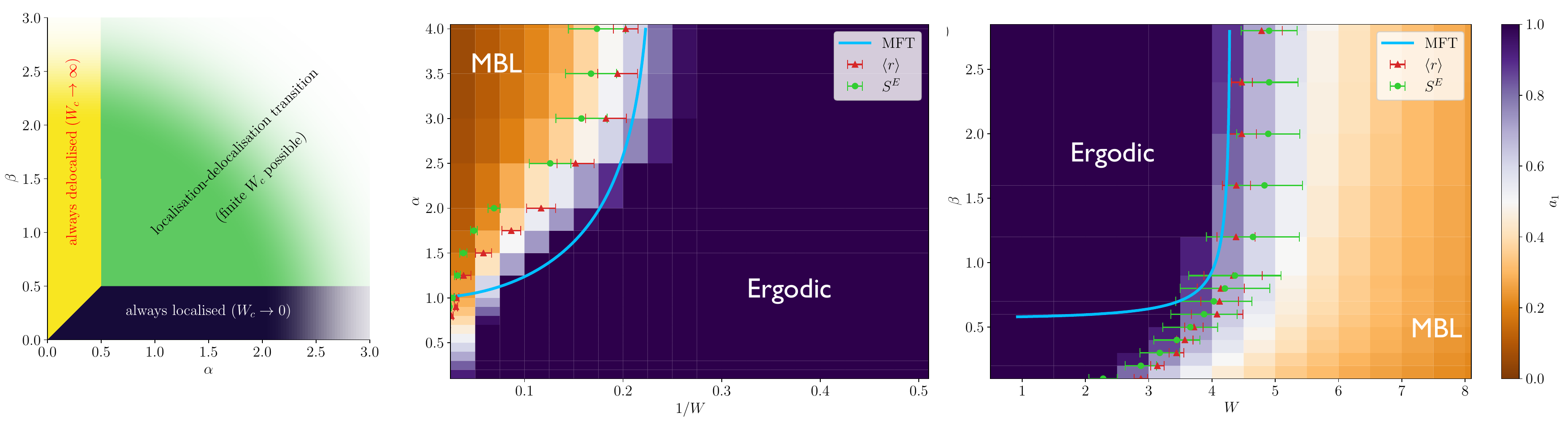}
\caption{Phase diagram of the long-ranged, disordered XXZ chain (Eq.~\ref{eq:hamLR-XXZ}). 
 Left panel shows the results from the self-consistent mean-field approach~\cite{roy2019self} as a function of $\alpha$ and $\beta$, the exponents which respectively govern the power-law decay of the transverse and longitudinal interactions. Middle (right) panel show the phase diagram in the $\alpha$-$W^{-1}$ ($\beta$-$W$) plane for fixed $\beta >1/2~(\alpha >1/2)$. The colour scale indicates the fractal dimension of eigenstates $a_1\equiv D_1$ (see Eq.~\ref{eq:PE}), whereas the two sets of data points indicate the 
transition obtained from level statistics and bipartite entanglement entropy. The blue line denotes the critical line obtained from the self-consistent mean-field theory. 
}
\label{fig:lrxxz}
\end{figure*}

The resultant phase diagram~\cite{roy2019self} is summarised in Fig.~\ref{fig:lrxxz}, as a function of the disorder strength $W$ and the exponents $\alpha$ and $\beta$  (which control respectively the decay of the transverse and longitudinal interactions in Eq.~\ref{eq:hamLR-XXZ}).
Quite strikingly, the theory predicts an MBL transition to be possible provided $\alpha, \beta$ are both $> 1/2$ (Fig.~\ref{fig:lrxxz} left panel), while by contrast, in the regime $\alpha<1/2$ with $\beta>\alpha$, an MBL phase is not possible. In addition, for fixed $\beta >1/2$, it was found that upon decreasing $\alpha$ -- {\it i.e.} making the transverse/spin-flip terms longer ranged --  the critical disorder $W_c$ grows (Fig.~\ref{fig:lrxxz} middle panel). In marked contrast, fixing $\alpha>1/2$ and decreasing the exponent $\beta$ ({\it i.e.} making the $\sigma^z$-$\sigma^z$ interactions longer ranged) is found to decrease $W_c$ (Fig.~\ref{fig:lrxxz} right panel), indicating an enhanced stability of the MBL phase with longer-ranged interactions. 
Predictions from the mean-field theory were also found to be in good overall agreement with 
the phase diagram obtained from exact diagonalisation studies of standard diagnostics, such as level statistics, and entanglement and participation entropies of eigenstates (see Fig.~\ref{fig:lrxxz} middle, right panels).

These results can be understood physically as follows. Recall that the disorder couples to the $\sigma^z$-component of  spins. Decreasing $\alpha$ thus increases the range of {\it transverse} (to $\sigma^z$) interactions. This facilitates spin-flips at longer ranges, which renders the system more prone to delocalisation  and thus increases the critical disorder. On the contrary, decreasing $\beta$ increases the range of {\it longitudinal} interactions, which provides the system with a rigidity against spin flips, as any spin flip has an energy cost involving several spins at longer distances. This cooperates with the external disorder and makes localisation more robust, thereby decreasing the critical disorder strength. Similar phenomenology was also observed in a model with disordered, long-ranged interacting fermions~\cite{2022fermionic}.


\subsubsection{Anderson localisation vs multifractality in mean-field theories \label{sec:al-v-mf-mft}}

Before concluding discussion of the self-consistent theory, it is useful to point out the issue of Anderson localisation {\it vs.} multifractality within it.
In the ergodic phase, the number of eigenstates which overlap a given Fock-space site is a finite fraction of the Fock-space dimension, whereas it is a vanishing fraction in the MBL phase.
In other words, for a finite system with Fock-space dimension $\nh$, the continuum nature of the LDoS sets in at scales $\propto \nh^{-1}$ set by the mean level spacing, whereas in the MBL phase the continuum sets in on a larger scale $\propto \nh^{-z}\gg\nh^{-1}$ ($0<z<1$).
  This can in fact be argued for quite generally for multifractal states~\cite{altshuler2016multifractal}.  The scale at which the LDoS is probed is set by the regulator $\eta$, and its 
dependence on $\nh$.

Within the self-consistent mean-field theory, however, the thermodynamic limit is taken first and from the outset, while the limit of $\eta\to 0^+$ is taken later. Consequently, while
the mean-field theory discussed above can  distinguish between ergodic and non-ergodic states, it cannot make the finer distinction within non-ergodic states between multifractal and Anderson localised states on Fock space. A self-consistent theory on the Fock-space which allows for $\eta$ to scale non-trivially with $\nh$, and also takes into account the Fock-space correlations, remains a work for the future.

 Notwithstanding these subtle issues about the order of limits, one general prediction of the self-consistent mean-field theory is the presence of L\'evy tails ($\propto y^{-3/2}$) in the distribution of the imaginary part of the self-energy (rescaled with $\eta$) in the MBL phase. This behaviour has been confirmed via exact numerical studies on finite-sized systems (with $\eta$ scaling appropriately with $\nh$),  such as exact diagonalisation~\cite{logan2019many} and the recursive Green's function method~\cite{sutradhar2022scaling}, 
the latter of which we turn to next.


\subsection{Recursive Green function method\label{sec:recursiveGF}}

While exact diagonalisation is often the method of choice for computing Fock-space propagators 
for finite-size systems, the associated computational costs can be severely constraining and are well documented.
In such a situation it is useful to have other numerically-exact methods available, even if they only offer quantitative improvement in terms of computational efficiency, and allow access to a few more spins or fermions.
The {\it recursive Green function method} provides one such example.

Originally introduced for Anderson localisation problems on regular lattices~\cite{lee1981anderson,mackinnon1980conductivity,mackinnon1983scaling}, the method can be adapted to the Fock-space graph, on which the local structure of the hoppings  (see Fig.~\ref{fig:fs-graphs}) again plays a crucial role~\cite{sutradhar2022scaling}. 
Two important points to remember here are that the topology of the Fock-space graph is such that, under hopping, any given Fock-space site (i) is connected to rows only one above and one below in the graph; and (ii) is {\it not} connected to any site on its own row. This structure then permits a {\it recursive} computation of the Green function, by iteratively adding each row of the Fock-space graph and updating the Green function for all sites up to that row. The size of the matrix requiring inversion at each step is thus proportional to the number of sites on the row added; at heart this is why the method is computationally cheaper than exact diagonalisation.

\begin{figure}
    \centering
    \includegraphics[width=\linewidth]{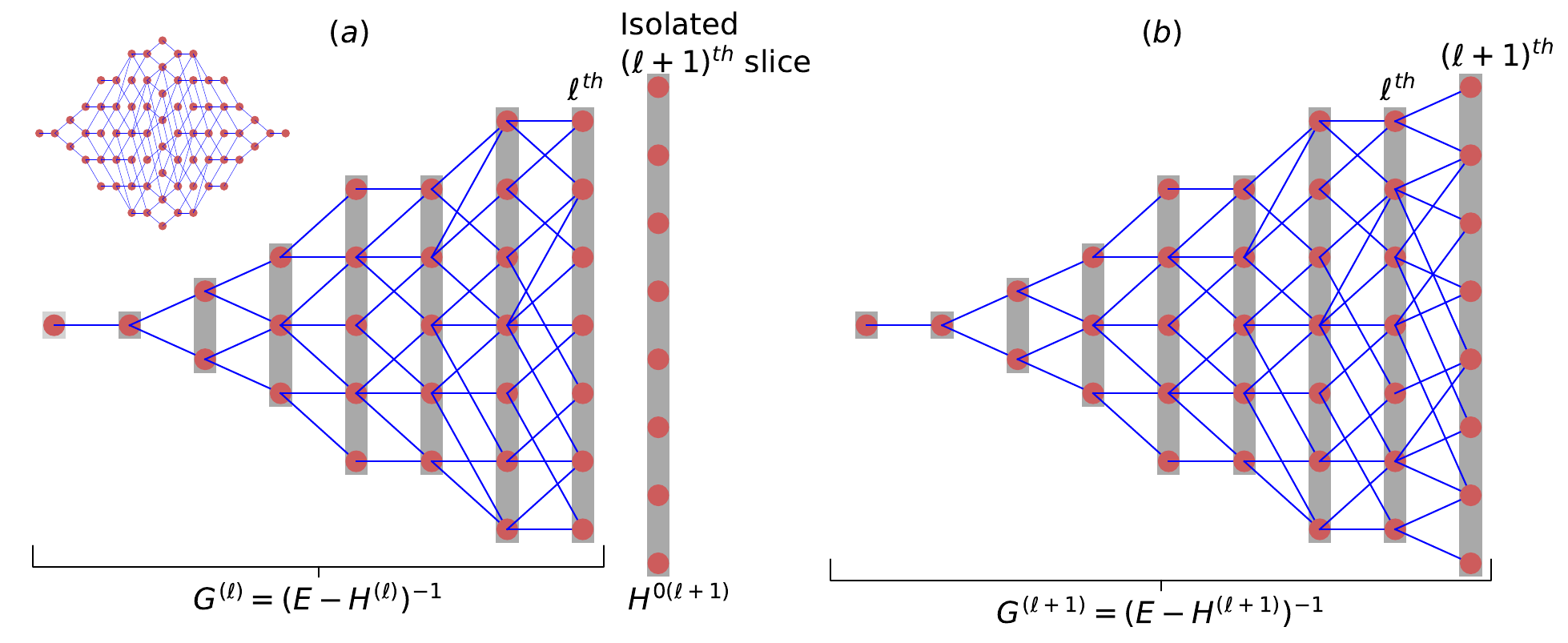}
    \caption{
		A graphical representation of the recursive Green function method. The partial Fock spaces are the same as in Fig.~\ref{fig:fs-graphs} (left) but rotated by $90^{\circ}$. (a) The situation at the start of the $(\ell+1)^{\rm th}$ step, where the $(\ell+1)^{\rm th}$ row/slice is added. (b) As slices are progressively added, the Green function of the left part is updated via the iterative equations in Eq.~\ref{eq:recursive-GF}. The figure is taken 
from Ref.~\cite{sutradhar2022scaling}.}
    \label{SFig:recursive_Gf}
\end{figure}
 
We sketch the basic steps of the formalism using the setting of the disordered XXZ chain (for further details 
see Ref.~\cite{sutradhar2022scaling}). Consider the Fock-space graph arranged as shown in Fig.~\ref{fig:fs-graphs}, with the two domain wall states forming the apices of the graph. Define $H^{(\ell)}$  as the Hamiltonian truncated at the $\ell^{\rm th}$ row of the Fock-space graph, with  $G^{(\ell)}=\left[\omega^+\mathds{1}-H^{(\ell)}\right]^{-1}$  the corresponding matrix Green function. The aforementioned structure of the Fock-space graph implies that $H^{(\ell+1)}$ can be written as the matrix
\eq{
H^{(\ell+1)}=\begin{pmatrix}
H^{(\ell)} & \mathcal{T}^{(\ell,\ell+1)}\\
{\cal T}^{(\ell+1,\ell)} & H^{(0)(\ell+1)}\end{pmatrix}\,,
}
where $H^{(0)(\ell+1)}$ contains solely the diagonal part of the Hamiltonian for the sites on row $\ell+1$, and $\mathcal{T}^{(\ell,\ell+1)}$ encodes the hoppings between the sites on rows $\ell$ and $\ell+1$. 
The iterative structure of the method means that $H^{(\ell+1)}$ is the Hamiltonian  at step $\ell+1$ of the process, and at this step the Green function can be written as 
\begin{equation}
G^{(\ell+1)}=\begin{pmatrix}
G^{(\ell+1)}_l & G^{(\ell+1)}_{lr}\\
G^{(\ell+1)}_{rl} & G^{(\ell+1)}_r
\end{pmatrix}\,.
\end{equation}
To understand the notation, it is convenient of think of a fictitious bipartition of the Fock-space graph, between all sites up to and including row $\ell$, and the sites on the row $\ell+1$ just added (see Fig.~\ref{SFig:recursive_Gf}). The subscripts $l$ and $r$ then denote sites to the left and right of this partition; for example, $G^{(\ell+1)}_l$ contains the propagagators for all sites up to and including row $\ell$, $G^{(\ell+1)}_{lr}$ contains the off-diagonal propagators between sites to the left and right of this bipartition, and so on.
In the same way, we denote ${\cal T}^{(\ell,\ell+1)}\equiv {\cal T}_{lr}$ since it connects the sites to the left of the bipartition to those on the right. With this, at the $(\ell+1)$-th step of recursion, the Green  function $G^{(\ell+1)}$ and the Hamiltonian $H^{(\ell+1)}$ satisfy the matrix equation
\eq{
\left[\omega^+\mathds{1}-H^{(\ell+1)}\right]G^{(\ell+1)}&=\mathds{1}\nonumber\\
\Rightarrow\begin{pmatrix}
\big(G^{(\ell)}\big)^{-1} & \mathcal{T}_{lr}, \\
\mathcal{T}_{rl} & \big(G^{0(\ell+1)}\big)^{-1}
\end{pmatrix}
\begin{pmatrix}
G^{(\ell+1)}_{l} & G^{(\ell+1)}_{lr} \\
G^{(\ell+1)}_{rl} & G^{(\ell+1)}_{r}
\end{pmatrix}
&=\mathds{1}\,
\label{Eq:2by2_Gr_fn_eqn_Fs}
}
where $\big(G^{(\ell)}\big)^{-1}=(\omega^+\mathds{1}-H^{(\ell)})$.  The action of this recursion is shown graphically in Fig.~\ref{SFig:recursive_Gf}. Equation \eqref{Eq:2by2_Gr_fn_eqn_Fs} can be simplied to yield the recursion relations
\eq{
    G^{(\ell+1)}_l &= G^{(\ell)}_l+G^{(\ell)}_l{\cal T}_{lr}G^{(\ell+1)}_r{\cal T}_{rl}{G}^{(\ell)}_l\nonumber\\
    G^{(\ell_+1)}_{lr}&= - G^{(\ell)}_l{\cal T}_{lr}G_r^{(\ell+1)}\nonumber\\
    G^{(\ell_+1)}_{rl}&= - G_r^{(\ell+1)}{\cal T}_{rl}G^{(\ell)}_l\nonumber\\
    G^{(\ell+1)}_r &= [(G^{0(\ell+1)})^{-1}-{\cal T}_{rl}G^{(\ell)}_l{\cal T}_{lr}]^{-1}\,\label{eq:recursive-GF}
}
where, crucially, in the final equation the inversion is performed over a matrix which is only as large as the dimension of the $(\ell+1)^{\rm th}$ row. This set of equations contains and enables access to all propagators, both local and non-local. Here we review some results for the local propagator; more precisely for the local self energy, which follows trivially from the local propagator using Eq.~\ref{eq:self-energy-def}.

\begin{figure}
\includegraphics[width=\linewidth]{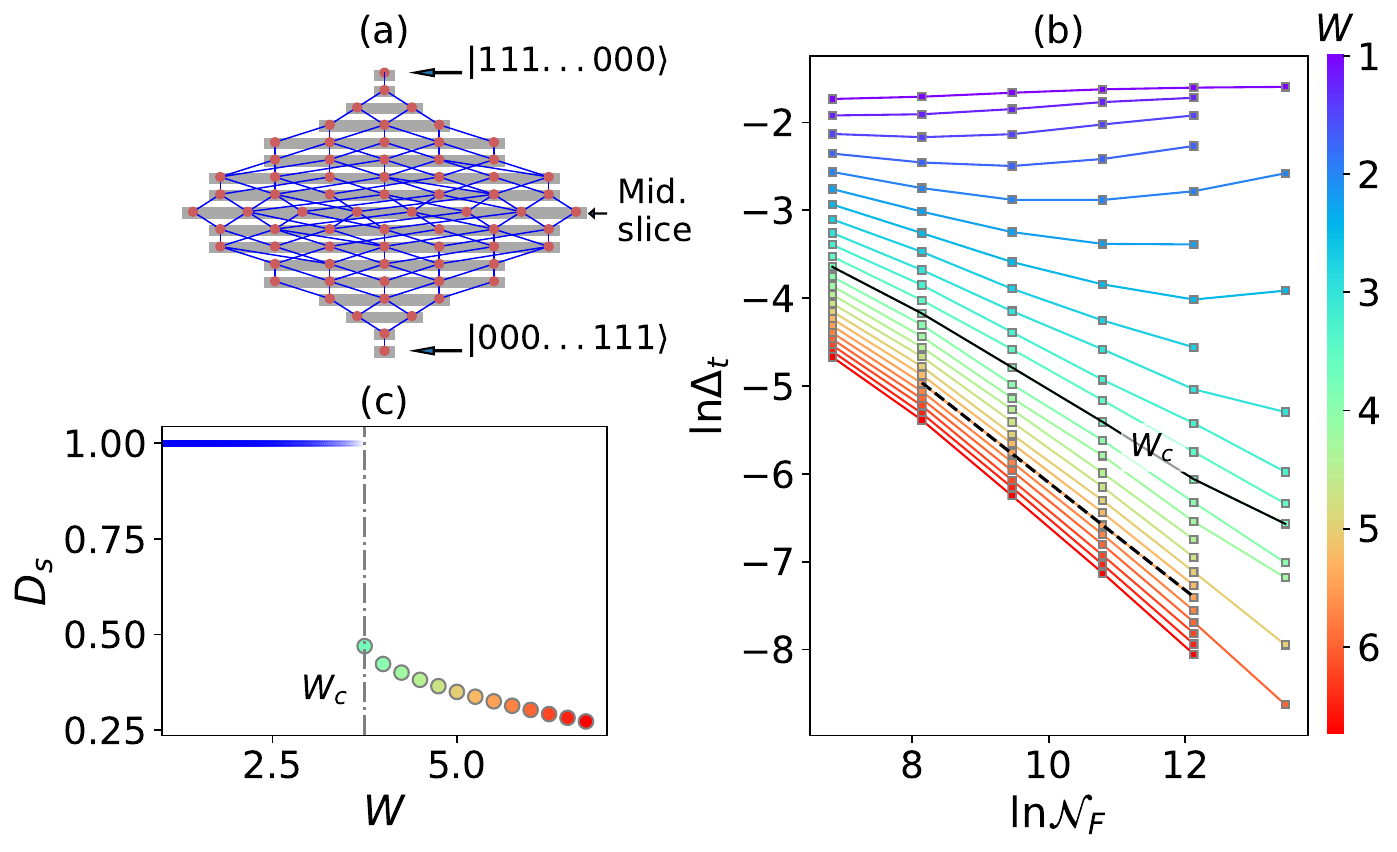}
\caption{(a) 
Fock-space graph of the disordered XXZ chain (as in Fig.~\ref{fig:fs-graphs})
for $L=8$; the top apex is $\ket{111..000}$, {\it i.e.} all spins up on the left half, and the bottom apex the state with all spins up on the right half. (b)  $\ln\tilde{\Delta}_t$ as a function of $\ln\nh \propto L$ for different $W$ (colour bar) across the MBL transition ($W_c$). $\tilde{\Delta}_t$  decays exponentially with $L$ for $W>W_c$, whereas it converges to an $L$-independent value for $W<W_c$.  (c) The fractal dimension $D_s$ 
obtained from the finite-size scaling theory (see Sec.~\ref{sec:scaling}) jumps discontinuously across the transition, from $D_s<1$ in the MBL phase to $D_s= 1$ in the thermal phase. At $W=W_c+$ (vertical dash-dotted line) 
$D_s\simeq 0.5$. The figure is taken from Ref.~\cite{sutradhar2022scaling}, and the notation ${\cal N}_F\equiv\nh$.}
\label{fig:recursive-GF-Deltat}
\end{figure}

In Fig.~\ref{fig:recursive-GF-Deltat}, results for the typical value of the imaginary part of the self-energy (scaled with $\sqrt{L}$ of course), are shown as a function of $\ln\nh\propto L$ for several disorder strengths for the XXZ chain~\cite{sutradhar2022scaling}. The behaviour is qualitatively different for weak and strong disorder. At weak disorder, $\tilde{\Delta}_t$ saturates with increasing $L$, such that it has a finite value in the thermodynamic limit. On the other hand, at strong disorder $\tilde{\Delta}_t$ decays as a power-law in the Fock-space dimension (exponentially with $L$),
\eq{
	\tilde{\Delta}_t \sim \nh^{-(1-D_s)},\quad D_s=\begin{cases}
	1 &\quad : W<W_c~{\rm ergodic}\\
	<1 & \quad : W>W_c~{\rm MBL}
	\end{cases}\,,
}
with $D_s$ referred to to as the spectral fractal exponent. The fact that $0<D_s<1$ in the MBL phase is intimately connected to the multifractality of MBL eigenstates reviewed  in Sec.~\ref{sec:anatomy}. In fact, for multifractal states $D_s$ can be shown to be the same as the fractal dimension $D$ extracted from the eigenstates under very general considerations~\cite{altshuler2016nonergodic}. 
A fractal dimension $D<1$ can be roughly understood via the estimate that the state has support on $\sim \nh^D$ Fock-space sites (a vanishingly small fraction of the total number of such).

It is also worth contrasting  one aspect of these results with those obtained from the self-consistent theory discussed in Sec.~\ref{sec:self-consistent}. In the latter, the thermodynamic limit $L\to\infty$ is taken before the limit $\eta\to 0$. As such, in the localised phase $\Delta_t \propto \eta$, and the theory 
cannot distinguish between multifractal and Anderson localised states, as mentioned in Sec.~\ref{sec:al-v-mf-mft}. 
But in any numerical calculation, such as the recursive Green function method (or exact diagonalisation), the computation is performed strictly for a finite-sized system. This allows $\tilde{\eta}$ to be slaved to $\nh$ in different ways, and the scaling of $\tilde{\Delta}_t$ with $\nh$ can in principle depend on $\tilde{\eta}$. In fact, in Ref.~\cite{altshuler2016nonergodic} it was argued that $\tilde{\Delta}_t$ saturates as a function of $\tilde{\eta}$ below an energy scale $\tilde{\eta}_c\sim \nh^{-z}$ ($0<z<1$); namely $\tilde{\Delta}_t\sim \tilde{\eta}_c^\theta\sim \nh^{-(1-D_s)}$ for $\tilde{\eta}\ll \tilde{\eta}_c$, whereas $\tilde{\Delta}_t\sim \tilde{\eta}^\theta$ for $\tilde{\eta}\gg \tilde{\eta}_c$.
Physically, $\tilde{\eta}_c$  is an energy scale characteristic of a multifractal state for any finite system, 
decreasing with increasing $\nh$ but being much larger than the mean level spacing $\propto \nh^{-1}$.
In the calculations of Ref.~\cite{sutradhar2022scaling}, $\tilde{\eta}\propto \nh^{-1}\ll \tilde{\eta}_c$ was employed. In this case, it is expected that $\tilde{\Delta}_t\sim \nh^{-(1-D_s)}$. That is indeed the behaviour for $\tilde{\Delta}_t(\nh)$  found in Fig.~\ref{fig:recursive-GF-Deltat} over the entire MBL range ($W>W_c$).


\subsection{Forward Scattering Approximation\label{sec:FSA}}

We now discuss 
briefly, following Ref.~\cite{pietracaprina2016forward},
the calculation of non-local propagators on the Fock-space within the Forward Scattering Approximation (FSA)~\cite{stern1973forward,nguen1985tunnel,medina1992quantum,pietracaprina2016forward,baldwin2016manybody,baldwin2017clustering}, and how this can be employed to gain
analytically-based insight into MBL.

The propagator between any two Fock-space sites $I, K$ is simply the matrix element of the resolvent operator,
\eq{
G_{IK}(\omega)  &= \braket{I|(\omega^{+} - H)^{-1}|K}\,.
}
It can be spectrally resolved as 
\eq{
G_{IK}(\omega) =\sum_{\psi}\frac{\psi_I^\ast \psi_K^{\pd}}{\omega^{+}-E_\psi}\,,
}
and in the vicinity of an eigenstate $\psi$ of energy $E_{\psi}$ has residue
\eq{\lim_{\omega\to E_\psi}(\omega-E_\psi)G_{IK}(\omega)=\psi_I^\ast\psi_K^{\pd}\,.\label{eq:residueGIJ}}

Any non-local propagator ($K\neq I$) can be expressed as an exact  renormalised  path expansion
(closely related to the Feenberg renormalised perturbation series discussed in Sec.~\ref{sec:self-consistent});
namely
\eq{
G_{IK}(\omega) & = \frac{1}{\omega^+-{\cal E}_I-S_I(\omega)}\times\nonumber \\&~~\sum_{p\in {\rm paths}^\ast(I,K)}\prod_{K_n \in p}\frac{{\cal T}_{K_{n-1} K_{n}}^{\pd}}{\omega^+-{\cal E}_{K_n}^{\pd}-S_{K_n}^{(p)}(\omega)}\,,
\label{eq:shortest-paths}
}
where the sum over ${\rm paths}^\ast$ runs over all \emph{self-avoiding} paths from  $I$ to $K$ on the Fock-space graph  (with $\cdots K_{n-1},K_{n},\cdots$ denoting the sequence of successive Fock-space sites on the path).
The first term in Eq.~\ref{eq:shortest-paths} is the exact local propagator $G_{II}(\omega)$ 
for the `starting' site $I$, with self-energy $S_{I}(\omega)$ (see Eq.~\ref{eq:self-energy-def}).
$S_{K_n}^{(p)}$ by contrast denotes the self-energy for site $K_{n}$ in which all sites traversed on the path
$p$  up to that point are excluded, with $G_{K_{n}K_{n}}^{(p)}(\omega)=(\omega^+-{\cal E}_{K_n}-S_{K_n}^{(p)}(\omega))^{-1}$ the corresponding local propagator. As first emphasised by Anderson~\cite{anderson1958absence},
this self-avoiding path expansion -- with its associated renormalised local propagators -- circumvents the 
pathologies of its unrenormalised counterpart (the bare locator expansion), in which `repeating' paths lead to spurious divergences reflecting trivial local resonances. 

The eigenvalue $E_{\psi}$ of an eigenstate $\ket{\psi}$ satisfies $E_\psi = {\cal E}_I + S_I(E_{\psi})$,
with $|\psi\rangle$ localised on site $I$ in the absence of hopping. This allows one to write the residue of the propagator in Eq.~\ref{eq:shortest-paths} at $\omega=E_\psi$ as
\eq{
\lim_{\omega\to E_\psi}&(\omega-E_\psi)G_{IK}^{\pd}(\omega)=
|\psi_{I}^{\pd}|^2\times \nonumber\\&\lim_{\omega\to E_\psi}\sum_{p\in {\rm paths}^\ast(I,K)}\prod_{K_n \in p}\frac{{\cal T}_{K_n K_{n-1}}^{\pd}}{\omega-{\cal E}_{K_n}^{\pd}-S_{K_n}^{(p)}(\omega)}\,
}
(the $|\psi_{I}|^{2}$ coming from Eq.~\ref{eq:residueGIJ} for the local propagator $G_{II}(\omega)$).
Comparing this to Eq.~\ref{eq:residueGIJ}   immediately gives
\eq{
\psi_K^{\pd} = \psi_{I}^{\pd}\sum_{p\in {\rm paths}^\ast(I,K)}\prod_{K_n \in p}\frac{{\cal T}_{K_n K_{n-1}}^{\pd}}{E_\psi^{\pd}-{\cal E}_{K_n}^{\pd}-S_{K_n}^{(p)}(E_\psi)}\,
\label{eq:jeremy}
}
for the eigenstate amplitude on Fock-space site $K$ of an eigenstate which in the absence of hoppings is localised on site $I$.

Eq.~\ref{eq:jeremy} remains exact. At this stage the key approximation is made in the FSA~\cite{pietracaprina2016forward}: to treat Eq.~\ref{eq:jeremy} perturbatively, exclusively to leading  order in powers of the hopping $\mathcal{T}$. For sites $I$ and $K$ separated by a given graph distance $d_{IK}$ on the Fock-space graph,
the minimum number of powers of $\mathcal{T}$ appearing in the product in Eq.~\ref{eq:jeremy} is
$d_{IK}$ (by its definition); with the corresponding set of paths between $I$ and $K$ being the
`shortest paths'. Given this, $S_{K_n}^{(p)}$ in Eq.~\ref{eq:jeremy} can be neglected, since
 $S_{K_n}^{(p)}\sim\mathcal{O}(|\mathcal{T}|^{2})$ to leading order (as evident e.g. from Eq.~\ref{eq:SI-RPS}),
while $\psi_{I}\to 1$ and $E_{\psi}\to \mathcal{E}_{I}$ to leading order. Eq.~\ref{eq:jeremy} thus becomes
\eq{
\psi_K^{\pd} = \sum_{p\in {\rm spaths}(I,K)}\prod_{K_n \in p}\frac{{\cal T}_{K_n K_{n-1}}^{\pd}}{\E_I^{\pd}-{\cal E}_{K_n}^{\pd}}\,,
\label{eq:fsa-final}
}
with the path sum over the set of shortest paths. This is the basic expression for the eigenstate amplitude within the FSA.  It was also argued in Ref.~\cite{pietracaprina2016forward} that the omitted self-energy corrections will serve to  weaken the bare resonances in Eq.~\ref{eq:fsa-final}, and as a result that the FSA is expected to overestimate 
the critical disorder for a localisation transition.

\begin{figure}
\includegraphics[width=\linewidth]{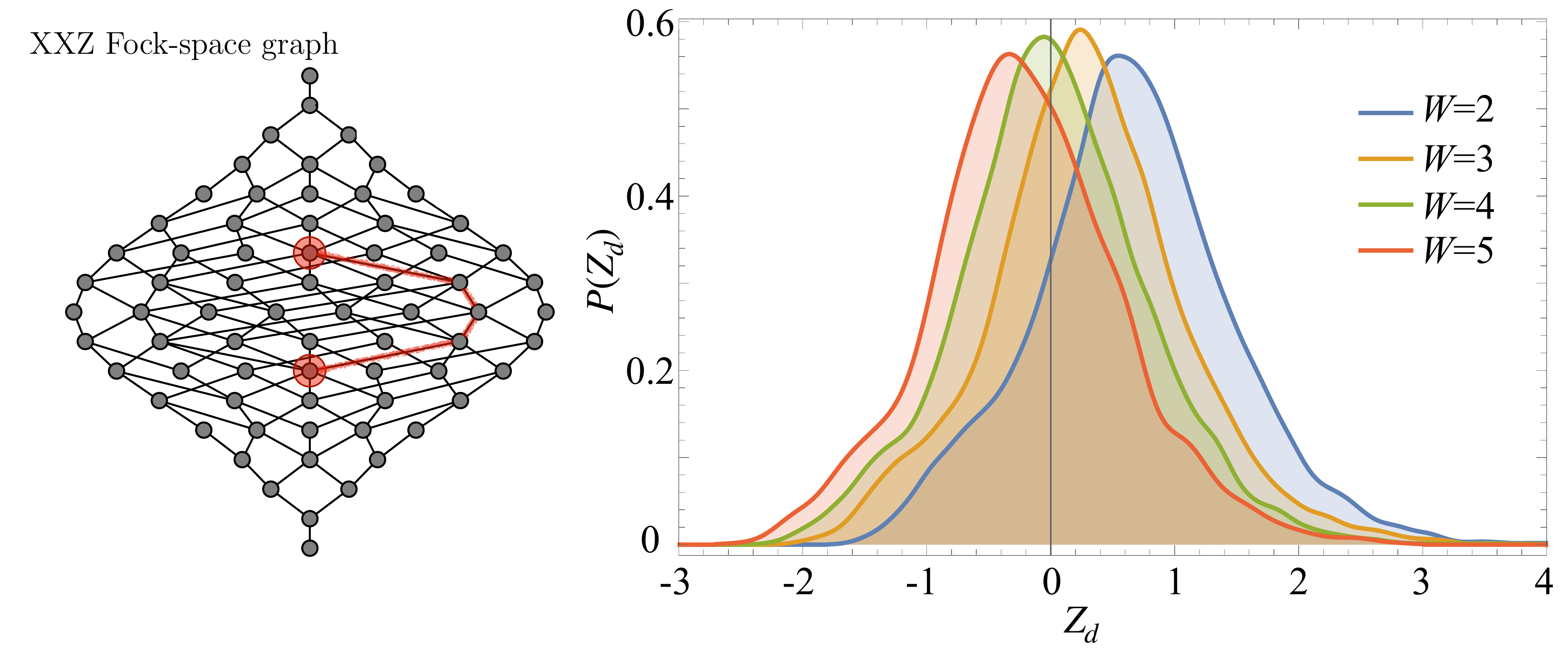}
\caption{Left panel: the XXZ Fock-space graph, shown explicitly for $L=8$. 
The shortest path length between the N\'eel and anti-N\'eel states is $L/2$ (illustrated by the red line). 
Right panel:  for $L=20$ and different values of the disorder strength $W$, showing the probability distribution of $Z_d$ (Eq.~\ref{eq:Zddef}) for the N\'eel and  anti-N\'eel states. The figure is taken and adapted from Ref.~\cite{pietracaprina2016forward}. }
\label{fig:fsa-Zd-dist}
\end{figure}

Given that the FSA provides an estimate for eigenstate amplitudes on arbitrary sites on the Fock-space graph, a localisation criterion can be defined as follows~\cite{pietracaprina2016forward}. Consider
\eq{
\psi_d^{\pd} = \max_{K: d_{IK}{\pd}=d}{\pd}\psi_K^{\pd}\,,
}
giving the largest eigenstate amplitude for those sites $K$ which lie a given graph distance
$d_{IK} =d$ from the site $I$ on which the eigenstate $\ket{\psi}$ is localised in the absence of hoppings.
$\ket{\psi}$ is then defined to be localised if the probability
\eq{
\lim_{d\to\infty}P\left(\frac{\ln |\psi_d^{\pd}|^2}{2d} \leq -\frac{1}{2\xi}\right)\to 1
\label{eq:fsa-loccrit}
}
for some finite $\xi$. This criterion ensures that all eigenstate amplitudes remain within an envelope which decays exponentially with the distance from $I$ on the graph. 
This condition can be turned  into a stronger condition for the existence of the delocalised phase~\cite{pietracaprina2016forward}. For arbitrarily small $\epsilon \equiv \xi^{-1}$, if there exists a site $K$ at an arbitrarily large distance $d$, for which $\ln(|\psi_K|^2)/2d$ exceeds $-\epsilon$ with a finite probability, 
then it is delocalised.  Formally, this condition for the delocalised phase can be expressed as 
\eq{
\lim_{d\to\infty}P\left(\frac{\ln |\psi_d^{\pd}|^2}{2d} \geq -\epsilon\right)\to 1\,,
\label{eq:fsa-crit}
}
for arbitrarily small, positive $\epsilon$.

\begin{figure}
\includegraphics[width=\linewidth]{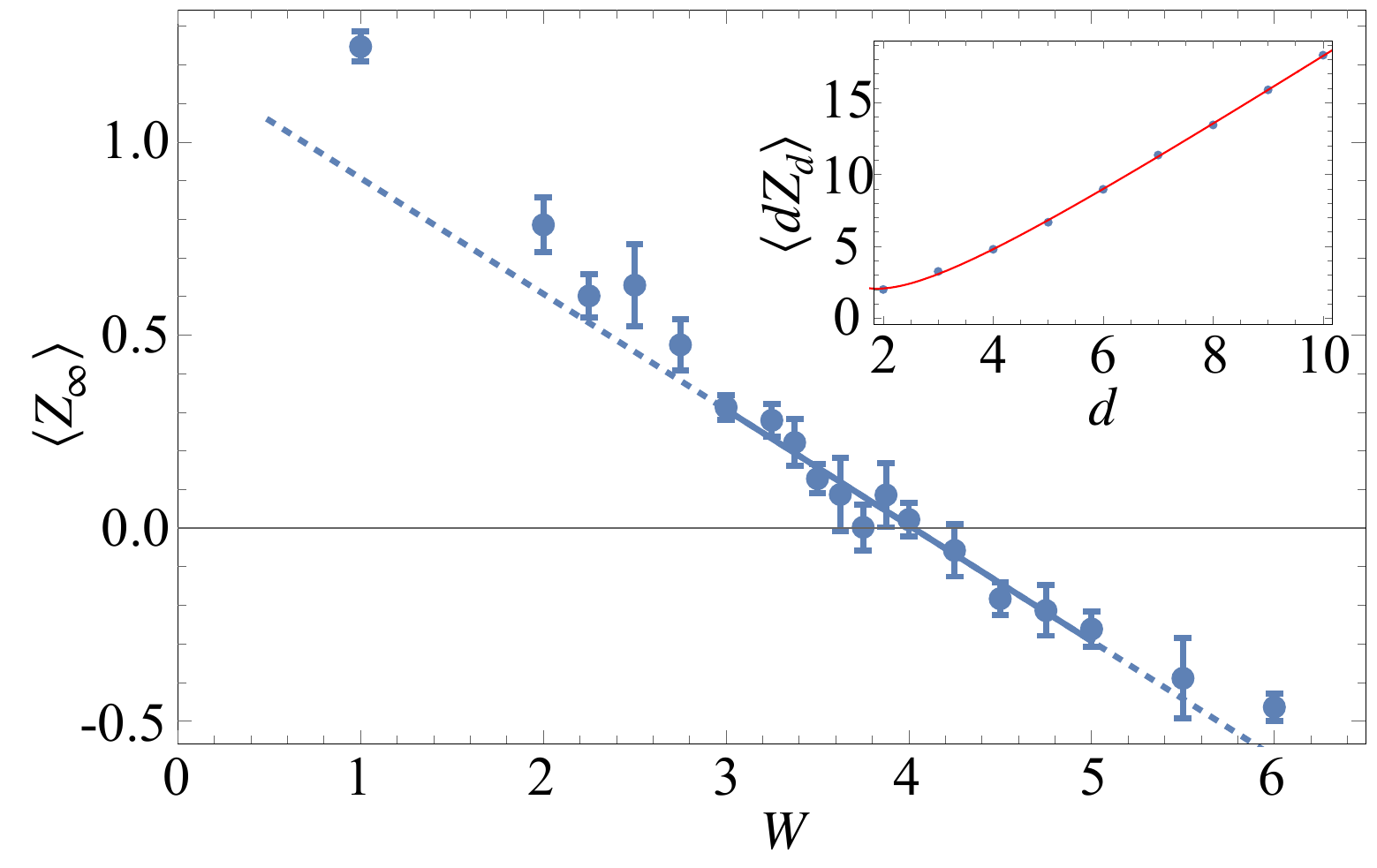}
\caption{Main panel shows the extrapolated value of the mean $\braket{Z_\infty}$ \emph{vs} disorder strength $W$. 
The crossing  $\braket{Z_\infty}=0$ signals the MBL transition, at an estimated critical disorder of $W_c= 4.0\pm 0.3$.  Inset  illustrates (for $W=1$) the finite-size scaling of $\braket{d Z_d}$ (see Eq.~\ref{eq:Zddef}) with the distance $d=L/2$ between the N\'eel and anti-N\'eel states; with the red line the fit to 
Eq.~\ref{eq:fsa-fit}. The figure is taken and adapted from Ref.~\cite{pietracaprina2016forward}.}
\label{fig:fsa-Zinf}
\end{figure}

Eqs.\ \ref{eq:fsa-crit},\ref{eq:fsa-loccrit} imply that the transition can  be addressed by analysing the statistics of the eigenfunction amplitudes as a function of the graph distance $d$, as embodied in the distribution
over disorder realisations,  $P(Z_{d})$, of $Z_{d}$ defined by
\eq{
Z_d^{\pd} = \frac{\ln|\psi_d^{\pd}|^2}{2d}\,.
\label{eq:Zddef}
}
The FSA is of course applicable to one-body as well as many-body localisation, and provides in the former case~\cite{pietracaprina2016forward} a rich description  of localisation in $D \geq 3$-dimensional hypercubic lattices (with an increasingly accurate estimate of the critical
disorder on increasing $D$).

In Ref.~\cite{pietracaprina2016forward}, the FSA was also used to study the MBL transition in the disordered
Heisenberg chain ($J_{i}=J \equiv 1$ in the XXZ model  Eq.~\ref{eq:ham-xxz}), in the $\sigma^z$-basis with the spin-flip terms constituting the hoppings on the Fock-space graph. Two Fock-space sites $I$ and $K$ were chosen, viz.\  the N\'eel state and the anti-N\'eel state. These two sites have a mutual graph distance of $d=L/2$, 
with $2(L/2)!$ shortest hopping paths connecting them. Varying the system size, $L$, allowed study of $Z_d$ 
(Eq.~\ref{eq:Zddef}) for varying values of $d=L/2$, and eventual extrapolation to $d\to\infty$ to extract the critical point. In such a setting, noting that $\xi$ in Eq.\ \ref{eq:fsa-loccrit} diverges on approaching the transition and that  $\epsilon \to 0$ in Eq.\ \ref{eq:fsa-crit}, the condition for the  MBL transition
can be taken as~\cite{pietracaprina2016forward}
\eq{
	\lim_{d\to\infty}\braket{Z_d^{\pd}(W_c^{\pd})} = 0\,,
}
with the mean over the distribution $P(Z_{d})$.

For a fixed $L$,  Fig.~\ref{fig:fsa-Zd-dist} shows the distributions $P(Z_{d})$ for different values of the
disorder strength $W$. As one expects, with increasing $W$ the distributions shift towards smaller values of $Z_d$.
In order to extrapolate to $d=L/2 \to\infty$, Ref.~\cite{pietracaprina2016forward} used a fitting function of form
\eq{
	d\braket{Z_d^{\pd}(W)}=c_1^{\pd}+\braket{Z_\infty^{\pd}(W)}d+c_2^{\pd}d^{-1}\,,
	\label{eq:fsa-fit}
}
the quality of which is illustrated in the inset to Fig.~\ref{fig:fsa-Zinf}.
The main panel in Fig.~\ref{fig:fsa-Zinf} shows the extrapolated $\braket{Z_\infty (W)}$ as a function of disorder strength, with the critical disorder given by the value of $W$ for which $\braket{Z_\infty (W)}=0$, yielding
$W_c= 4.0 \pm 0.3$~\cite{pietracaprina2016forward}.

The FSA analysis on the Fock-space graph of the disordered Heisenberg chain has a further interesting outcome. From the basic structure of Eq.~\ref{eq:fsa-final}, it is clear that  the distribution of amplitudes over all the 
shortest paths must be fat-tailed, reflecting simply the possibility of small denominators due to bare resonances. One might therefore guess that  (as is the case for one-body localisation) the path sum 
is dominated by an optimal path, {\it i.e.} the path with the largest amplitude. In Ref.~\cite{pietracaprina2016forward} it was however shown that this is not the case. In fact the participation ratio of the path amplitudes scaled linearly with the number of paths, indicating that a finite fraction of the paths contributed to the sum in Eq.~\ref{eq:fsa-final}. This can be  understood as a signature of the strong correlations on the Fock-space graph~\cite{pietracaprina2016forward}. Using the ideas in Ref.~\cite{ros2015integrals}, it can be argued that the strongest correlations are between paths in Eq.~\ref{eq:fsa-final} which share the same spins that are flipped,  
but with the spin flips occurring in a different order, and that the underlying locality of the model directly implies strong correlations between the energy denominators arising in such paths.


\subsection{Decimation on Fock space\label{sec:aoki}}

We now discuss an exact renormalisation procedure on the Fock space, which consists of systematically `decimating' Fock-space sites. The method is exact in that, at every step, it preserves {\it exactly} the Green functions of the remaining  set of Fock-space sites.  Originally introduced by Aoki~\cite{aoki1980real,aoki1982decimation}, and later employed in other works on single-particle 
localisation~\cite{monthus2009statistics,monthus2009statistics2}, this procedure was also implemented on the Fock-space graph in Ref.~\cite{MonthusGarel2010PRB}. It yields the statistics of an effective, renormalised matrix element between arbitrary pairs of Fock-space sites at an energy scale $\omega$.

To set up the decimation, first write the Schr\"odinger equation for the eigenstate amplitudes $\{\psi_{K}\}$,
using the Fock-space Hamiltonian (Eq.~\ref{eq:fs-ham-general}), 
\eq{
	\omega \psi_K^{\pd} = \mathcal{E}_K^{\pd}\psi_K^{\pd} + \sum_{J(\neq K)}{\cal T}_{KJ}^{\pd}\psi_J^{\pd}\,.
	\label{eq:psiI-schrod}
}
This holds for all Fock-space sites $K$. For $K=I$, it gives
\eq{
	\psi_I^{\pd} = \frac{\sum_{J(\neq I)}{\cal T}_{IJ}^{\pd}\psi_J^{\pd}}{\omega-\E_I}\,.
	\label{eq:psiI-aoki}
}
The Fock-space site $I$ can now be decimated, by eliminating $\psi_I$ in Eq.~\ref{eq:psiI-schrod}
 in favour of Eq.~\ref{eq:psiI-aoki}. With this, the Schr\"odinger equation remains of form Eq.~\ref{eq:psiI-schrod},  but now with $K=I$ excluded from the set $\{\psi_{K}\}$, and with the `bare' Fock-space site energies and hoppings replaced by  renormalised versions,
\begin{equation}
{\cal E}_K^{\rm new} = {\cal E}_K^{\pd} +\frac{|\mathcal{T}_{IK}^{\pd}|^2}{\omega-\E_I},
~~~~
{\cal T}_{KJ}^{\rm new}={\cal T}_{KJ}^{\pd} + \frac{{\cal T}_{KI}^{\pd}{\cal T}_{IJ}^{\pd}}{\omega-\E_I}.
\end{equation}
This procedure can be continued until only two arbitrarily chosen Fock-space sites remain, and the resulting matrix element at the final step is the renormalised matrix element. Given a choice of two final remaining sites, note that because every step of the decimation is formally exact, all other sites can be eliminated in any order and the final matrix element will be the same.

In Ref.~\cite{MonthusGarel2010PRB}, Monthus and Garel implemented this decimation for the 
spinless fermion model Eq.~\ref{eq:ham-fermion} (taking the $\{\epsilon_{i}\}$ as independent Gaussian
variables with vanishing mean and variance $W^{2}$,  and with a minor modification of also having next-nearest neighbour hoppings). The N\'eel and anti-N\'eel states  were chosen as the final two Fock-space sites to retain. As with the FSA discussed in Sec.~\ref{sec:FSA},  the graph distance between these two states is macroscopic,
$\propto L$,  whence running the decimation for different system sizes $L$ allows one to study the matrix element as a function of the distance between the Fock-space sites;  in the following, the magnitude of this  renormalised matrix element is denoted as $V_L$.

\begin{figure}
\includegraphics[width=\linewidth]{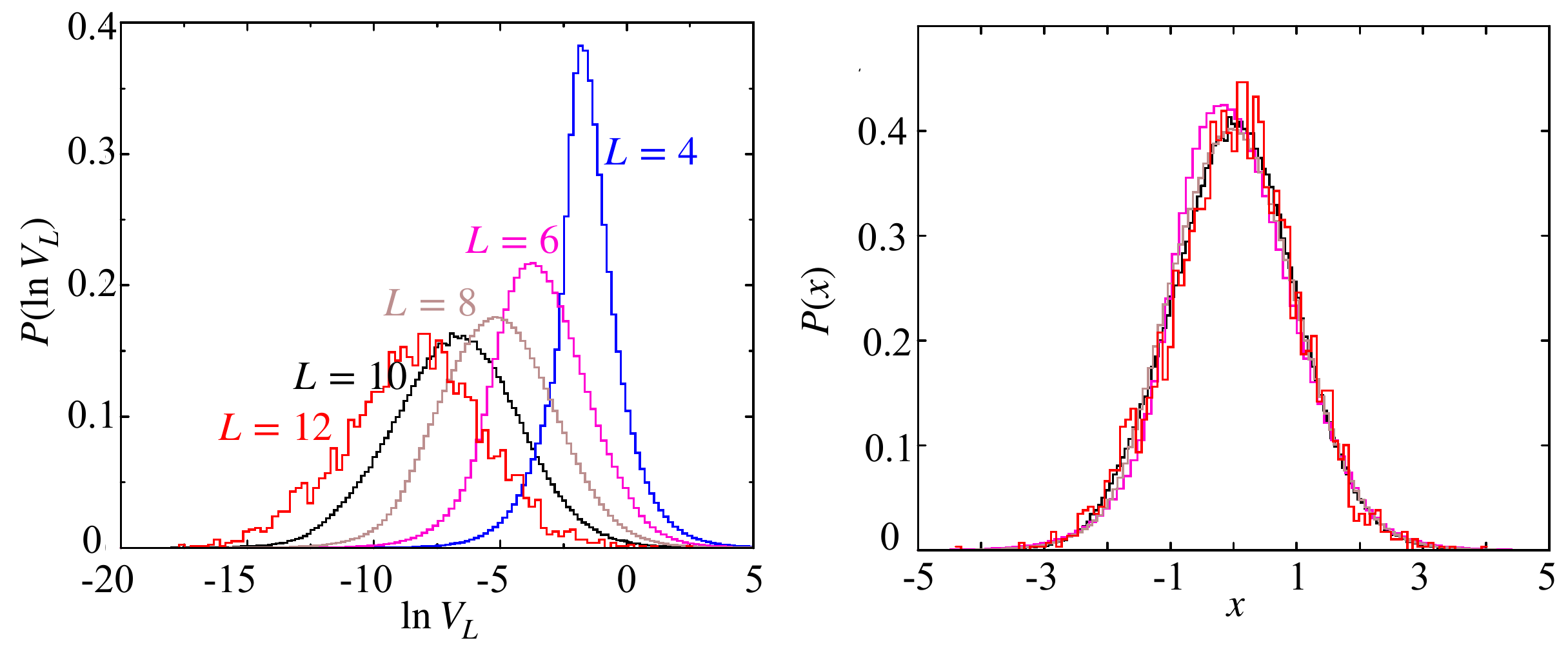}
\caption{Deep in the MBL phase, distribution of the logarithm of the matrix element, $P(\ln V_L)$, for different $L$. Left panel shows the bare distributions. Right panel:  when shifted by their mean $\braket{\ln V_L}$ and scaled by their standard deviation (Eq.~\ref{eq:xL-aoki}), the distributions show an $L$-independent collapse.
The figure is taken and adapted from Ref.~\cite{MonthusGarel2010PRB}.}
\label{fig:aoki-loc1}
\end{figure}

\begin{figure}
\includegraphics[width=\linewidth]{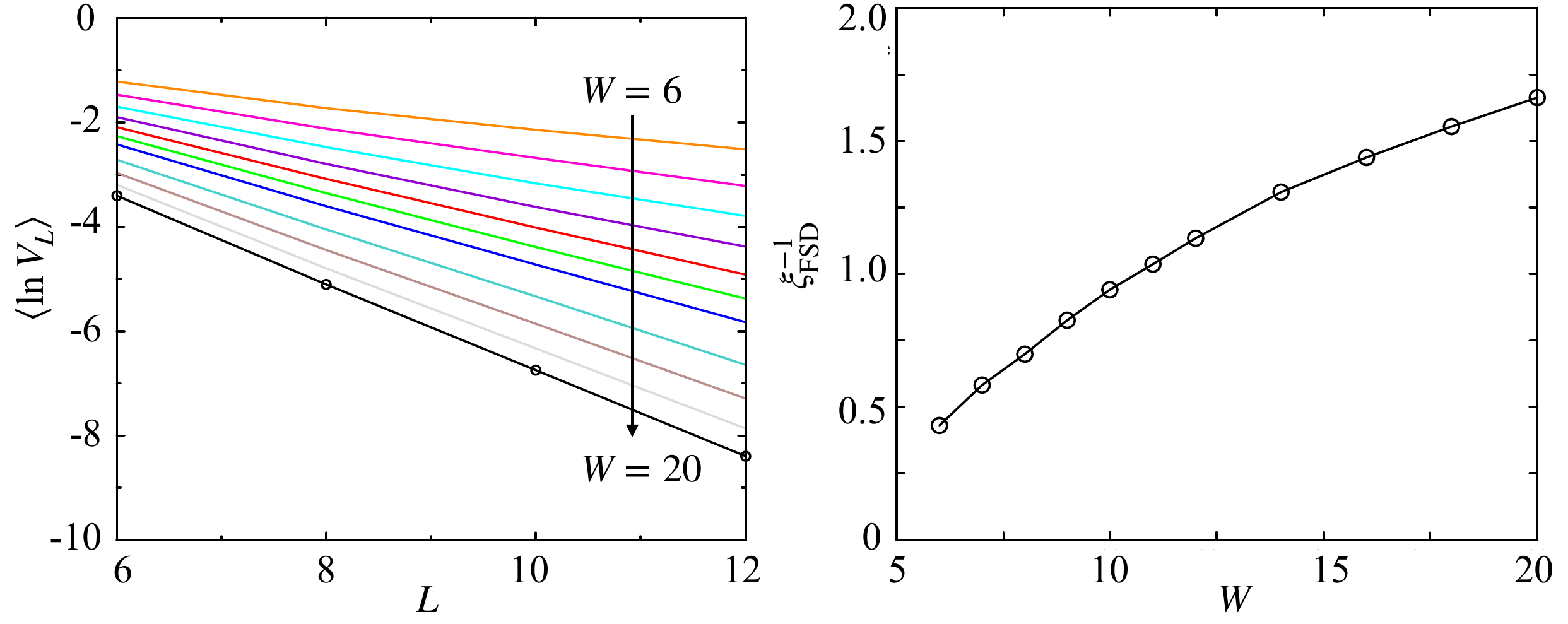}
\caption{Left: In the MBL phase, result for $\braket{\ln V_L}$ as a function of $L$, for different values of $W$.
 The data shows a linear behaviour with $L$, with a slope proportional to $\xi_{\rm FSD}^{-1}$ (see main text). Right: Extracted $\xi_{\rm FSD}^{-1}$ as a function of disorder strength, $W$. The figure is taken and adapted from Ref.~\cite{MonthusGarel2010PRB}.}
\label{fig:aoki-loc2}
\end{figure}

In Fig.~\ref{fig:aoki-loc1}, for a disorder strength $W$ deep in the MBL phase, resultant distributions of $\ln V_L$ are shown for different $L$~\cite{MonthusGarel2010PRB}. For linearly increasing $L$, the distributions $P(\ln V_{L})$ move linearly towards the left, suggesting that $V_L$ typically decays exponentially with $L$.
Further,  as seen in Fig.~\ref{fig:aoki-loc1}  right panel, the distributions of 
\eq{
	x_L^{\pd} = \frac{\ln V_L^{\pd} - \braket{\ln V_L^{\pd}}}{\sigma_{V_L^{\pd}}}\,
	\label{eq:xL-aoki}
}
(with $\sigma_{V_L}$ the standard deviation of $\ln V_L$) for different $L$ show an $L$-independent collapse.
Assuming that $\sigma_{V_L}$ scales sufficiently slowly with $L$, the behaviour of $\braket{\ln V_L}$ with $L$ for different values of $W$ is a natural quantity of interest. In Fig.~\ref{fig:aoki-loc2}, the left panel shows results for $\braket{\ln V_L}$ as a function of $L$ for a range of $W$ in the MBL 
phase~\cite{MonthusGarel2010PRB}. These show a linear behaviour with $L$, implying
\eq{
	V_L^{\rm typ} = \exp[\braket{\ln V_L^{\pd}}] ~\simeq ~ \exp\left[-\frac{L/2}{\xi_{\rm FSD}^{\pd}(W)}\right]\,,
}
where $\xi_{\rm FSD}$ can be interpreted as a localisation length on the Fock space (the subscript ${\rm FSD}$ denoting that it is obtained from Fock-space decimation). The right panel in Fig.~\ref{fig:aoki-loc2} shows
the evolution of $\xi_{\rm FSD}^{-1}$ as a function of $W$. To understand its critical behaviour 
as the transition is approached from the MBL phase,  a three-parameter fit was employed~\cite{MonthusGarel2010PRB},
\eq{
	\xi_{\rm FSD} \simeq c (W-W_c^{\pd})^{-\nu_{\rm FSD}^{\pd}}\,,
}
the best fit yielding an estimate of the critical disorder in the range  
$5.2\leq W_c \leq 5.9$, and exponent $\nu_{\rm FSD}\simeq 0.45$.

\begin{figure}[!t]
\includegraphics[width=\linewidth]{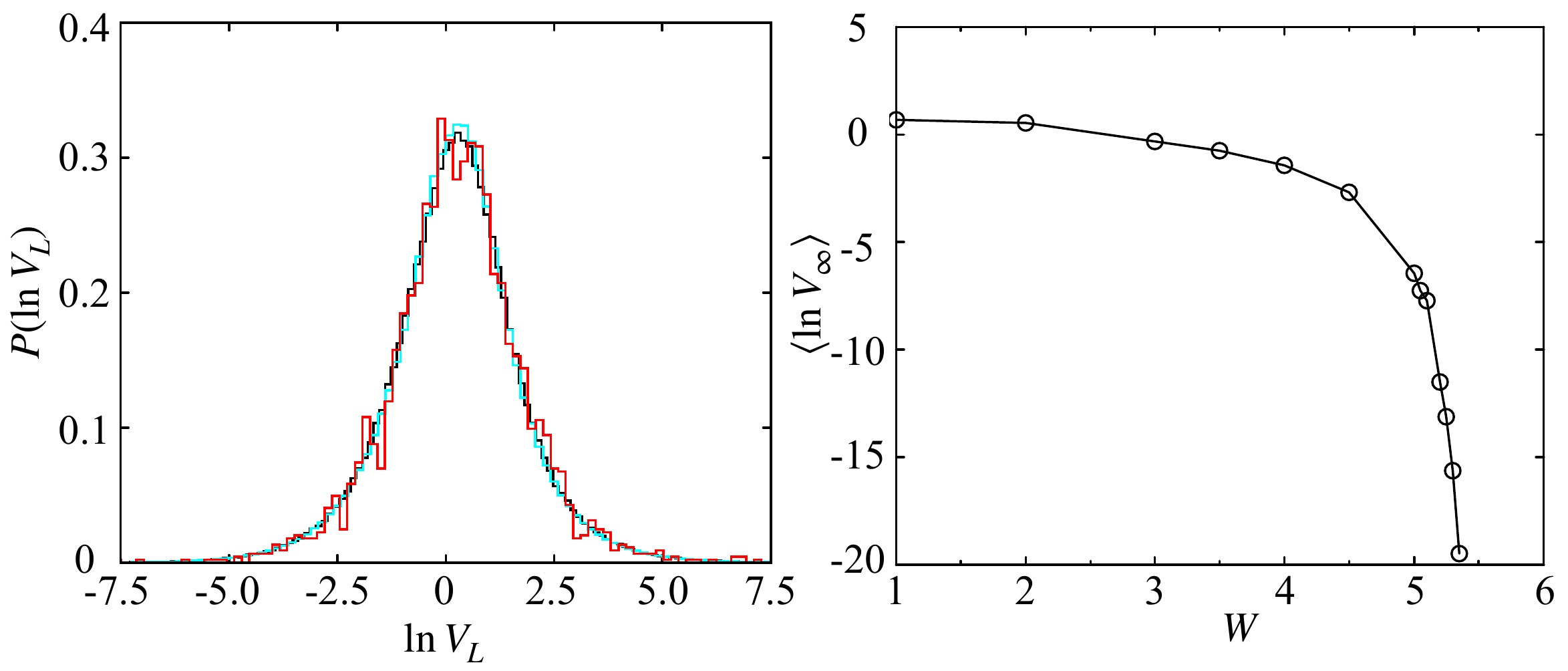}
\caption{Left: Deep in the ergodic phase, the distribution $P(\ln V_L)$ for different $L$  is converged with $L$. Right: Result for $\braket{\ln V_\infty}$ obtained from the converged distribution as a function of $W$. The figure is taken and adapted from Ref.~\cite{MonthusGarel2010PRB}.}
\label{fig:aoki-deloc2}
\end{figure}

The above behaviour pertains to  the MBL phase, and the approach to the transition from that phase. 
Turning to the ergodic phase at weaker $W$, the situation is found to be qualitatively different from that at strong $W$. As shown in Fig.~\ref{fig:aoki-deloc2} (left panel), the distributions for $\ln V_L$ converge with 
$L$. Hence, $\braket{\ln V_L}$ takes on a finite value in the thermodynamic limit,  denoted by $\braket{\ln V_\infty}$. Results for $\braket{\ln V_\infty}$ as a function of $W$ are shown in the right panel of Fig.~\ref{fig:aoki-deloc2}~\cite{MonthusGarel2010PRB}. The behaviour on approach to the transition from the ergodic phase was
found to be compatible with an essential singularity; a fit to the form
\eq{
	V_{\infty}^{\rm typ} = \exp[\braket{\ln V_{\infty}^{\pd}}]
	~\simeq ~ \exp[-\tilde{c}(W_c^{\pd}-W)^{-\kappa}]\,,
}
yielding~\cite{MonthusGarel2010PRB} a critical point in the range $5.5\leq W_c \leq 5.7$ -- consonant with the 
estimate above on approach from the MBL side -- and an exponent $\kappa\simeq 1.4$. 
Essential singularities are also known to occur for the one-body Anderson transition on a Bethe lattice~\cite{mirlin1991localisation,monthus2008anderson}.


\section{Scaling at the many-body localisation transition \label{sec:scaling}}

In this section, we review a much more  systematic scaling theory of the MBL transition, based on Fock-space quantities; in which essential singularities and diverging correlation lengths again appear  on approaching the transition from the delocalised and MBL sides, respectively.

The approach draws on insights from the scaling theory of Anderson transitions on high-dimensional graphs~\cite{garciamata2017scaling}. The latter problem is of course fundamentally different from the many-body problem on Fock space. While eigenstates are ergodic in the delocalised phase, in an Anderson localised phase they are exponentially localised, and multifractality is present only at the critical point~\cite{evers2008anderson}. In the MBL problem by contrast, multifractality occurs both at the transition and throughout the entire MBL phase~\cite{deluca2013ergodicity,mace2019multifractal}. Notwithstanding this key difference, however, the scaling theories in the two cases are quite similar.  Let us briefly discuss the physical picture of the eigenstates, which motivates a scaling ansatz for the participation entropies (PEs, Eq.~\ref{eq:PE}), or equivalently, for the inverse participation ratios (IPRs, Eq.~\ref{eq:IPR}). 

In the ergodic phase, there exists a  {\it correlation volume}, $\Lambda$, such that the eigenstate is non-ergodic inside that volume. The eigenstate can then be interpreted as an ergodic tiling of such non-ergodic volumes over the entire graph~\cite{garciamata2017scaling,garciamata2022critical}, such that there is global ergodicity. The non-ergodicity, within the volume $\Lambda$,  is  in fact the incipient multifractality, characteristic of the critical point. One thus expects the correlation volume $\Lambda$ to diverge as the MBL transition is approached from the delocalised side, such that at the critical point the eigenstates are globally multifractal.

Turning now to localised phases, in a one-body Anderson localised phase eigenstates are understood to
have significant weight on a few rare branches of the graph, on which the weights decay exponentially away from the localisation centre, with a localisation length $\xi$ which is finite. As the transition is approached, the localisation length on these rare branches diverges, although the localisation length on typical branches remains finite~\cite{garciamata2020two,garciamata2017scaling,garciamata2022critical}.
This physical picture is also relevant for the MBL problem on the Fock space. The key difference is that in the MBL phase the weight of the eigenstate is significant not only on a few but on  many branches, the number of which scales non-trivially with system size, leading to the multifractality throughout the MBL phase.

The existence of a correlation volume $\Lambda$ in the ergodic phase, and a correlation length $\xi$ in the localised phase, suggest a scaling ansatz for the PEs (Eq.\ \ref{eq:PE}) of form
\eq{
\overline{S^\mathrm{PE}_q}-\overline{S^\mathrm{PE}_{q,c}}=\begin{cases}
{\cal G}^{\rm vol}\left(\frac{\nh}{\Lambda}\right)&\ : W<W_c ~{\rm ergodic}\\
{\cal G}^{\rm lin}\left(\frac{\ln \nh}{\xi}\right)&\ :W>W_c ~{\rm MBL},
\end{cases}\,
\label{eq:PE-scaling}
}
or equivalently for the IPRs, 
 \eq{
\frac{\overline{{\cal L}_q}}{\overline{{\cal L}_{q,c}}}=\begin{cases}
{\cal F}^{\rm vol}\left(\frac{\nh}{\Lambda}\right)&\ :W<W_c ~{\rm ergodic}\\
{\cal F}^{\rm lin}\left(\frac{\ln \nh}{\xi}\right)&\  :W>W_c ~{\rm MBL},
\end{cases}\,
\label{eq:IPR-scaling}
}
with $\overline{S^\mathrm{PE}_{q,c}}$ $(\overline{{\cal L}_{q,c}})$ the PE (IPR) at the critical point.
The scaling functions  depend universally on $\nh/\Lambda$ in the ergodic phase, and on
$(\ln \nh)/\xi \propto L/\xi$ in the MBL phase.

\begin{figure}[!b]
\includegraphics[width=\linewidth]{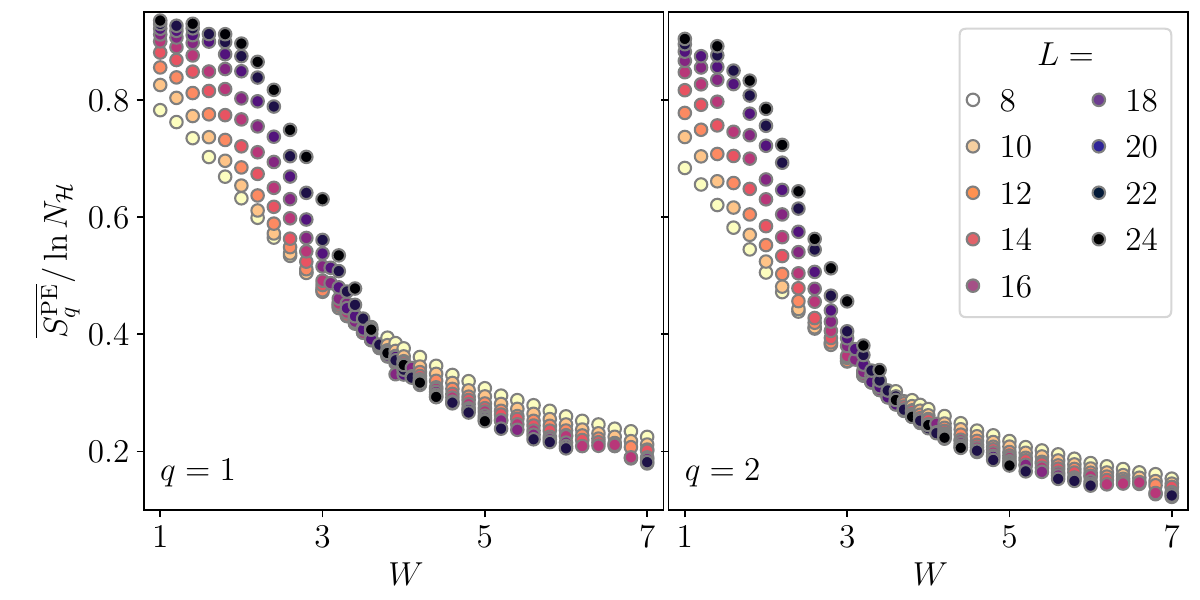}
\caption{
First ($q=1$) and second ($q=2$) participation entropies scaled by $\ln \nh$ (left and right panels respectively),  as a function of disorder strength $W$ for a range of $L$, and for the disordered Heisenberg chain (isotropic XXZ model).  The data and figure are taken and adapted from Ref.~\cite{mace2019multifractal}.}
\label{fig:sq-v-w}
\end{figure}

We next review some of the numerical results which provide  compelling evidence for the above scaling ansatz. Recall from Sec.~\ref{sec:anatomy} that the PEs and IPRs are expected to behave as 
\eq{
\begin{split}
\overline{S^\mathrm{PE}_q} &= D_q\ln \nh + b_q\\
\ln\overline{{\cal L}_q} &= -\tau_q\ln \nh + a_q~,
\end{split}\,
}
with $D_q=1$ (or $\tau_{q}=q-1$) in the ergodic phase and $0<D_q <1$ in the MBL phase.\footnote{ 
$q=1,2$ are considered for the PEs, and $q=2$ for the IPR (since $\tau_{q=1}=0$ trivially, from
wavefunction normalisation, see Eq.~\ref{eq:IPR}).} 
For the disordered Heisenberg chain (isotropic XXZ model), $\overline{S^\mathrm{PE}_q}/\ln \nh$ for $q=1,2$
are shown in Fig.~\ref{fig:sq-v-w} as a function of $W$, for a range of system sizes $L$~\cite{mace2019multifractal}.  In the ergodic phase, the data drifts towards 1 with increasing $L$, with the subleading correction $b_q<0$.  In fact, the data has a crossing point, indicating that the subleading correction $b_q$ changes sign from negative to positive across the putative critical point. From the asymptotics of the scaling functions, discussed shortly, it is readily shown  that the negative $b_q$ (or equivalently positive $a_q$) in the ergodic phase is related to the correlation volume  sufficiently far from the critical regime, 
whereas the positive subleading correction in the MBL phase is attributed to rare events leading to tails in the distribution of the PEs~\cite{mace2019multifractal}. An important fallout of  identifying the critical point to be where $b_q$  changes sign is that
\eq{
\overline{S^{\rm PE}_{q,c}} = D_{q,c}\ln\nh\,,\quad 0<D_{q,c}<1\,,
}
where $\overline{S^{\rm PE}_{q,c}}$ is the PE at the critical point and the inequalities for $D_{q,c}$ are strict 
(equivalently, $0<\tau_{2,c}<1$, again with strict inequalities). This provides evidence that at the MBL transition itself the eigenstates are multifractal, just like the entire MBL phase, and hence that the critical point is a part of the MBL phase itself.

\begin{figure}
\includegraphics[width=\linewidth]{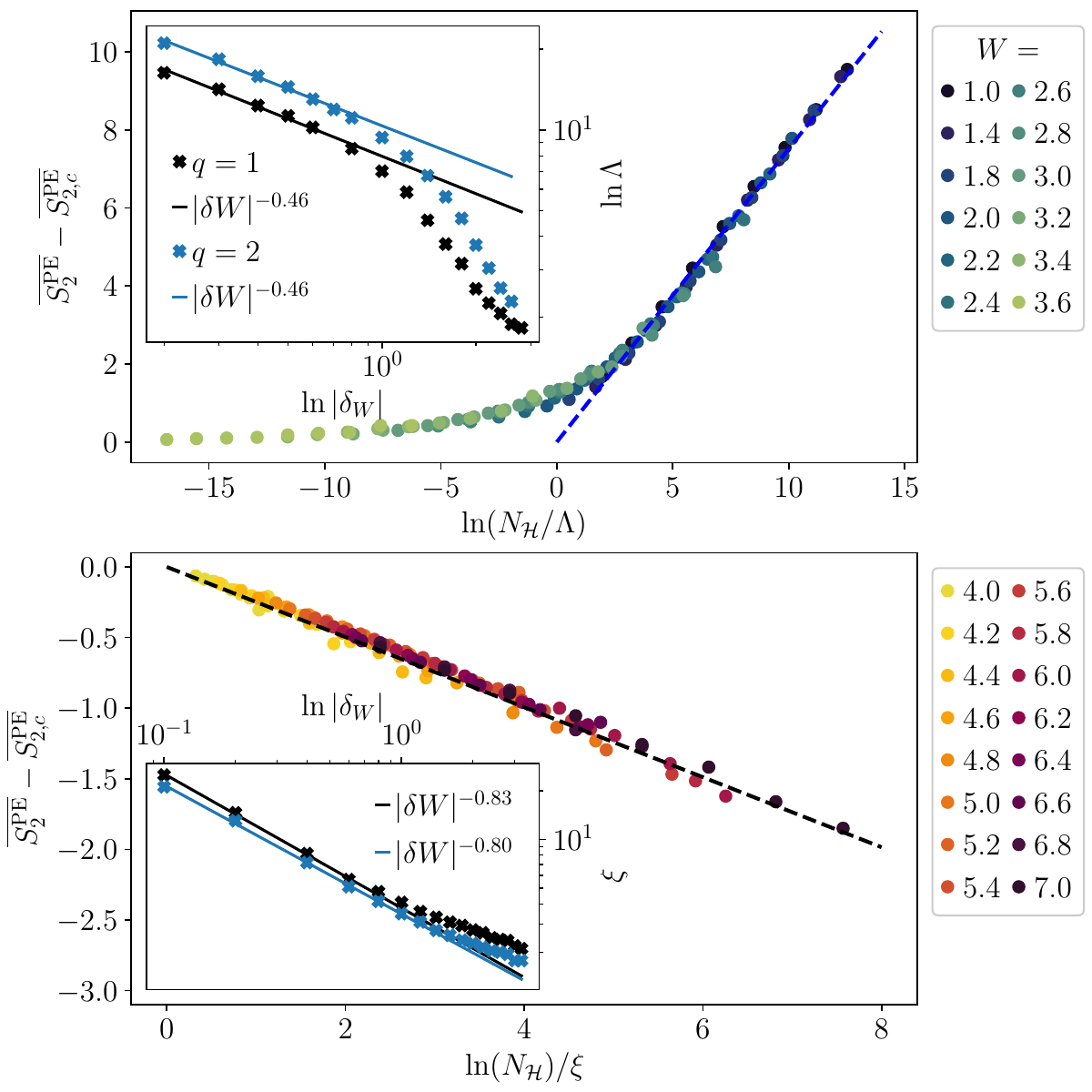}
\caption{Scaling curves for the second PE, $\overline{S_{2}^{PE}}-\overline{S_{2,c}^{PE}}$,
following the volumic scaling ansatz in the ergodic phase (top panel) and the linear scaling ansatz in the MBL phase (bottom), for the disordered Heisenberg chain. Data collapse over a range of $W$ and $L$ 
validates the ans\"atze in Eq.~\ref{eq:PE-scaling}. Inset in top panel shows (for $q=2$ and $1$)
the divergence of the correlation volume, $\Lambda$, with an essential singularity; while inset to the bottom panel shows the power-law divergence of the correlation length $\xi$. The data and the figure are taken and adapted from Ref.~\cite{mace2019multifractal}.}
\label{fig:ipr-scaling-xxz}
\end{figure}

\begin{figure}
\includegraphics[width=\linewidth]{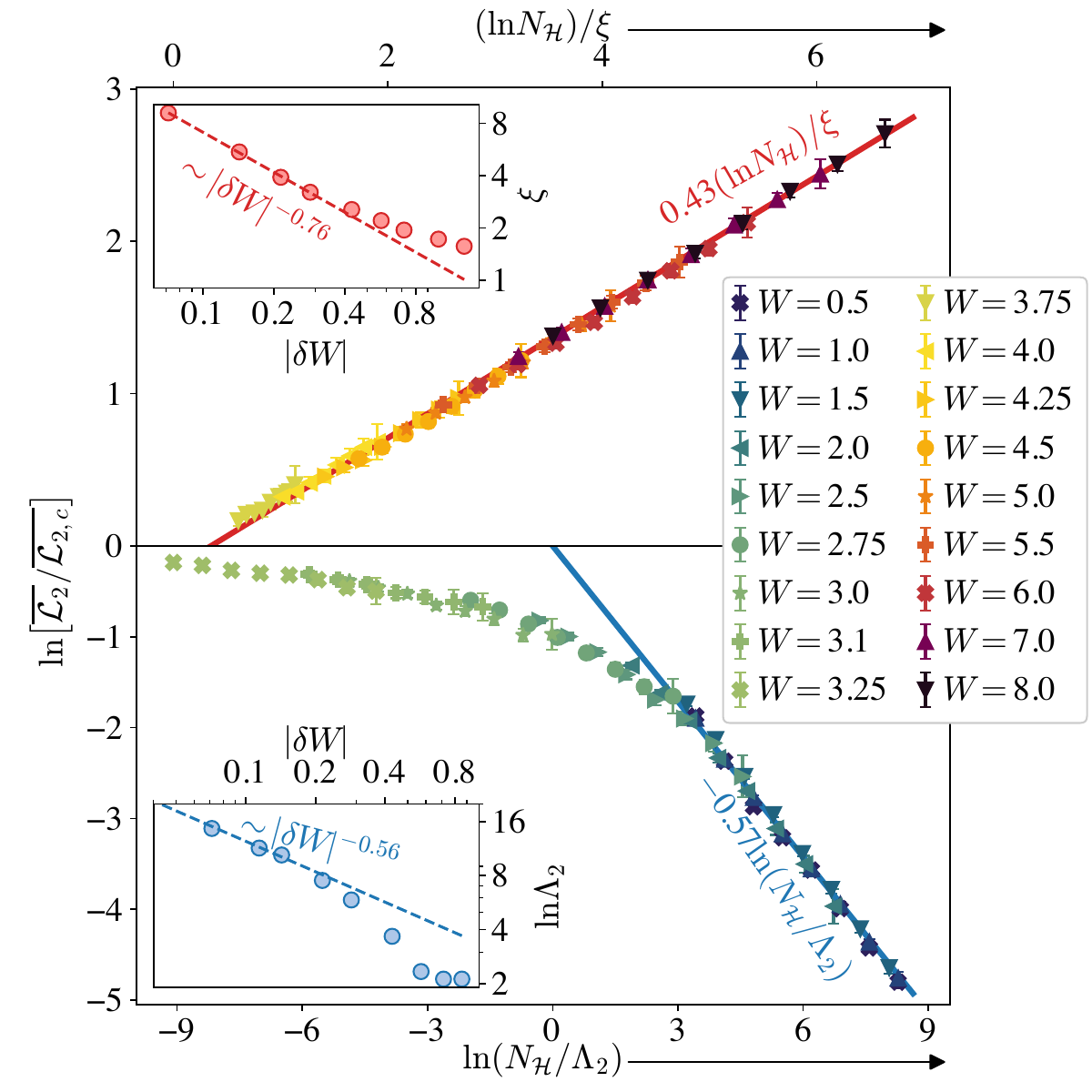}
\caption{Analogous scaling collapse to that in Fig.~\ref{fig:ipr-scaling-xxz}, but for the second ($q=2$) IPR of the disordered TFI chain (shown as $\ln(\overline{\mathcal{L}_{q}}/\overline{\mathcal{L}_{q,c}})$).
In this figure, the top panel corresponds to the MBL phase and the bottom  to the ergodic phase. The data again validates the volumic and linear scaling ans\"atze in Eq.~\ref{eq:IPR-scaling}. 
Insets also show the essential-singularity [power-law] divergence of the correlation volume $\Lambda$ [correlation length $\xi$] on approaching the transition from the ergodic [MBL] phase. The figure is taken and adapted from Ref.~\cite{roy2021fockspace}.}
\label{fig:ipr-scaling-tfi}
\end{figure}

Fig.~\ref{fig:ipr-scaling-xxz} shows data (adapted from Ref.~\cite{mace2019multifractal}) for the second PE of the disordered Heisenberg chain, which validates  the scaling ansatz in Eq.~\ref{eq:PE-scaling}: in the ergodic phase, plotting  $\overline{S^\mathrm{PE}_2}-\overline{S^\mathrm{PE}_{2,c}}$ as a function of $\nh/\Lambda$ --
the Fock-space dimension scaled with the correlation volume -- for  each $W$, collapses the data for different $W$ onto a common scaling curve. Similarly, in the MBL phase, $\ln \nh \propto L$ scaled with the $W$-dependent correlation length, $\xi$, again  yields a scaling collapse.
In the same spirit, Fig.~\ref{fig:ipr-scaling-tfi} shows analogous scaling plots (adapted from Ref.~\cite{roy2021fockspace}), now shown for the IPR of the disordered TFI chain
(Eq.~\ref{eq:ham-tfi}); and which likewise evince the validity of the scaling ansatz in Eq.~\ref{eq:IPR-scaling}.

Let us now discuss the asymptotics of the scaling functions, beginning with the ergodic phase, in the asymptotic limit of $\nh\gg \Lambda$. Physically, this is the regime where for a given $L$ 
 we are sufficiently far away from the critical point, or for a given $W$ the system size is large enough that it lies outside the critical regime. In this case one expects $\overline{S^{\rm PE}_{q}} = \ln \nh + b_q$, or equivalently $\ln{\overline{\cal L}_{q}}=-\ln\nh+a_q$. On the other hand, for $\nh\ll\Lambda$ the system is in the critical regime, whence we expect $\overline{S^{\rm PE}_{q}}\to \overline{S^{\rm PE}_{q,c}}$ and $\overline{\cal L}_{q}\to \overline{\cal L}_{q,c}$. This implies that the asymptotics of the scaling functions are 
\eq{
{\cal G}^{\rm vol}(x)=\begin{cases}
(1-D_{q,c})\ln x\ & : x\gg 1\\
~0\ & : x\to 0 
\end{cases}\
}
with $x \equiv \nh/\Lambda$, or equivalently
\eq{
{\cal F}^{\rm vol}(x)=\begin{cases}
x^{\tau_{q,c}-1}\, & : x\gg 1\\
1\, & : x\to 0 .
\end{cases}\,
}
These asymptotic behaviours are indeed clearly evident in Figs.~\ref{fig:ipr-scaling-xxz},\ref{fig:ipr-scaling-tfi}
(recall that Fig.~\ref{fig:ipr-scaling-tfi} shows $\ln(\overline{\mathcal{L}_{2}}/\overline{\mathcal{L}_{2,c}})=\ln\mathcal{F}^{\rm vol}$).

Turning to the MBL phase, in the asymptotic limit $(\ln \nh)/\xi \gg 1$  where the system size $L$  exceeds
the correlation length $\xi$, we expect $\overline{S^{\rm PE}_{q}}\sim D_q\ln\nh$  or equivalently $\ln\overline{{\cal L}_{2}}\sim -\tau_2\ln\nh$; while for $(\ln \nh)/\xi \ll 1$ the system is in the critical regime, so again
$\overline{S^{\rm PE}_{q}}\to \overline{S^{\rm PE}_{q,c}}$ and $\overline{\cal L}_{2}\to \overline{\cal L}_{2,c}$.
This suggests the asymptotic behaviour
\eq{
{\cal G}^{\rm lin}(x)=\begin{cases}
-D_{q,c} x\, &  : x\gg 1\\
0\, & : x\to 0 
\end{cases}\
}
(with $x\equiv (\ln \nh)/\xi$),  or equivalently
\eq{
{\cal F}^{\rm lin}(x)=\begin{cases}
\exp[{\tau_{q,c} x]}\, & : x\gg 1\\
1\, & : x\to 0 .
\end{cases}\,
}
These asymptotics are likewise  evident in Figs.~\ref{fig:ipr-scaling-xxz},\ref{fig:ipr-scaling-tfi}.

A further outcome of these scalings is that the correlation length $\xi$ can be identified as
$\xi =(1- D_{q}/D_{q,c})^{-1}$ (or equivalently $\xi =(1-\tau_{2}/\tau_{2,c})^{-1}$).
 The divergence of $\xi$ therefore controls how the fractal dimension $D_q$, or the fractal exponent $\tau_2$,
approaches its critical value as the transition is approached from the MBL phase. Recall moreover from Eq.~\ref{eq:tau2-xi} that the Fock-space localisation length, $\xi_F$, is directly related to the fractal exponent $\tau_2$. That $\tau_{2,c}<1$ strictly, automatically means that $\xi_F$ goes to a finite value, $\xi_{F,c}$ at the critical point. In fact, it follows  from Eq.~\ref{eq:tau2-xi} and $\xi =(1-\tau_{2}/\tau_{2,c})^{-1}$ that
\eq{
\frac{1}{\xi_F}-\frac{1}{\xi_{F,c}}~\overset{\xi\gg 1}{\sim}~&\frac{1}{\xi}\left(\frac{\tau_{2,c}\ln 2}{1-2^{-\tau_{2,c}}}
\right)\,,
}
so the divergence of $\xi$  directly controls how $\xi_F$ approaches its finite value at the transition.

Finally, note from the insets in Figs.~\ref{fig:ipr-scaling-xxz},\ref{fig:ipr-scaling-tfi} that on approaching the transition from the ergodic phase the correlation volume diverges as an essential singularity,
\eq{
\Lambda \sim \exp[c/{|\delta W|}^{\mu}]\,\quad : \mu\simeq 0.5\,
}
(where $\delta W = W-W_c$); and on approaching the transition from the MBL side, the correlation length diverges as a power-law,
\eq{
\xi \sim |\delta W|^{-\nu}\,\quad : \nu\simeq 0.8\,.
}
While we do not attach any special significance to the values of the exponents, one fact is by now self-evident: 
consistent with phenomenological real-space approaches~\cite{goremykina2019analytically,dumitrescu2018kosterlitz,morningstar2019renormalization,morningstar2020manybody} discussed in Sec.~\ref{sec:introphenom}, all results from the detailed scaling theory show the MBL transition to be of Kosterlitz-Thouless (KT) type.\footnote{A generic KT transition is one in which the critical point is the terminal point of a line of fixed points characteristic of one phase (in this case the MBL phase).}

As a matter of completeness, we note that an  equivalent scaling theory for the MBL transition can be developed in terms of the imaginary part of the Feenberg self-energy on the Fock-space graph, as was studied in Ref.~\cite{sutradhar2022scaling} using the recursive Green method discussed in Sec.~\ref{sec:recursiveGF}. 
 An identical scaling theory also appears to hold for a quasiperiodic MBL transition~\cite{ghosh2024scaling},  a phenomenon arguably quite different from disorder-induced MBL transitions. Another interesting connection 
is to results obtained from the Aoki decimation on Fock space, discussed in Sec.~\ref{sec:aoki}. 
Indeed, the behaviour of the renormalised  Fock-space matrix elements~\cite{MonthusGarel2010PRB} already carried signatures of their vanishing with an essential singularity on the ergodic side of the transition, and yielded a
power-law diverging  lengthscale on the MBL side of the transition.

Before closing this section, let us simply summarise the important outcomes of the scaling theory.
(i) Eigenstates are fully extended throughout the ergodic phase, but are multifractal both {\it at} the transition and throughout the MBL phase. 
(ii) The ergodic phase is characterised by  a correlation volume, which diverges with an essential singularity on approaching the transition.
(iii) The MBL phase is characterised by a correlation length,  which diverges as a power law on approaching the transition. This divergence  controls how the Fock-space localisation length $\xi_{F}$ (Sec.~\ref{sec:anatomy}) approaches its finite critical value.
(iv)  Fractal exponents such as $D_{q}$ and $\tau_{2}$ jump discontinuously across the MBL transition.
	 As a result of the latter, notice further
the  mean polarisation $\overline{\mathcal{M}_{S}}$ 
 -- a real-space diagnostic for the MBL transition -- is also discontinuous across the transition 
(as follows 
from the relation between $\overline{\mathcal{M}_{S}}$  
and the fractal exponent
$\tau_{2}$, embodied in Eqs.~\ref{eq:tau2-xi},\ref{eq:MSMBL}).


\section{Concluding remarks \label{sec:conclusion}}

Disorder and interactions have long provided cornerstones for understanding a vast range of 
condensed matter phenomena. More recent interest has centred on the dynamics of quantum systems far out of equilibrium, fuelled in no small part by experimental advances, as well as by the fundamental theoretical
and conceptual questions arising.
This has led to the dynamical phase diagram of disordered, interacting systems emerging as a centrepiece of modern condensed matter and statistical physics.
In particular, understanding the MBL phase and the associated quantum phase transtion between ergodic and MBL phases have been two central questions to which  much work has been dedicated. To say that the questions are challenging would be an understatement; not least because the relevant systems are far from equilibrium, 
so the applicability of  `traditional' techniques of quantum many-body theory is not {\it a priori} a given. This has called for new approaches, both numerical and analytical, some microscopic and some phenomenological in nature. One such approach,  strongly rooted in microscopics, has been to address the questions from the 
viewpoint of the Fock space of such systems; this article reviews some of the progress made in recent years
from that perspective.

The genesis of this approach is that the Hamiltonian of an interacting system can be expressed as that of a fictitious single particle on the associated Fock-space graph  (reviewed in Sec.~\ref{sec:models} for some standard models of MBL). With this mapping,  wavefunctions on the Fock-space graph are naturally
insightful quantities to study, particularly in regard to how they break ergodicity in the MBL phase. Section~\ref{sec:anatomy} reviewed several results on the multifractal properties of the eigenstates, and their finer structures, in terms of Fock-spatial correlations. However, as discussed in Sec.~\ref{sec:fscorrs},
this reduction to an effective single particle problem comes at a cost:
the effective disorder on the high-dimensional Fock-space graph is very strongly correlated, making it fundamentally different from Anderson localisation on high-dimensional graphs. 
The complex topology of the Fock-space graph, and the correlations thereon, are manifest in the Fock-space propagators, both local and non-local.  These carry signatures of the MBL phase and the transiton; Sec.~\ref{sec:props} reviewed a range of techniques, analytical and numerical, to calculate them at different levels of approximation. The statistical properties of eigenstate amplitudes on the Fock-space graph point towards the existence of a scaling theory of the transition in terms of Fock-space-based quantities, 
which was discussed in Sec.~\ref{sec:scaling}.

The range of techniques and results reviewed in this article show that the Fock-space landscape of MBL is rather rich and insightful.  Yet the landscape is also vast,  and much remains to be explored and understood. 
The great majority of works so far have studied the structure of  correlations on the Fock space, and their manifestations on the anatomy of individual eigenstates. However, since the MBL phase and the MBL transition are fundamentally dynamical in nature, understanding real-time dynamics on the Fock space is an obvious  and important
direction to explore. This  entails understanding the spreading of the wavepacket of the fictitious single particle  -- in other words probability transport -- on the Fock space~\cite{creed2023probability,pain2023connection,sun2024characterizing,biroli2017delocalised,biroli2020anomalous,yao2023observation}.  A complementary but parallel question would also be to understand  how dynamics on the Fock space are manifest in those of real-space local observables, and vice versa. Such questions are relevant not  only for the MBL phase, but also for the disordered but ergodic phase preceding the MBL transition, where transport is known to be anomalous~\cite{luitz2016anomalous,agarwal2015anomalous,roy2018anomalous}. This may also lead to the emergence of dynamical time or energy scales corresponding to various relaxation processes, which might show universal scaling behaviour near the MBL transition.

While the most basic questions pertaining to dynamics on Fock space concern probability transport, it will also be interesting to understand higher-point dynamical correlations, which might shed light on aspects of 
anomalously slow information scrambling in MBL systems. In particular, it will be  necessary
to understand the class of higher-point dynamical correlations between eigenstate amplitudes, which carry information about quantities such as out-of-time ordered correlators (OTOCs) and entanglement entropies. One can then try to develop a microscopically-based theory for the logarithmic in time growth of entanglement, and the similar spread of the OTOC front~\cite{znidaric2008many,bardarson2012unbounded,serbyn2013universal,nanduri2014entanglement,chen2016universal,chen2017out,huang2017out,deng2017logarithmic}.

Finally, one of the most  obvious yet central challenges is to develop a microscopic theory for the MBL transition. Phenomenologically, it is well understood that the transition is driven by rare many-body resonances eventually leading to an avalanche instability. A question thus arises: how are such processes manifest in 
eigenstate correlations on the Fock space, dynamical or otherwise; and can they provide a microscopic understanding of how multifractality in the MBL phase gives way to ergodicity 
on crossing the transition? The fact that MBL eigenstates are strongly inhomogeneous on the Fock space~\cite{roy2021fockspace}, and that methods from statistical mechanics can be borrowed to better understand the anomalously large amplitudes stemming from the resonances~\cite{tarzia2020manybody,biroli2024largedeviation},
 hints at this direction holding considerable promise.


\begin{acknowledgements}
We thank S. Banerjee, J. T. Chalker, I. Creed, A. Duthie, S. Garratt, S. Ghosh, A. Lazarides, S. Mukerjee, B. Pain, J. Sutradhar and S. Welsh for collaborations on several works, and many insightful discussions 
related to the subject of this review. We are also grateful to N. Laflorencie and F. Alet for sharing with us
their raw data from Ref.~\cite{mace2019multifractal}.
SR acknowledges support from SERB-DST, Government of India, under Grant No. SRG/2023/000858, from the Department of Atomic Energy
under Project No. RTI4001, and from an ICTS-Simons Early Career Faculty Fellowship via a grant from the Simons Foundation (Grant No. 677895, R.G.).
DEL acknowledges support from the Infosys Foundation, and from EPSRC (UK) under Grant No. EP/S020527/1.
\end{acknowledgements}

\bibliography{refs}

\end{document}